\documentclass[final,3p,times]{elsarticle}

\usepackage{graphicx}

\usepackage{amssymb}
\usepackage{amsmath}
\usepackage{subfig}
\usepackage{graphicx}
\usepackage{url}
\usepackage{algorithm}
\usepackage{algpseudocode}
\usepackage{multirow}
\usepackage{wasysym}
\usepackage{marvosym}
\usepackage{color, xcolor, colortbl}
\usepackage{array, booktabs}
\usepackage{xspace}
\usepackage{enumitem}

\usepackage{bm}
\definecolor{light}{rgb}{0.8,0.8,0.8}
\definecolor{medium}{rgb}{0.6,0.6,0.6}
\definecolor{dark}{rgb}{0.4,0.4,0.4}
\definecolor{darkmed}{rgb}{0.3,0.3,0.3}
\definecolor{darkest}{rgb}{0.2,0.2,0.2}
\definecolor{Black}{rgb}{0,0,0}
\definecolor{White}{rgb}{1,1,1}
\definecolor{lightpurple}{rgb}{0.78823,0.709803,0.74509}
\definecolor{lightpurpletext}{rgb}{0.788235,0.5529411,0.658823}
\definecolor{skyblue}{rgb}{0.80392,0.866666,0.92941}
\definecolor{skybluetext}{rgb}{0.61568627,0.7647058,0.913725}
\definecolor{darkgreen}{rgb}{0.3137254,0.458823,0.18431}
\definecolor{foliagegreen}{rgb}{0.188,0.415,0.105}
\definecolor{steelbluegrey}{rgb}{0.1961,0.2353,0.2392}
\definecolor{highlightblue}{rgb}{0.4078,0.6431,0.85}
\definecolor{matlabblue}{rgb}{0,0.2705,0.85}
\definecolor{darkred}{rgb}{0.8,0.1725,0}
\definecolor{fireenginered}{rgb}{0.505,0.1411,0}
\definecolor{darkpurple}{rgb}{0.6431,0.3137,0.8509}
\definecolor{gaylordpurple}{rgb}{0.416,0.204,0.549}
\definecolor{deludedorange}{rgb}{0.7409,0.4392,0}
\definecolor{darksalmon}{rgb}{0.9137,0.411,0.706}

\newcommand{\secref}[1]{Section \ref{#1}}
\newcommand{\fref}[1]{Fig.~\ref{#1}}
\newcommand{\tref}[1]{Table~\ref{#1}}

\renewcommand{\eqref}[1]{Eq.~(\ref{#1})}
\newcommand{\eref}[1]{(\ref{#1})}

\newcommand{\mat}[1]{\textrm{\textbf{#1}}}

\newcommand{\psif}{\bm{\psi} (\bm{r}, \bm{\Omega})}
\newcommand{\psifd}{\bm{\psi} (\bm{r}, \bm{\Omega}')}

\newcommand{\xten}[1]{$\times$ 10$^{\text{#1}}$}

\newcolumntype{a}{>{\columncolor{light}}c}
\newcolumntype{b}{>{\columncolor{skyblue}}c}

\begin{document}

\begin{frontmatter}



\title{Scalable angular adaptivity for Boltzmann transport}
\author[AMCG]{S. Dargaville}
\ead{dargaville.steven@gmail.com}
\author[QMU]{A.G. Buchan}
\author[AWE,AMCG]{R.P. Smedley-Stevenson}
\author[AMEC,AMCG]{P.N. Smith}
\author[AMCG]{C.C. Pain}
\address[AMCG]{Applied Modelling and Computation Group, Imperial College London, SW7 2AZ, UK}
\address[QMU]{School of Engineering and Material Sciences, Queen Mary University of London, E14 NS, UK}
\address[AWE]{AWE, Aldermaston, Reading, RG7 4PR, UK}
\address[AMEC]{ANSWERS Software Service, Wood PLC, Kimmeridge House, Dorset Green Technology Park, Dorchester, DT2 8ZB, UK}
\begin{abstract}
This paper describes an angular adaptivity algorithm for Boltzmann transport applications which for the first time shows evidence of $\mathcal{O}(n)$ scaling in both runtime and memory usage, where $n$ is the number of adapted angles. This adaptivity uses Haar wavelets, which perform structured $h$-adaptivity built on top of a hierarchical P$_0$ FEM discretisation of a 2D angular domain, allowing different anisotropic angular resolution to be applied across space/energy. Fixed angular refinement, along with regular and goal-based error metrics are shown in three example problems taken from neutronics/radiative transfer applications. We use a spatial discretisation designed to use less memory than competing alternatives in general applications and gives us the flexibility to use a matrix-free multgrid method as our iterative method. This relies on scalable matrix-vector products using Fast Wavelet Transforms and allows the use of traditional sweep algorithms if desired. 
\end{abstract}


\end{frontmatter}
\section{Introduction}
\label{sec:Introduction}
Calculating deterministic solutions to Boltzmann-type equations that govern particle transport through an interacting medium is important in many multi-physics applications, including simulating the distribution of neutrons in nuclear reactors, determining radiative heat transfer in the atmosphere, charged particle transport and modelling waves on the sea surface (spectral wave modelling). These transport processes are characterised in terms of temporal (1D), spatial (1D, 2D or 3D), angular (1D or 2D) and energy/frequency dimensions (1D), giving between 5 to 7 dimensions depending on the application. These transport problems are written as propogation/advection terms in each of the dimensions, along with a combination of (possibly) non-linear interaction/source terms. 

These problems are challenging to solve for several reasons. If the advection terms in each space are dominant (e.g., neutron propogation in a vacuum, long-distance wave propogation in deep water), the system is hyperbolic and stable discretisations must be used in each space. If the interaction terms are large, the system can becomes parabolic/elliptic and hence the discretisations used must be robust to this, given the relative sizes of these terms varies across space/angle/energy/time. Of particular importance is the large size of the phase-space when discretised (formed from the tensor product of the spaces mentioned above); this means that underesolved discretisations are very common in Boltzmann transport and showing that a discretisation has converged can be laborious. Depending on the nature of the interaction terms and discretisations used, very low memory methods for solving the discretised system can be used, which store $\ll 1$ copy of the solution in memory. Unfortunately these methods cannot be used for general Boltzmann transport problems, and the large phase-space means that storing even one copy of the solution for common discretisations can exceed the available memory on modern computers. 

Given the hyperbolic advection terms, many Boltzmann transport problems have strong discontinuities in the angular dimension, forcing the use of low order angular discretisations for stability combined with $h$-refinement in order to adequately resolve. Obtaining accurate solutions can therefore require millions/billions of discretised cells/points in the angular domain. A large class of these problems only require this high resolution on limited areas of the angular domain (or have differing resolution requirements across space/energy). Adaptivity can therefore play an important role in helping overcome the size of the phase-space by focusing resolution only where important in a problem, reducing the size of the discretised system while providing accurate solutions.  Our goal in the Applied Modelling and Computation Group (AMCG) has been to develop automated adaptive methods for general deterministic Boltzmann transport problems that are truly practical. Previously in the AMCG, we have investigated the use of spatial adaptivity, angular adaptivity, and combined space/angle adaptivity \cite{Buchan2005, Buchan2006, buchan_linear_2008, Buchan2008244, Baker2013, goffin_minimising_2013, Goffin2014, Goffin2015, Goffin2015a, Adam2016a, Adam2016, Soucasse2017}. 

There are two key innovations we describe in this paper, tested on a range of Boltzmann transport problems in two and three spatial dimensions, with a 2D angular domain, drawn from nuclear/radiative transfer applications. The first is the use of anisotropic angular adaptivity, with a form of structured $h$-adaptivity in the angular domain using Haar wavelets, that focuses resolution only where required. This anisotropic resolution varies across space, allowing very high accuracy solutions to be computed with very few unknowns. This technology is very flexible and shares important similarities with existing discretisation technology in angle, allowing it to be used with established space/energy discretisations and low-memory iterative methods that are available in some applications. The use of wavelets gives the adaptivity a direct connection to importance and smoothness (through regular and goal-based error metrics). The second innovation is the use of a stable spatial discretisation that has been designed to use less memory than competing alternatives, advancing the state-of-the-art in applications which cannot use the low-memory methods mentioned above.

The combination of our anisotropic wavelet adaptivity and spatial discretisation therefore produces a general ``low-memory'' discretisation technology. This provides the capability to solve very challenging problems to high accuracy and pushes the boundary of numerical technology for general Boltzmann transport problems. We should note, that for problems with anisotropic features, ``high accuracy'' can often mean one decimal place of accuracy at best, after heavy angular refinement. Such problems are not purely hypothetical; particle accelerators for example, feature transport of particles down narrow ducts, with a radius in centimetres but lengths of 10-100 metres. The solid angle subtended by the target on the source can be made arbritrarily small by extending the length of the accelerator, requiring arbritrarily fine resolution in angle. Applying insufficient resolution in angle means that significant streaming paths will be completely ignored, as ``ray-effects''/``the garden-sprinkler effect'' mean the solution will only be non-zero in spatial regions which align with the angular discretisation. 

Furthermore, traditional techniques for resolving these problems are highly specialised and can break down. To help ameliorate ray-effects without increasing the angular resolution, additional diffusion can be added in space \cite{Booij1987}, dependent on the discrete resolution applied in angle (see \cite{Tolman2002} for related techniques). This is only viable over small distances with pure advection, as the spatial distance between two ray-effects increases as the rays propogate, and hence requires knowing the maximum distance a particle can travel, or keeping track of the age of each particle. 

Problems where the solution is required everywhere throughout the domain, or where there are numerous sources prohibit the use of direct integration through techniques like ray-tracing or Monte-Carlo. Two examples of such problems include radiative transfer where the entire domain is above absolute-zero and therefore emitting photons, or in a spectral-wave problem where waves are constantly created throughout the entire domain by wind. Direct integration through techniques like Monte-Carlo methods can be very efficient when integrating over a small part of the phase-space (e.g., determining the flux in a small volume at the end of a duct), however they requiring biasing the sampling of particles from parts of the phase-space ``important'' to the answer. \textit{A priori} information can sometimes be used, for example, in accelerators, where a direct line of sight provides the major contribution to the flux at the end of the duct. This can be used to bias the sampling in both space and angle in Monte-Carlo methods. This is similar to using a refined spatial mesh and a biased/rotated discretisation in angle with deterministic methods.

For problems with a limited number of directional changes for the particles (e.g., scattering/reflections off the edges of an ``L'' shaped duct), calculations can be run in stages, where the solution from the first straight section of the duct is used to form a source for the second section (this still relies on biasing in direction to send particles down each section of the duct preferentially). Some particle accelerators feature these types of ``kinks'' in the beamline in order to discriminate between particles of different energies. 

For sufficiently complex problems however, one cannot \textit{a priori} determine important regions in the phase-space, or seperate a problem into stages. Monte-Carlo methods typically use ``variance-reduction'' methods to bias their sampling in these cases, however these rely on using a determinstic solution to compute an ``importance-map'' (very similar to goal-based error metrics) to inform this biasing. Unfortunately, this puts us back where we started, relying on the ability to compute a deterministic solution in problems with significant directionality, without using \textit{a priori} information on ``important'' regions of the phase-space, material properties or relative sizes of interaction/source terms. As such, for difficult problems, there is no practical solution algorithm besides using a deterministic code with angular adaptivity that can perform arbitrary refinement, with error metrics that are robust in the presence of ray-effects.

As such, this paper presents a first-step towards this goal and begins with a review highlighting the unique features in deterministic Boltzmann transport that combine to form scalable (runtime and memory) algorithms for solving the BTE with uniform angular resolution. The work we present is then guided by these concerns and focuses on the practical aspects of building a scalable adaptive algorithm in angle. Previously, we have used both adaptive Haar wavelets in angle and the aforementioned spatial discretisation in Boltzmann transport problems \cite{Adam2016a, Adam2016, Soucasse2017}, but those works have focused on the application to specific problem domains and did not use the non-standard Haar discretisation and Fast-Wavelet-Transforms that are key features of this work. Very little of the existing literature on angular adaptivity for Boltzmann-type problems shows either the runtimes or memory use; those that do feature only a few levels of refinement in their angular discretisations. As such, we focus not only on describing the key aspects of our methods, but also how they are practically implemented; to our knowledge this work is the first that shows scalable, angular adaptivity for solving Boltzmann transport problems, with fixed refinement and both regular and goal-based error metrics. 
\section{Boltzmann Transport Equation}
\label{sec:Boltzmann Transport Equation}
In this work we consider the transport of neutral particles (neutrons and photons) governed by the  Boltzmann Transport Equation (BTE), which is commonly used in nuclear/radiative transfer applications. In many applications, the BTE is more complex (e.g., advection in angle/energy, non-linear source terms), but the form below contains the basic terms included in many applications; where needed we discuss difficulties caused by more general forms of the BTE throughout the paper. Without loss of generality, we write the mono-energetic steady-state BTE in first-order form as the integro-differential equation
\begin{equation}
\bm{\Omega} \cdot \nabla_{\bm{r}} \psif + \Sigma_\textrm{t} \psif - S(\psif)  = S_{\textrm{e}}(\bm{r}, \bm{\Omega}),
\label{eq:bte}
\end{equation}
where $\psif$ is the angular flux in direction $\bm{\Omega}$, at spatial position $\bm{r}$. The macroscopic cross sections define the material that particles are moving through and $\Sigma_\textrm{t}$ is the total cross-section. The interaction/source terms have been separated into, $S$, which is dependent on $\psif$, and those which can be considered purely ``external'', $S_\textrm{e}$. We only consider the first-order form of the BTE in this work; other forms (e.g., second order even-parity, FOSILS) are often used but their use is limited to specific parameter regimes (e.g., second-order forms cannot handle vacuum regions).

For the majority of neutron/photon transport problems, the interaction/source term $S$ is linear (making \eref{eq:bte} linear) and describes the scattering from angle $\bm{\Omega}'$ into angle $\bm{\Omega}$ as particles interact with the medium they are propogating in, normally written as
\begin{equation}
S(\psif) = \int_{\bm{\Omega}'} \Sigma_\textrm{s} (\bm{r}, \bm{\Omega}' \rightarrow \bm{\Omega}) \psifd \textrm{d}\bm{\Omega}',
\label{eq:scatter}
\end{equation}
where $\Sigma_\textrm{s}$ is the macroscopic scatter cross-sections. Importantly, these scatter cross-sections, $\sigma_{\textrm{s}}$, are typically tabulated in extensive nuclear data libraries. These libraries contain cross-sections for many different elements/isotopes/materials, expressed as the coefficients of Legendre polynomials on the sphere. Computing the interaction/source term in \eref{eq:bte} therefore requires mapping the angular flux $\bm{\psi}$ into Legendre moments. We denote this mapping operator as $\mat{D}_{\textrm{m}}$ and write 
\begin{equation}
\bm{\phi} = \mat{D}_{\textrm{m}} \bm{\psi}; \qquad \bm{\psi} = \mat{M}_{\textrm{m}} \bm{\phi}
\label{eq:map}
\end{equation}
where $\bm{\phi}$ is the solution expanded in Legendre moments and $\mat{M}_{\textrm{m}}$ is the mapping operator from moments back to $\bm{\psi}$. The first moment is simply a constant function on the sphere and is known as the scalar flux (this is the average flux across the sphere). We can write the discretised form of \eref{eq:bte} as
\begin{equation}
\begin{pmatrix}
  \mat{L} & \mat{M}_{\textrm{m}} \mat{S} \\
  \mat{D}_{\textrm{m}} & \mat{I}
 \end{pmatrix}
\begin{pmatrix}
  \bm{\psi} \\
  \bm{\phi} 
 \end{pmatrix}  
 =
\begin{pmatrix}
  \mat{b} \\
  \mat{0} 
 \end{pmatrix} 
\label{eq:disc_bte}
\end{equation}
where $\mat{L}$ is the streaming/collision term, $\mat{S}$ contains the scatter cross-sections (as Legendre coefficients) and $\mat{b}$ is the external source. In general, solving \eref{eq:disc_bte} can be very difficult and storing even one copy of $\bm{\psi}$ can easily exceed available memory given the size of the phase-space. We can rewrite \eref{eq:disc_bte} explicitly in terms of $\bm{\phi}$ to give
\begin{equation}
(\mat{I} - \mat{D}_{\textrm{m}} \mat{L}^{-1} \mat{M}_{\textrm{m}} \mat{S}) \bm{\phi}  = -\mat{D}_{\textrm{m}} \mat{L}^{-1} \mat{b}.
\label{eq:moment_bte}
\end{equation}
Typically, a Richardson iteration (known as source iteration in the nuclear community) or Krylov methods are used to solve \eref{eq:moment_bte}. In this work, we only consider fully-implicit methods for solving \eref{eq:moment_bte}; many communities that use BTE-type systems have spent considerable time developing efficient operator-splitting methods (e.g., operational models in spectral wave modelling), but these require considerable tuning.

Solving for the moments, $\bm{\phi}$, in \eref{eq:moment_bte} requires far less memory than solving for $\bm{\psi}$ and the angular flux, $\bm{\psi}$ can be easily reconstructed using \eref{eq:moment_bte} and \eref{eq:map}. Importantly, both \eref{eq:moment_bte} and this reconstruction rely on inverting $\mat{L}$. Careful choice of the space/angle discretisations are therefore essential to ensure efficient inversion of $\mat{L}$. Furthermore, the discretisation used in \eref{eq:disc_bte} must be stable in the presence of strong discontinuities in space/angle (e.g., in hyperbolic regions) and robust in diffusive regions. The method used to invert $\mat{L}$ must be matrix-free and use very little memory (typically $\ll 1$ copy of $\bm{\psi}$) and exhibit sufficient parallelism to scale to the largest supercomputers. Perhaps the only existing technology that satisfies these constraints uses Discontinuous-Galerkin finite-element methods (DG) to discretise in space with upwinding on a structured grid and S$_n$ in angle (method of characteristics can also be competitive, but storing tracks can become prohibitively expensive in 3D).
\section{Background}
\label{sec:Background}
A number of authors have investigated using angular adaptivity in Boltzmann transport applications. We loosely classify these into four main categories: sparse-grid methods \cite{Grella2013, Grella2013}; adaptive S$_n$ quadratures \cite{Stone2008, Jarrell2010, Jarrell2011, Lau2016, Lau2017, Zhang2018}; $h$ adaptive FEMs \cite{Koeze2012, Kophazi2015} and adaptive wavelet methods \cite{Watson2005Thesis, Buchan2006, Buchan2008244, Goffin2015, Goffin2015a, Adam2016, Adam2016a, Soucasse2017}. We do not discuss $p$ adaptive FEMs here, as we are focused on problems with discontinuities in the angular domain.
\subsection{Sparse-grid methods}
\label{sec:Sparse-grid methods}
Sparse grid methods are an attempt to overcome the ``curse of dimensionality'' that plague discretisations formed from tensor product spaces. They assume that only parts of the tensor product space contribute significantly to the solution, namely those formed by independent refinement of each space. For example, the combination of fine spatial discretisation and coarse angular discretisation, along with coarse spatial and fine angular are considered important, but not fine spatial discretisation and fine angular discretisation. \cite{Grella2013} gives an excellent explanation of sparse-grid methods along with theoretical and experimental convergence results for P$_n$ and P$_0$ FEM in angle, though they do not adapt their sparse grids. \cite{Widmer2008} however, combined sparse-grid methods with Haar wavelet adaptivity for non-scattering problems in radiative transfer.

Unfortunately, sparse-grid methods assume a significant degree of regularity; many Boltzmann transport problems feature step functions in space/angle simultaneously. If we define a simple problem with particles streaming in from a boundary, but with an obstruction along half of the incoming boundary, the solution features a step function that persists in both space/angle (i.e., a shadow). This makes sparse grid methods not well suited in problems that this paper targets. 
\subsection{Adaptive S$_n$}
\label{sec:Adaptive S$_n$}
Much of the existing literature use locally refinable S$_n$ quadratures (as traditional quadratures face difficulties like negative weights with refinement) combined with high-order interpolation between (a limited number) of regions with differing angular refinement. Naturally this leads to difficulties when trying to refine around regions of angular discontinuities, which is what this work targets. \cite{Stone2008} form a product quadrature and use regular adaptivity performed by thresholding based on the magnitude of the flux values. They divide their spatial domain into a limited number of regions which use quadratures of different order. Their interpolation strategy leads to Gibbs oscillation, which can lead to spurious unlimited refinement in discontinuous problems. 

\cite{Jarrell2010, Jarrell2011} use locally refinable quadratures that can adapt anisotropically on the sphere with fixed angular refinement, built from several different orders of basis functions. They also perform adaptivity only in a limited number of regions; when the authors allow each cell to have its own adapted quadrature, they found their interpolation error becomes significant. \cite{Lau2016,Lau2017} follows on from this work and try to improve the interpolation scheme used between areas of differeing angular resolution. This interpolation is designed to preserve a certain number of moments in smooth regions of the solution. Outside of these smooth regions, they employ an optimisation scheme to try and ``fix-up'' the interpolation to preserve 0th order moments. Finally \cite{Zhang2018} also divide their spatial domain into regions and perform goal-based angular adaptivity based on a nested quadrature. They perform interpolation by fitting high-order spherical harmonics to portions of each octant and show results up to 6 levels of refinement.  
\subsection{Adaptive FEMs}
\label{sec:Adaptive FEMs}
The application of adaptive FEMs (where a wavelet basis is explicitly not used) is limited to only a few authors. \cite{Koeze2012} built a combined space/angle adaptive FEM algorithm, using both P$_0$ and linear basis functions. The authors only investigated problems with one spatial dimension, though they do use both regular and goal-based error metrics. \cite{Kophazi2015} is perhaps the most sophisticated paper in the literature which uses angular adaptivity in Boltzmann-transport applications. They studied the use of 5 different basis on a hierarchical triangular discretisation of the sphere in neutronics applications, including P$_0$ and linear. Rather than use regular or goal-based adaptivity, they simply show their method working with fixed angular refinement, up to 6-7 levels of refinement. They perform block-sweeps on their octants as part of their iterative method, and they do not show the runtime of their algorithm. \cite{Favennec2019} allow unstructured adaptivity on their angular domain, however they use the same adapted angular discretisation across their entire spatial domain.
\subsection{Adaptive wavelets}
\label{sec:Adaptive wavelets}
A handful of authors have investigating using adaptive wavelets in angle for Boltzmann transport problems. \cite{Watson2005Thesis} investigated Haar wavelets in azimuth and discrete ordinates in the polar dimension for neutronics applications. Given this, the authors can only adapt in azimuth and investigated problems limited to two spatial dimensions. 

The other authors using wavelet discretisations have been based in the AMCG over the past decade; \cite{Buchan2006, Buchan2008244} used adaptive spherical wavelets, also in neutronics applications, but only up to 4 levels of refinement. These wavelets are continuous across octants, making it difficult to scalably apply boundary conditions. \cite{Goffin2015, Goffin2015a} followed on from this work by using adaptive linear octahedral wavelets, based of the uniform description in \cite{Buchan2006, Buchan2008244}. They also only used up to 4 levels of refinement, and again these wavelets are continuous across the octant. \cite{Goffin2015, Goffin2015a} however first introduced the use of goal-based angular adaptivity in these wavelet problems. \cite{Adam2016, Adam2016a} used adaptive Haar wavelets in spectral wave modelling, where the angular domain is one-dimensional. They showed up to 8 levels of refinement, but their iterative method is matrix-based, making scaling studies difficult with high levels of refinement. Finally, \cite{Soucasse2017} used the same adaptive Haar framework we describe here in a coupled radiative transfer application. This work focused largely on a novel goal-based metric for heat deposition by photons in coupled problems and used the standard Haar decomposition described in \secref{sec:Wavelets}. They also only showed up to 4 levels of refinement. These works have all showed the potential of adaptive wavelet discretisations in angle, but have not focused on their scalable application. 

The first clear advantage of using an adapted wavelet space is that we never need to perform interpolation between angular/energy regions with differing resolution, as wavelet basis are hierarchical. The second and key feature which motivates our use of a wavelet space is the ``norm-equivalence'' and ``cancellation properties'' (using the nomenclature of \cite{Cohen}) of common wavelet expansions. Norm-equivalence is a consequence of the orthonormality of the wavelets and can be used to show a direct relationship between the norm of the wavelet coefficient and the norm of the function represented by the expansion. This means that wavelets with small coefficients only contribute small perturbations to the function norms. Furthermore, the cancellation property results in wavelet coefficients that are small if the function is smooth over the support of the wavelet (smooth up to the order the scaling functions can represent exactly and small proportional to the order of smoothness). The combination of these properties means that a simple thresholding of our wavelet coefficients is sufficient to drive our angular adaptivity. Wavelets are removed from the expansion if their coefficient is small, with the children of a wavelet added to the expansion if the coefficient is large. This results in refinement of the angular domain in areas of large importance and in areas close to discontinuities.

The ability to threshold removes the need for a heuristic to guide the adaptivity. For example, if we have smooth regions in angle with large flux that are well represented by a small number of angular elements and a highly discontinuous region with equivalently large flux, an adaptive S$_n$ or FEM/FVM scheme driven by the magnitude of the flux would chose to refine both areas equally. Preventing this would require some measure of smoothness (e.g., a derivative) combined with the magnitude of the flux across the sphere. In a sense, one can view the mapping to an equivalent wavelet space and subsequent thresholding as a metric that combines both importance and smoothness for an adaptive FEM/FVM scheme. 

Wavelet methods can be considered as FEM methods which use wavelet basis functions in their expansion (this is discussed further in Section \ref{sec:Wavelets}). This fact is key to the scalability of the algorithm in this paper. If we consider the scalability of a general FEM/FVM angular discretisation, we must also consider several points. First, we write the entry in the $i^\textrm{th}$ row and $j^\textrm{th}$ column of the angular tables/matrices for our angular expansion as $\mat{A}_{i,j} = \int_{\bm{\Omega}} G_i G_j \, \textrm{d}\bm{\Omega}$. For general angular basis functions, $G$, across the sphere, these angular matrices may not be sparse and hence their storage may not scale linearly with angle size. Furthermore, there is coupling between unknowns on angular elements with FEM/FVM which presents problems when determining inflow/outflow across the boundaries of spatial elments. Typically some form of discontinuous formulation in angle is required to decouple the angular elements and ameliorate this point. This coupling also affects the sweepability of FEM/FVM schemes; the individual points used in a collocation scheme (S$_n$) can always be independently swept, whereas the coupling between general basis functions results in (at best) block-sweeps of the unknowns on an angular element (if DG functions are used). In regimes with strong discontinuities on the sphere, this is not a large disadvantage, as low-order basis functions would be used with large numbers of elements for stability. These points are discussed further below; we now introduce the wavelet basis that we use to discretise the angular domain.
\section{Wavelets in angle}
\label{sec:Wavelets}
\begin{figure}[th]
\centering
\includegraphics[width=0.4\textwidth]{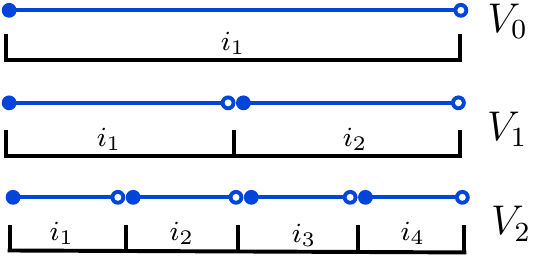}
\caption{Nested spaces $V_j$ with constant scaling functions (in blue) on each partition, $i_n$.}
\label{fig:v_j}
\end{figure}
\begin{figure}[th]
\centering
\subfloat[][$u=0$ and hence $V_2 = V_0 \oplus W_0 \oplus W_1$]{\label{fig:v_l_0}\includegraphics[width =0.425\textwidth]{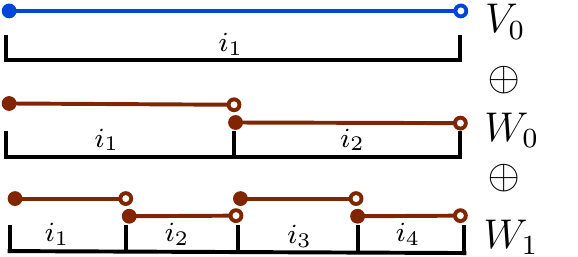}}\hspace{0.5cm}
\subfloat[][$u=1$ and hence $V_2 = V_1 \oplus W_1$]{\label{fig:v_l_1}\includegraphics[width =0.425\textwidth]{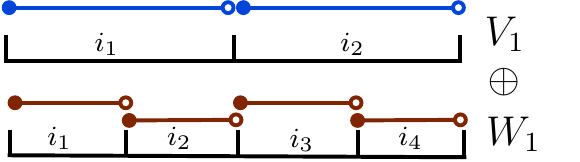}}\\
\caption{Equivalent representations of $V_2$ (see Figure \ref{fig:v_j}) with a Haar wavelet expansion given different values of $u$. The constant scaling functions are shaded in blue, the wavelet functions are in red}
\label{fig:v_l}
\end{figure}

We begin with a basic overview of wavelet theory and how wavelets are constructed, through a multiresolution analysis (MRA). Consider a function $f \in L^2(\mathbb{R})$, on the real line, and a nested sequence of subspaces $V_j$ (where $V_j \subseteq V_{j+1}$ and $\cup_j V_j$ is dense in $L^2$). If we write $n$ scaling functions on each level $j$, $\omega_{j,n}$, (whose set forms a Riesz basis of $V_j$), we can approximate $f$ in $V_j$ as a linear combination of these scaling functions. Thus far, we have simply approximated a function on a series of nested partitions of the real line, where we will denote each of the individual partitions on level $j$ as $i_n$ (where $n \in [1, 2^j]$), see Figure \ref{fig:v_j}. We will now consider our wavelet space, $W_j$ that complements $V_j$, i.e., $V_{j+1} = V_j \oplus W_j$. Given that $W_j \subset V_{j+1}$, we must be able to construct the $k$ wavelet basis functions in $W_j$, $\tau_{j,k}$, from the scaling functions $\omega_{j+1,n}$. We can recursively apply the definition of our wavelet space on the nested partition to give a hierarchy of wavelet spaces, down to whichever level of $V_j$ we chose, namely $u$, and hence $V_j = V_u \oplus_{n=u}^{j-1} W_n$. Our wavelet representation of $f$ on level $j$, with $k$ wavelet functions on a given level $m$, with $n$ scaling functions on a ``base'' level $V_u$ is therefore given by
\begin{equation}
f \approx f_j = \sum_n \alpha_{u, n} \omega_{u,n} + \sum_{m=u}^{j-1} \sum_k \beta_{m,k} \tau_{m,k}, 
\label{eq:wavelet_expand}
\end{equation}
where $\alpha_{u, n}$ and $\beta_{m,k}$ are the expansion coefficients for the scaling and wavelet functions, respectively. The framework for building wavelets described above is very general; we have not actually specified the scaling or wavelet functions. There are many choices available, however in this work, we are deliberately going to restrict ourselves to using Haar wavelets, whose scaling functions are given by the constant function on each partition, or 
\[
\omega_{u,n}(x) =     
\begin{cases}
 1, & \text{if}\ x \in i_n \\
 0, & \text{otherwise}
\end{cases},
\]  
and the wavelet functions are defined as
\[
\tau_{m,k}(x) =     
\begin{cases}
 1, & \text{if}\ x \in i_{2k-1} \\
 -1, & \text{if}\ x \in i_{2k} \\
 0, & \text{otherwise}
\end{cases}.
\]  
Haar wavelets are perhaps the simplest wavelet one could construct and have a number of key properties. Both the scaling and wavelet functions have compact support (above $V_u$ and $W_u$), while the wavelet functions have a zero integral over the domain. Furthermore, the coefficient associated with the scaling function on an individual partition is the average of the $f$ over that partition. Figure \ref{fig:v_l} shows how we can use two different Haar wavelet expansions to represent $V_2$ exactly (see Figure \ref{fig:v_j}), each of which has been built on top of a different base level of $V_u$. In the rest of this work, we refer to the ``children'' of a wavelet on level $j$ as the wavelets on level $j+1$ that share both the same support of the ``parent'' (in 1D our wavelets can have 2 children) and the same wavelet function.

One of the key features of a wavelet representation is that the expansion is hierarchical given their multiresolution construction; moving between coarser/finer representations is as simple as adding/removing wavelets on a given level. For example, if we wish to project the representation $V_2 = V_0 \oplus W_0 \oplus W_1$ (see \fref{fig:v_l_0}) onto $V_1$ (i.e., coarsen it), we can simply set the wavelet coefficients in $W_1$ to zero (i.e., $\beta_{1,k}=0$), giving $V_2 = V_0 \oplus W_0$. Performing this coarsening directly from $V_2$ to $V_1$ without a wavelet representation would require some form of interpolation.

To form an approximation on the surface of a circle (for applications with a 1D angular domain, like spectral waves), we can consider a (periodic) wavelet expansion on the circumference of the circle. We must choose a ``base'' level, $u$, for our expansion; a natural choice is the space $V_2$, which is simply the circle discretised with a constant basis function on each quadrant. \cite{Adam2016a, Adam2016} used Haar wavelets to discretise the 1D angular domain found in spectral wave modelling, though they also allowed the base level of their expansion to change. This can be a useful parameter in spectral wave modelling, as many of the (non-linear) source terms are actually simplified terms derived from integral expressions across the phase-space. These source terms are often parameterised in terms of a fixed angular discretisation, and hence it can be useful to match $u$ with the parameterisation used. This allows these source terms to be included without modification, with refinement capturing further detail in other operators (e.g., the advection operator) and the hierarchical structure of wavelet formulations allowing both to be combined without interpolation.

With a 2D angular domain, there are several choices in equivalent Haar wavelet formulations. The ``standard'' decomposition is formed from the tensor product of two 1D expansions, which feature long, thin wavelets (that are local only in each of the two angular dimensions independently and hence span an increasing number of partitions with refinement) in each dimension as it is refined. This is in comparison to the ``non-standard'' decomposition that is more efficient to compute and has fixed sparsity (each wavelet spans 4 partitions) on each level. ``Interleaved'' decompositions can also be formed, see \cite{Kopp1998} for a detailed comparison of these different wavelet expansions. Each decomposition can exactly represent the same $V_j$ space, but the schemes can differ in compressive ability if used as part of a compression/adaptive scheme. For example, if used on a 2D angular domain with a 2D spatial domain, one would expect the long, thin wavelets found in the standard decompositon to be able represent azimuthal symmetry with fewer wavelet functions than the non-standard. In contrast, for a 2D angular domain with a 3D spatial domain, with highly localised flux in angle, the long, thin wavelets of the standard decomposition are a disadvantage, as the fixed sparsity of the non-standard decomposition lend themselves towards localised refinement.

In this work, we focus mostly on the use of the non-standard decomposition, as we are focused on problems with highly localised flux in angle. Considering the tensor product of two 1D expansions, we apply the ``base'' level $V_2$ in the azimuthal dimension and $V_1$ in the polar (or $V_0$ in the polar to give the half-sphere if needed); we denote this lowest level wavelet discretisations as H$_1$. Using the quadrants/octants as our coarsest discretisation also preserves the angular symmetries found when using a structured spatial mesh. Adding wavelets of increasing levels corresponds to subdivision of the polar and azimuthal dimensions at the mid-way point of each partition, for example, we denote a single level of refinement from H$_1$ as H$_2$. This is the simplest form of hierarchical refinement and is traditionally considered a poor discretisation of the sphere, given elements will cluster around the poles. We will see in later sections this is not a significant disadvantage in an adaptive scheme, and also has performance implications when considering anisotropic scatter.

\begin{figure}[th]
\centering
\subfloat[][Equivalent P$_0$ DG-FEM/cell-centred FVM hierarchical discretisation. The boundaries of angular elements are defined with constant azimuthal/polar values. Grey dots indicate a constant function in each angular element.]{\label{fig:pq_sphere}\includegraphics[width =0.4\textwidth]{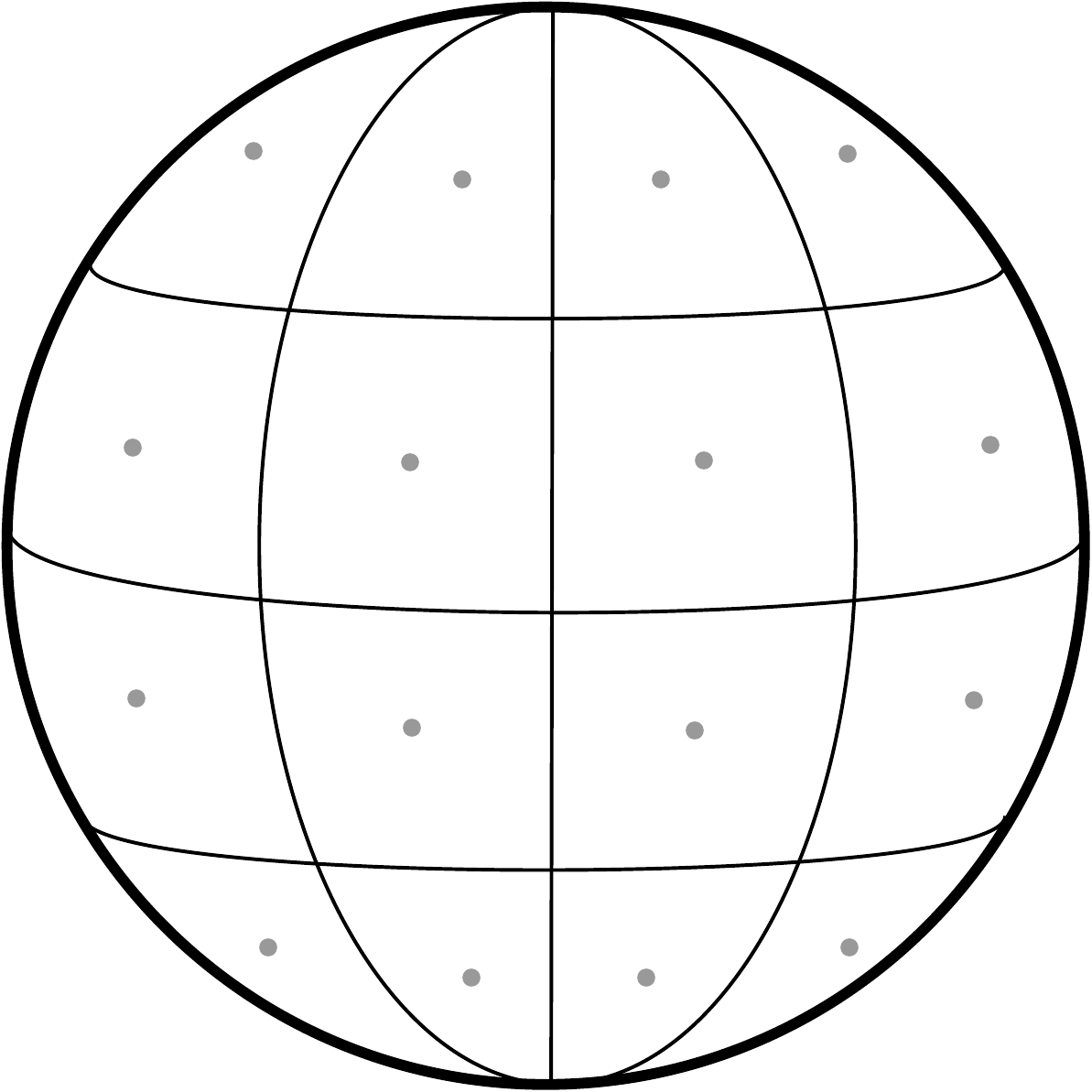}} \hspace{0.5cm}
\subfloat[][Haar wavelet angular basis functions for H$_{2}$ on a single octant (mapped to a square for visibility). We have one constant scaling function, and three wavelet functions. Given the hierarchical basis, our expansion could be coarsened down to H$_{1}$ by simply removing all but the top left function (i.e., the constant scaling function) from an expansion of the octant.]{\label{fig:wavelet_functions}\includegraphics[width =0.3\textwidth]{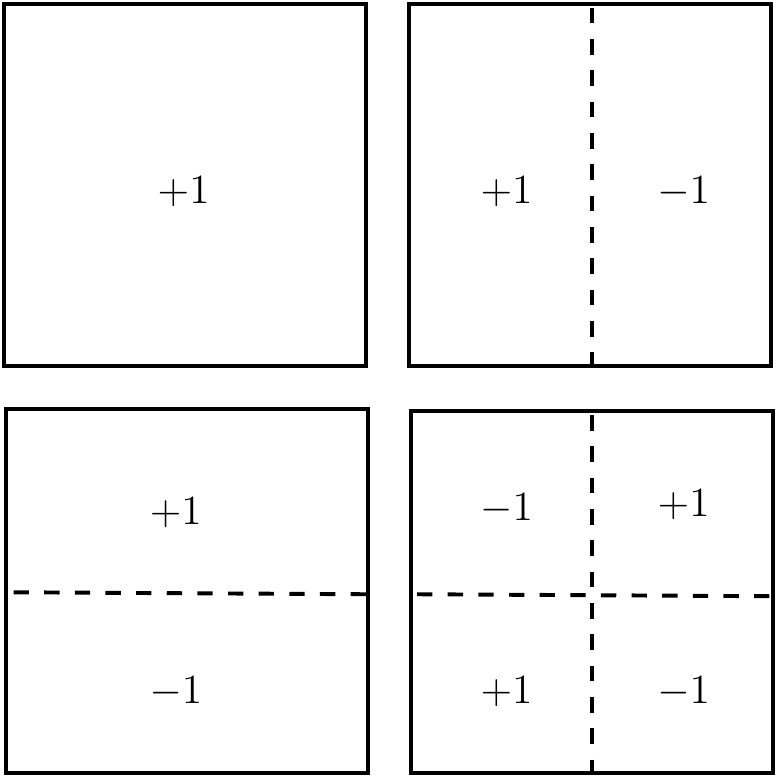}}\\
\caption{Visualisation of our P$_0$ DG-FEM/cell-centred FVM hierarchical discretisation and the equivalent Haar wavelet expansion on an octant given a 2D angular domain.}
\label{fig:pq}
\end{figure}

Considering a FEM applied to the angular domain, we can discretise up to level $j$ using our constant scaling functions as basis functions on the finest partitions/elements of $V_j$ (i.e., applying P$_0$ DG-FEM/cell-centred FVM). We could also use our scaling function and wavelets as basis functions on the base partition/elements of $V_u$. By definition, these two angular discretisations are exactly equivalent. This is a trivial statement given the construction of $V_j$ and $W_j$, but it is an important one that seems to have been largely overlooked in Boltzmann transport and we discuss this further below. We are deliberately using Haar wavelets in this work to exploit the equivalence between our wavelet approximation and P$_0$ DG-FEM in angle (see Figure \ref{fig:pq}), which shares many important properties with S$_n$ methods. Furthermore, our P$_0$ DG-FEM and wavelet discretisations are very similar to that formed using S$_n$ with a product quadrature built from two 1D Newton-Cotes quadratures (see \cite{Hanus2014} for more on the exact similarities between the methods). Figure \ref{fig:pq_sphere} shows the P$_0$ FEM/FVM discretisation equivalent to H$_2$ on a 2D angular domain. 

In moving from representing our angular domain with a P$_0$ discretisation to an equivalent wavelet discretisation, we have traded a hierarchy in the angular mesh for a hierarchy in the basis functions. In order to exploit the equivalence between these spaces, we first must consider how to map between our P$_0$ and wavelet spaces. Given the Haar wavelet functions, we can form a mapping matrix $\mat{M}$ whose entries are simply a patchwork of $\pm 1$. We can then map from P$_0$ space to Haar space by computing $\mat{y} = \mat{M} \mat{x}$, where $\mat{x}$ is a vector containing the values in P$_0$ space (i.e., $f_j$ on each partition/element) and $\mat{y}$ is a vector containing the coefficients $\alpha_{u,n}$ and $\beta_{m,k}$. If we consider one of the diagonal angular matrices in P$_0$ space, say $\mat{P}$, we can easily compute the equivalent angular matrix in Haar space with the matrix-triple product $\mat{M} \mat{P} \mat{M}^{\textrm{T}}$. 

\begin{figure}[th]
\centering
\subfloat[][Sparsity of $\mat{A}_x$ for H$_{\textrm{4}}$ ordered by level]{\label{fig:ax_sparsity_44}\includegraphics[width =0.35\textwidth]{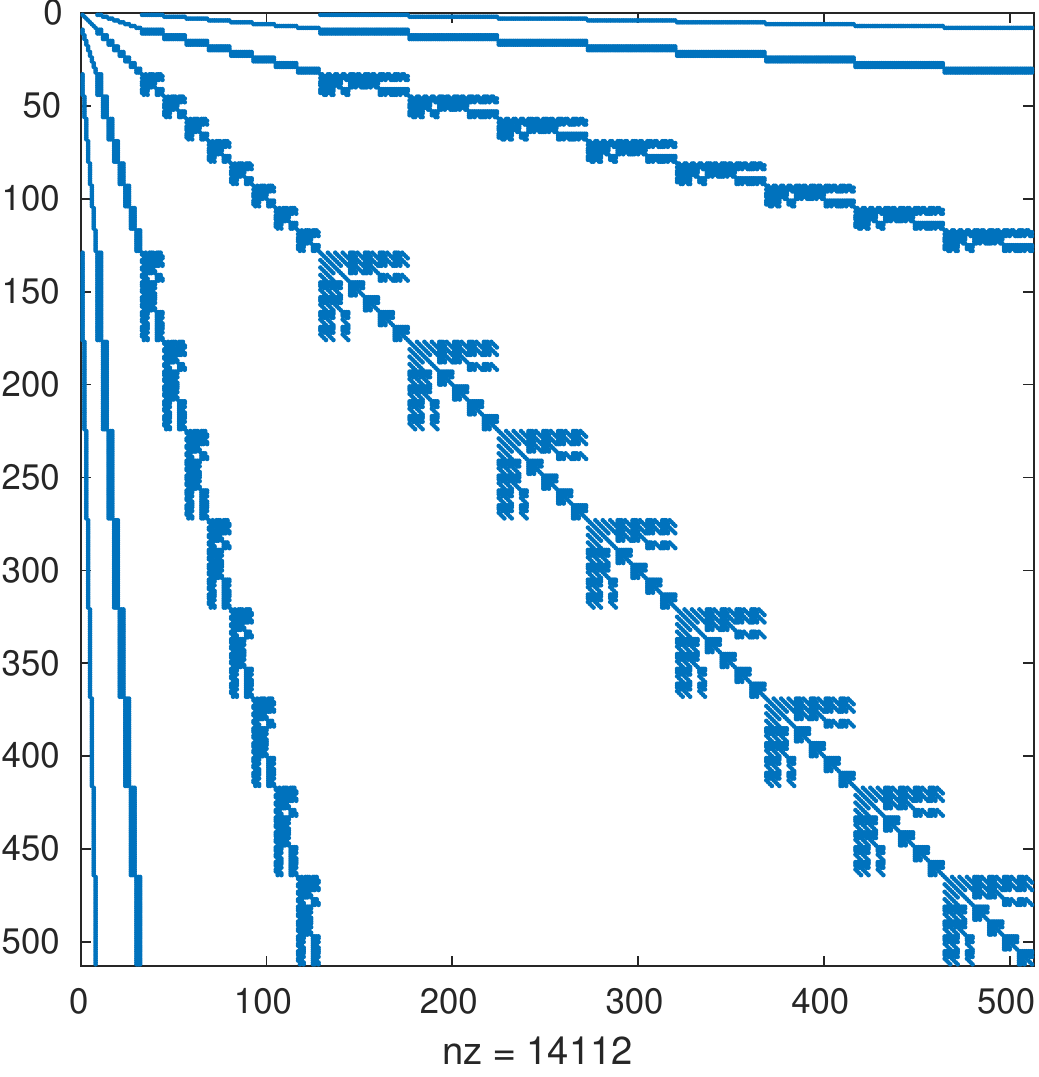}} \hspace{0.5cm}
\subfloat[][Sparsity of $\mat{A}_x$ for H$_{\textrm{4}}$ ordered by octant]{\label{fig:ax_sparsity_44_block}\includegraphics[width =0.35\textwidth]{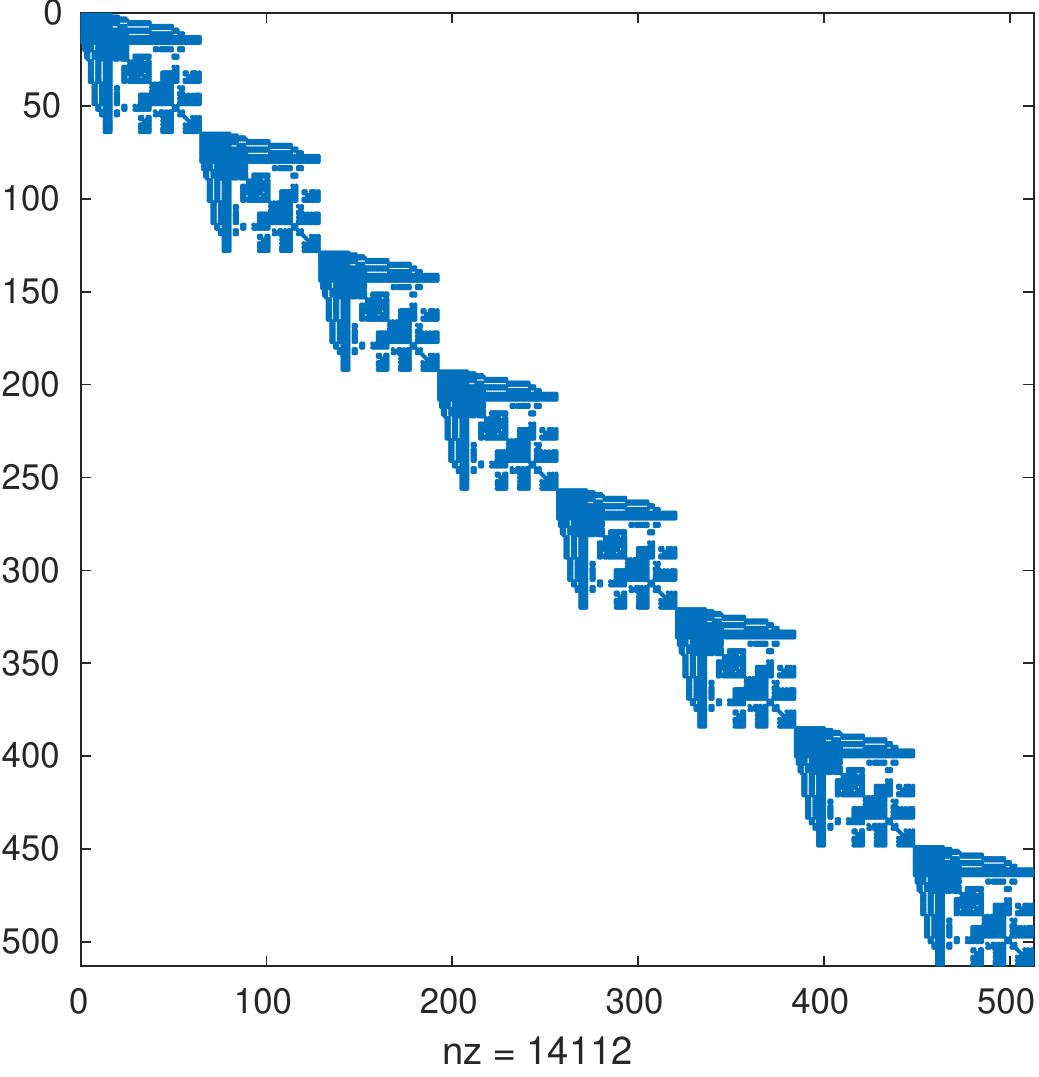}}\\
\caption{Sparsities of the uniform $\mat{A}_x$ angular matrix on a 2D angular domain (8 octants across the sphere) for the standard Haar decomposition. Wavelets are numbered by both order in these figures (i.e., the functions present in H$_{1}$ are first in the ordering, followed by H$_{2}$, etc.) and by octant. To form a scalable wavelet algorithm, these sparse matrices must never be formed/used.}
\label{fig:ax_sparsity}
\end{figure}

The obvious disadvantage to operating in wavelet space is that the angular matrices are no longer diagonal, as the product of any two wavelet functions with overlapping support may be non-zero (e.g, any wavelet function and its children). \tref{table:sparsity} shows the sparsity of the angular matrices generated using a standard Haar decomposition on a 2D angular domain and we can see that the sparsity does not decrease linearly; only the mass-matrix is diagonal due to the orthogonality of the wavelet basis. This is further illustrated in Figure \ref{fig:ax_sparsity} which shows the sparsity structure in $\mat{A}_{x}$ for a H$_{4}$ expansion. Some authors in the Boltzmann transport literature \cite{Koeze2012, Kophazi2015} have criticised wavelet discretisations in angle as impractical given these non-diagonal matrices, as the increased sparsity impacts many of the scalable features Boltzmann transport algorithms rely on (e.g., sweeps, determining inflow/outflow across faces for DG jump terms). The sparsity of the angular matrices does increase with the order of the expansion (given the decreasing support of each higher-order wavelet), but the global support of the scaling functions over each partition in $V_u$ (given the hierarchy) also means that storing/computing either $\mat{M}$ or the angular matrices in Haar space is not practical as the angular mesh is refined.

We can however exploit both the equivalence between our P$_0$ and wavelet spaces and the hierarchical nature of their construction. The well-known Mallat algorithm \cite{Mallat1989,Mallat1989a} allows $\mathcal{O}(n)$ mapping between our P$_0$ and wavelet spaces (this telescoping algorithm is often called the Fast Wavelet Transform (FWT) in signal processing). This algorithm is the key to constructing a scalable wavelet-based discretisation, as it performs the matrix-vector product $\mat{y} = \mat{M} \mat{x}$ and its inverse without storing $\mat{M}$, in $\mathcal{O}(n)$ operations. Importantly, this allows us to scalably:
\begin{enumerate}
\item Map to/from P$_0$ and wavelet spaces
\item Perform matrix-free matrix-vector products with the wavelet angular matrices - by mapping to P$_0$ space ($\mathcal{O}(n)$), performing the matvec on the diagonal P$_0$ angular matrices ($\mathcal{O}(n)$) and then mapping back to wavelet space ($\mathcal{O}(n)$)
\item Perform matrix-free matrix-vector products with the Riemann decompositions and hence apply boundary conditions/compute upwind contributions in wavelet space - again by mapping to P$_0$ space, computing in P$_0$ space and then mapping back
\item Compute the diagonal of matrices in wavelet space (e.g., angular matrices, Riemann decomposition) - computing the diagonal of $\mat{M} \mat{P} \mat{M}^{\textrm{T}}$ reduces to calculating, for each wavelet, the sum of each entry in $\mat{P}$ over the wavelets support (as the entries in $\mat{M}$ are $\pm 1$), which can easily be computed in the same hierarchical fashion as the Mallat algorithm.
\end{enumerate}

To form a scalable, Boltzmann transport algorithm based on wavelets, we must therefore take care to only require the use of these scalable components. We should note that we can also map between our angular flux in wavelet space directly to moment space to compute our scatter contributions using \eref{eq:map}. We use a fixed quadrature order on each P$_0$ element to calculate exact \textit{discrete} scatter contributions which form $\mat{D}_{\textrm{m}}$ and $\mat{M}_{\textrm{m}}$ in our (adapted) P$_0$ space and then just map each column into wavelet space. As we are in a hierarchical space, we need only compute the rows of these mapping matrices corresponding to the wavelets present in our adapted discretisation. This means the construction/application of our mappings are linear in the number of angles present, but quadratic in scatter order. In neutronics, we can therefore provide exact representations of interaction/source terms for a given \textit{discrete} representation of $\bm{\psi}$, hence allowing the moment mapping which enables low-memory algorithms. We discuss this further in \secref{sec:Solver technology}.

In other fields where $S$ is not a simple polynomial and cannot be represented by a truncated series (or in neutronics with heavily forward-peaked scattering for example), the mapping to a truncated form of $\bm{\phi}$ would be inexact and using an S$_n$ method would result in a loss of conservation. Removing the truncation would be pointless, as the Legendre series would require the same number of terms as angular unknowns used to represent the \textit{discrete} form of $\psi$, making $\mat{D}_{\textrm{m}}$ and $\mat{M}_{\textrm{m}}$ square and dense. Their storage and application would therefore scale quadratically with angle size. Hence we cannot practically use the low-memory iterative methods given arbitrary interaction/source terms, but our scheme (like other FEM/FVM schemes) will at least always remain conservative.
\begin{table}[htbp]
\caption{Percentage of non-zero entries in the uniform angular matrices for a standard Haar decomposition on a 2D angular domain (the non-standard decomposition is more sparse, but shows the same behaviour).}
\centering
\begin{tabular}{c c c c c c}
\toprule
Expansion & Angle size & $\mat{A}_x$ & $\mat{A}_y$ & $\mat{A}_z$ & Mass matrix\\
\midrule
H$_{1}$ & 8 & 12.5\% & 12.5\% & 12.5\% & 12.5\% \\
H$_{2}$ & 32 & 12.5\% & 12.5\% & 6.25\% & 3.13\% \\
H$_{3}$ & 128 & 9.57\% & 9.57\% & 2.73\% & 0.78\% \\
H$_{4}$ & 512 & 5.38\% & 5.38\% & 1.18\% & 0.195\% \\
H$_{5}$ & 2048 & 2.48\% & 2.48\% & 1.05\% & 0.049\% \\
H$_{6}$ & 8192 & 1.0\% & 1.0\% & 0.36\% & 0.01\% \\
\bottomrule
\end{tabular}
\label{table:sparsity}
\end{table}
\section{Angular adaptivity}
\label{sec:Angular adaptivity}
\begin{figure}[th]
\centering
\includegraphics[width=0.35\textwidth]{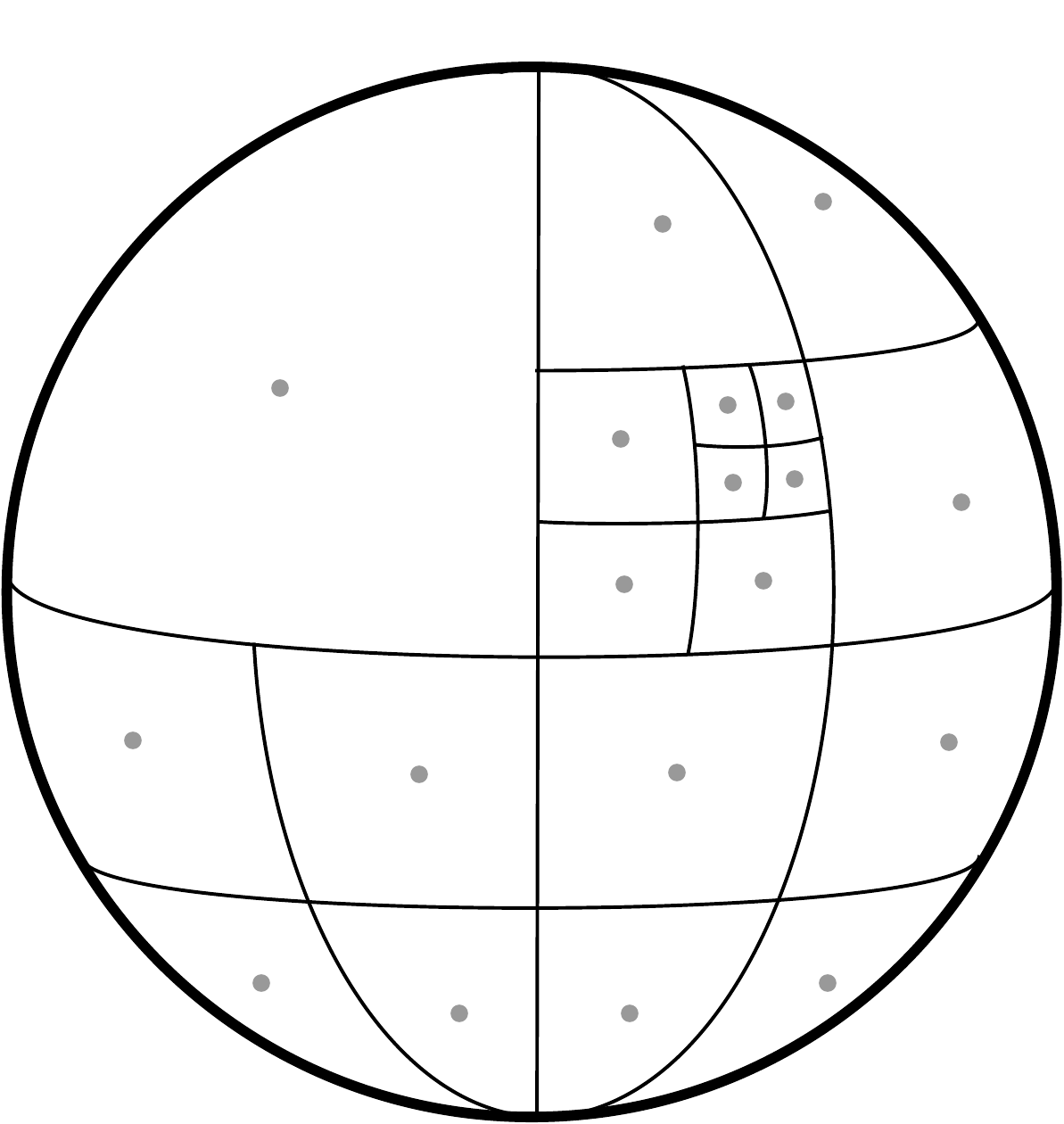}
\caption{Adapted P$_0$ DG-FEM/cell-centred FVM discretisation on the sphere formed from hierarchical refinement. As in the uniform case, we can construct an equivalent Haar wavelet expansion on the sphere.}
\label{fig:adapt_sphere}
\end{figure}
To begin, let us consider an angular discretisation with anisotropic resolution applied across the angular domain. In this work we consider the hierarchical refinement described in the previous section, where now each individual element on the angular domain is free to refine/coarsen (see \fref{fig:adapt_sphere}), guided by an error metric. This angular adaptivity is performed differently at each point in space and energy. Importantly, the scalable wavelet components detailed above (based around the Mallat algorithm) remain scalable when the wavelet discretisation has been adapted. The $\mathcal{O}(n)$ operations in angle now refers to the number of wavelets present, never the number of wavelets in a uniform discretisation. No aspect of the adaptive algorithm can rely on the uniform discretisation at any point. Our only restriction with the wavelet adaptivity is that we enforce the parents of wavelets are always present if their children are, to simplify the implementation. We should note that once adapted, our angular discretisation may not be symmetric and this impacts the implementation of reflective boundary conditions. Given our hierarchical wavelet angular discretisation, we can always capture a reflected ray in our coarser scales. Much like the misalignment between unstructured spatial grids and a given angular discretisation, or the exact \textit{discrete} representation of scatter, these features will be refined by the adaptivity if important, as the coarse wavelet coefficient will be large and hence targetted by the thresholding.   
\subsection{Data structures and ordering}
\label{sec:Data structures}
Before discussing our error metrics, we must consider that although the uniform/adapted wavelet operations described above are $\mathcal{O}(n)$, care must be taken to ensure the constants involved are not large. Typically that results in careful choice of the data structures used to represent values in both P$_0$ and wavelet spaces, in order to respect memory locality and cache hierarchies. This involves balancing both memory use and the speed of data access, particularly when applying the matrix-free mapping operator. Thankfully, we benefit from the decades of research on wavelet methods in these topics; in particular \cite{Latu2011} provides a good review of some important considerations. We prioritise the speed of access to our P0/wavelet values during our mapping, as we solve in wavelet space and perform this mapping at every spatial node, during every matvec in our iterative method. 
\begin{figure}[th]
\centering
\subfloat[][Angle numbers (not indices if adapted) for active wavelets, shown on a square for visibility. The shaded squares correspond to the support of each non-standard wavelet over the octant (this support covers 4 P$_0$ values on level 2 and greater).]{\label{fig:wavelet_ordering}\includegraphics[width =0.8\textwidth]{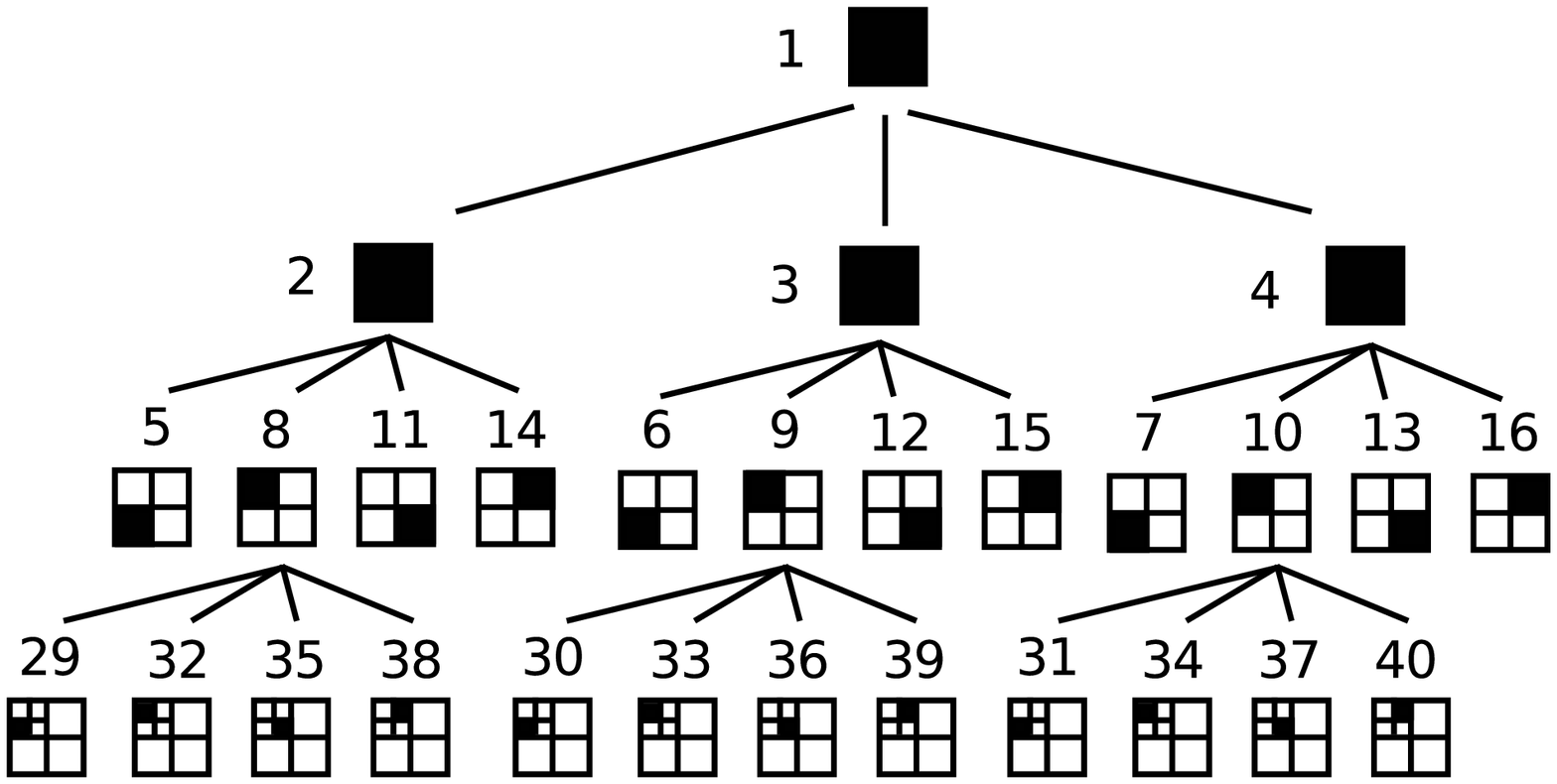}} \\
\subfloat[][Equivalent P$_0$ space on each level in \fref{fig:wavelet_ordering}, with the recursive indices of unknowns in P$_0$ space shown.]{\label{fig:P0_ordering}\includegraphics[width =0.6\textwidth]{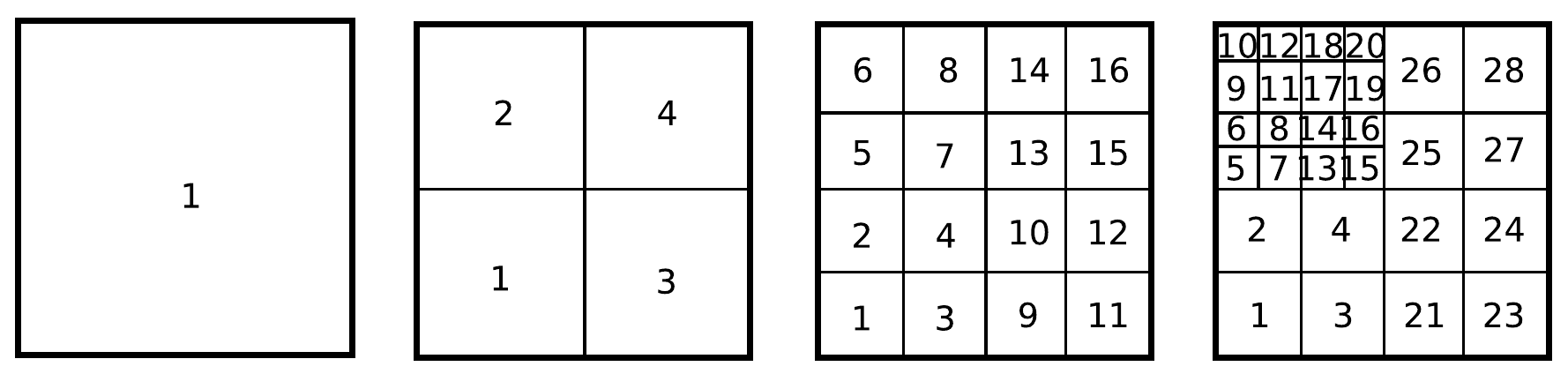}}
\caption{Complementary ordering of our unknowns in both wavelet and P$_0$ space on a single octant, given an adapted, 2D non-standard Haar wavelet discretisation with a maximum order of H$_4$. The discretisation up to H$_3$ is uniform.}
\label{fig:eg_probs}
\end{figure}
This requires careful and complementary ordering of unknowns in both our wavelet and P$_0$ spaces. For 1D wavelets, the ordering is trivial, wavelets with neighbouring support are adjacent. In 2D, the standard and non-standard Haar decompositions require different orderings, depending on how the mapping operator is implemented. A standard Haar decomposition is quite simple; if we consider the P$_0$ coefficients in a single octant as the entries in a $n \times n$ matrix, the mapping can be computed by performing a 1D mapping along all the rows first, followed by the all columns. An efficient ordering is therefore as simple as ensuring each 1D mapping respects the row/column ordering. 

An efficient ordering for a 2D non-standard Haar decomposition is more difficult. Figure \ref{fig:wavelet_ordering} shows the schematic of an adapted octant up to a maximum refinement of H$_4$ for a non-standard Haar discretisation. We impose a level-by-level ordering in wavelet space, with the further constraint that the three different wavelet functions (see Figure \ref{fig:wavelet_functions}) that share the same support are numbered sequentially. This numbering allows us to use simple arithmetic operations to determine many key relationships between wavelets. This saves us costly lookups/computation of these relationships during the mappings. 

Furthermore, in P$_0$ space, we define the indices on each level recursively, see \fref{fig:P0_ordering}. For example, the P$_0$ value at position 1 on level 2 has children in positions 1--4 on level 3. Both these orderings give arrays that corespond to flattened quad-trees, and give good data locality during the mapping (and are similar to cache-oblivious trees (e.g., see \cite{Bender2007}). We should also note these orderings and data structures are very similar to those used commonly in astrophysics, where the night sky is discretised using a $P_0$ formulation known as HEALPix \cite{Gorski2005}, where efficient mappings to spherical harmonic (we discuss this in \secref{sec:sub-grid}) and hierarchical wavelet spaces are also essential. We do not use HEALPix in this work as it does not feature octant symmetry and we would like to preserve this when using structured spatial grids. 
\subsection{Error metrics}
\label{sec:Error metrics}
We consider two forms of angular adaptivity in this work, regular and goal-based adaptivity. We will refer to \eref{eq:bte} as the ``forward'' problem, with exact solution $\bm{\psi}_{\textrm{exact}}$ and residual $\mathcal{R}$, hence $\mathcal{R}(\bm{\psi}_{\textrm{exact}}) = 0$. In this section, we are trying to compute an approximation, $\mat{e}$, to the exact error, $\bm{\epsilon} = \bm{\psi}_{\textrm{exact}} - \bm{\psi}$, in order to guide our adaptivity.
\subsubsection{Regular adaptivity}
\label{sec:Regular adaptivity}
Regular adaptivity targets reductions in the global error in the problem and is suitable even in non-linear Boltzmann transport problems. Regular adaptivity is simple with wavelets, as discussed in \secref{sec:Angular adaptivity}. Given the wavelet coefficients can be thresholded, with small coefficient guaranteeing small contribution to the norm of the function we are representing, our only job is picking a thresholding tolerance, $\tau$. We therefore define our regular error metric as $\mat{e} = \bm{\epsilon} \approx |\bm{\psi}|/\tau$; this metric can clearly be computed scalably. For convenience, in the results presented below, we also scale $\mat{e}$ by the maximum scalar flux across the problem; this simply helps make the choice of $\tau$ more problem agnostic. 
\subsubsection{Goal-based adaptivity}
\label{sec:Goal-based adaptivity}
If we can compute the adjoint of our equation (like in many linear Boltzmann transport problems), we can use goal-based adaptivity. Rather than reducing the error across the entire phase-space, goal-based adaptivity focuses resolution wherever needed to reduce the error in some arbritrary functional. Goal-based methods have a long history in Boltzmann transport problems, given the prevelance of problems where the flux/dose in a small target is of interest, and vitally is many orders of magnitude smaller than the flux elsewhere in the problem. An example of this is the the particle accelerators described in Section \ref{sec:Introduction}. Goal-based methods are essential in such problems, as regular adaptivity will reduce the error in a global norm by refining anywhere the solution is large (i.e., only at the source) and ignore regions in the phase-space with small flux (i.e., down the duct).

In this section, we briefly review the formulation of goal-based error metrics through a dual-weighted residual method, described by \cite{Goffin2015, Goffin2015a}. We can write the goal of the calculation in terms of a functional, $F$, of the solution as
\[
F(\bm{\psi}) = \int_P f(\bm{\psi}) \, \textrm{d}P,
\]
where $f$ is an arbritrary function of the solution and $P$ represents the phase-space. Functionals can be easily defined for quantities such as the average flux over a region, current over given surfaces, reaction rates and even eigenvalues. If we expand both $f$ and $\mathcal{R}$ about the exact solution using a Taylor series, discarding second-order terms and higher, we can write an approximation for the error in the goal functional as 
\begin{equation}
F(\bm{\psi}_{\textrm{exact}}) - F(\bm{\psi}) \approx - \int_P \mathcal{R}(\bm{\psi}) (\bm{\psi}^*_{\textrm{exact}} - \bm{\psi}^*) \, \textrm{d}P,
\label{eq:forward_error}
\end{equation}
where $\bm{\psi}^*$ and $\bm{\psi}^*_{\textrm{exact}}$ are the approximate and exact solutions of the adjoint equation with source term derived from the response function, respectively. Equivalently we can write
\begin{equation}
F(\bm{\psi}_{\textrm{exact}}) - F(\bm{\psi}) \approx - \int_P \mathcal{R}^*(\bm{\psi}^*) (\bm{\psi}_{\textrm{exact}} - \bm{\psi}) \, \textrm{d}P,
\label{eq:adjoint_error}
\end{equation}
where $\mathcal{R}^*(\bm{\psi}^*)$ is the residual of the adjoint equation. Using similar ideas we can also derive an expression for the discrete error. If we approximate the error in both the forward, $\bm{\epsilon}$, and adjoint solutions, $\bm{\epsilon}^*=\bm{\psi}^*_{\textrm{exact}} - \bm{\psi}^*$, by expanding both errors and residuals in terms of the basis used to compute the forward/adjoint solutions (hence the name dual-weighted) and using \eref{eq:forward_error} and \eref{eq:adjoint_error}, then we can write
\begin{equation}
|F(\bm{\psi}_{\textrm{exact}}) - F(\bm{\psi})| \approx \bm{\epsilon}^{\textrm{T}} \mat{R}^*
\label{eq:error}
\end{equation}
or equivalently
\begin{equation}
|F(\bm{\psi}_{\textrm{exact}}) - F(\bm{\psi})| \approx \bm{\epsilon}^{*\textrm{T}} \mat{R}
\label{eq:adjoint_err}
\end{equation}
where $\bm{\epsilon}^{\textrm{T}}$ and $\bm{\epsilon}^{*\textrm{T}}$ are the discrete forward and adjoint solution error, respectively, with $\mat{R}$ and $\mat{R}^*$ the discrete forward and adjoint residuals computed using $\bm{\psi}$ and $\bm{\psi}^*$, respectively. Care must be taken here, as the error expressions \eref{eq:error} and \eref{eq:adjoint_err} are always zero, due to Galerkin orthogonality. Commonly, either the error or the residual are modified, often by computing the solution on a refined grid, or by using higher order local interpolation to approximate the true error. We use neither of those approaches, as increasing the level of refinement of every ``leaf'' wavelet present and solving another forward/adjoint problem at each adapt step is very expensive. For problems with very anisotropic refinement, the number of wavelets added in this approach is often higher than the total number on a node. Using higher order interpolation locally to approximate error (or in general like many authors described above when using angular adaptivity) is also not a good choice for Boltzmann transport, as we use low-order discretisations to resolve strong discontinuities in angle. We would therefore have to take care our interpolation would not introduce significant Gibbs oscillations into the error metric (\cite{Stone2008} found this with high-order interpolation).

We therefore follow \cite{Goffin2015, Goffin2015a} and take a more conservative approach, by further approximating \eref{eq:error} and \eref{eq:adjoint_err} in order to ensure a non-zero error, by computing ``reduced-accuracy'' discrete residuals $\hat{\mat{R}}$ and $\hat{\mat{R}}^*$. We must also pick a target error for our goal-based adaptivity, similar to the thresholding tolerance in Section \ref{sec:Regular adaptivity}; we denote this tolerance again as $\tau$. We form our approximate error metric by computing the pointwise maximum of both the forward and adjoint pointwise errors, and scaling by the target error in each DOF, namely
\begin{equation}
\mat{e} = \frac{\max\{|\bm{\epsilon} \odot \hat{\mat{R}}^*|, |\bm{\epsilon}^* \odot \hat{\mat{R}}|\} N_{\textrm{DOF}}}{\tau}, 
\label{eq:gb_metric}
\end{equation}
where $\odot$ denotes pointwise multiplication. The use of the $\max$ operator in \eref{eq:gb_metric} ensures that features present in both the forward and adjoint solutions are resolved by the adaptivity (we define the orientation of our adjoint angular space to be the opposite of the forward, so we can easily compute products involving both our forward and adjoint wavelet coefficients). We are now left to define both the solution errors $\bm{\epsilon}$ and $\bm{\epsilon}^*$ and the reduced-accuracy residuals.

Similar to the regular adaptivity, we choose $\bm{\epsilon} \approx |\bm{\psi}|$ and $\bm{\epsilon}^* \approx |\bm{\psi}^*|$. For the reduced-accuracy residuals, if we write our discretised linear system as $\mat{G} \bm{\psi} = \mat{b} $ (we write this generically as we have not yet introduced our spatial discretisation), we therefore define a new reduced-accuracy discrete residual as , $\hat{\mat{R}}$, as
\begin{equation}
\hat{\mat{R}} = \textrm{diag}(\mat{G}) \bm{\psi}.
\label{eq:disc_resid}
\end{equation}

Equation \eref{eq:disc_resid} is a very simple expression for our reduced accuracy residual, and given that we can form the diagonal of our wavelet angular matrices (as described in Section \ref{sec:Wavelets}), we can compute our goal-based metric \eref{eq:gb_metric} scalably. As such, we now have a scalable angular adaptivity algorithm, with both regular and goal-based error metrics. We do not necessarily expect our goal-based metric to have a good effectivity index (the ratio of true error to predicted) given it is a simple diagonal matrix-vector product and does not include our discretised right-hand side $\mat{b}$ (and hence we do not expect our reduced accuracy residual to approach zero as the discretisation is refined), but we hope it is still enough to guide our adaptivity. We will examine this further in \secref{sec:Results}.

\begin{figure}[th]
\centering
\includegraphics[width=0.65\textwidth]{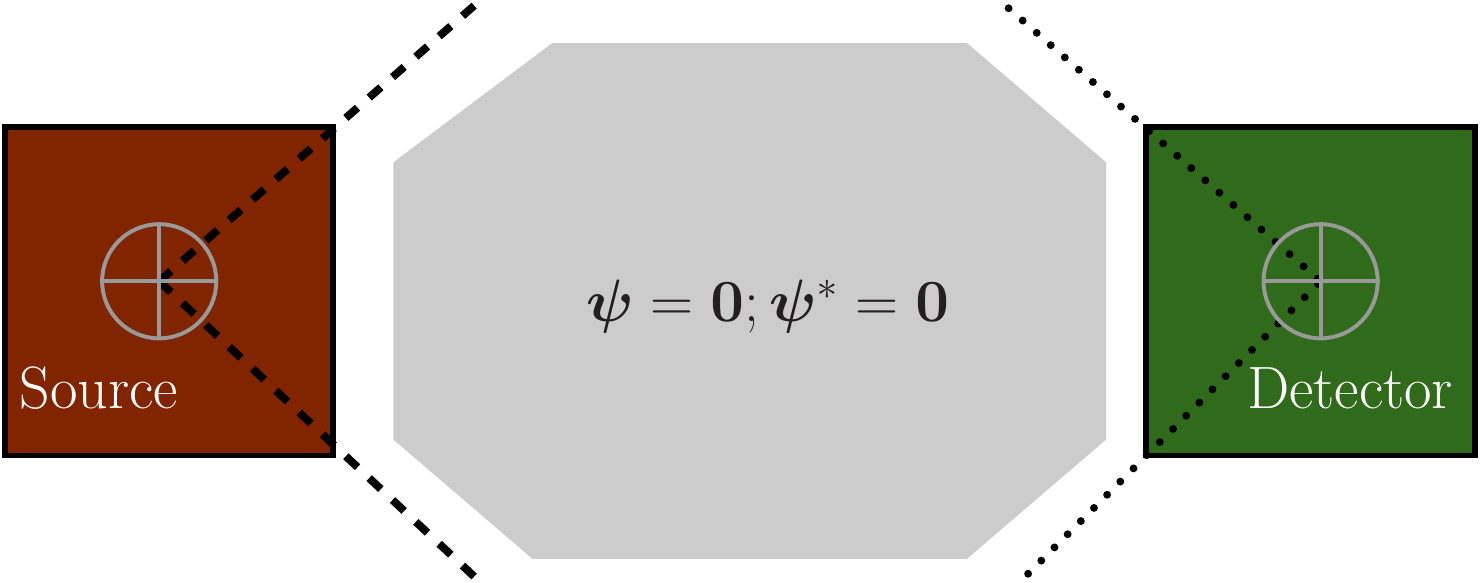}
\caption{Schematic of a source/detctor problem in a vacuum with a coarse angular discretisation. Both the solution of the forward (dashed lines) and adjoint (dotted lines) problems suffer from ray effects; the shaded grey region has zero flux in both the forward and adjoint solutions.}
\label{fig:ray_effect}
\end{figure}

However, we must also consider the fundamental problem introduced by ray-effects in our goal-based metric. \fref{fig:ray_effect} shows a schematic of the problem, where in a perfect vacuumm a source and detector are placed some distance apart. When using a coarse angular discretisation, both the forward and adjoint problem will not have enough angular resolution to ``see'' each other, resulting in large regions of zero flux (and also zero residual) throughout the domain. This means our error metrics \eref{eq:adjoint_error} and \eref{eq:error} will also be zero and hence angular adaptivity will not be triggered. This is a problem faced by all goal-based error metrics in the presence of advection, not just in Boltzmann transport problems. For example, goal-based metrics in time with advection operators face the same problem, where the response in a given spatial region is zero until a wave reaches the region. We leave tackling this problem to future work, and like \cite{Goffin2015, Goffin2015a}, we simply ensure that any goal-based problem we run has a non-zero response even with a coarse angular discretisation. 
\section{Spatial discretisation}
\label{sec:sub-grid}
Now we move to discussing our spatial discretisation. We should note that the discussion on scalable wavelet components above does not explicitly depend on our specific discretisation, it could easily be applied to a standard DG discretisation. Given the advection operators in \eref{eq:bte}, care must be taken to use a scheme that provides some form of stabilisation, like upwinding, Galerkin-least-squares \cite{hughes_new_1989} or SUPG \cite{brooks_streamline_1982}. We use a sub-grid scale (SGS) finite element formulation to discretise in space \cite{hughes_variational_1998, hughes_multiscale_2006, candy_subgrid_2008, buchan_inner-element_2010}. A brief overview of the spatial discretisation is given below, with a focus given to the aspects of the discretisation that contribute to scalability; for more details please see \cite{buchan_inner-element_2010}. To begin, we decompose the solution to \eref{eq:bte} as $\psi = \phi + \theta$, where $\phi$ and $\theta$ are the solutions on the ``coarse'' and ``fine'' scales, respectively. We then chose to represent the solution on each scale with a different finite element representation (like \cite{hughes_multiscale_2006}); on the coarse scale we use a continuous representation spanned by $\eta_N$ basis functions, while on the fine scale we use a discontinuous representation spanned by $\eta_Q$ basis functions, or
\begin{equation}
\phi(\bm{r}, \bm{\Omega}) \approx \sum_{i=1}^{\eta_N} N_i(\bm{r}) \tilde{\phi}_i(\bm{\Omega}); \qquad \theta(\bm{r}, \bm{\Omega}) \approx \sum_{i=1}^{\eta_Q} Q_i(\bm{r}) \tilde{\theta}_i(\bm{\Omega}),
\label{eq:space}
\end{equation}
where $N_i$ and $Q_i$ are the basis functions for the continuous and discontinuous spaces respectively, with $\tilde{\phi}_i$ and $\tilde{\theta}_i$ being the associated expansion coefficients. 

To discretise in angle, we use the Haar wavelets described in detail in \secref{sec:Wavelets}, though here we refer to an arbitrary angular discretisation. We can represent the expansion coefficients $\tilde{\phi}_i$ and $\tilde{\theta}_i$ in \eref{eq:space} with an arbitrary angular discretisation, with the basis functions $G_j(\bm{\Omega})$ and different numbers of basis functions $\eta_A^i$ and $\eta_D^i$ for each expansion coefficient in space, on the coarse and fine scales respectively. Finally if we write the expansion coefficients $\tilde{\phi}_{i,j}$ and $\tilde{\theta}_{i,j}$, we have 
\begin{equation}
\tilde{\phi}_i(\bm{\Omega}) \approx \sum_{j=1}^{\eta_A^i} G_j(\bm{\Omega}) \tilde{\phi}_{i,j}; \qquad \tilde{\theta}_i \approx \sum_{j=1}^{\eta_D^i} G_j(\bm{\Omega}) \tilde{\theta}_{i,j}.
\label{eq:angle}
\end{equation}
If we apply the FEM as usual to \eref{eq:bte}, by integrating and applying Green's theorem, we can recover the linear system
\begin{equation}
(\mat{A} - \mat{B} \mat{D}^{-1} \mat{C}) \tilde{\bm{\Phi}} = \mat{S}_{\bm{\Phi}} - \mat{B} \mat{D}^{-1} \mat{S}_{\bm{\Theta}}.
\label{eq:SGS}
\end{equation}
where $\mat{S}_{\bm{\Phi}}$ and $\mat{S}_{\bm{\Theta}}$ are the discretised source and $\tilde{\bm{\Phi}}$ and $\tilde{\bm{\Theta}}$ are vectors containing the coefficients of the coarse and fine discretised solutions, $\tilde{\phi}_{i,j}$ and $\tilde{\theta}_{i,j}$, respectively. The definition of the matrices $\mat{A}, \mat{B}, \mat{C}$ and $\mat{D}$ are given in \cite{buchan_inner-element_2010}, though we should note that \eref{eq:SGS} has the same number of DOFs as the continuous problem; indeed $\mat{A}$ is simply the linear system that would result from discretising \eref{eq:bte} with continuous finite elements. Once we have solved \eref{eq:SGS}, we can reconstruct the fine solution with
\begin{equation}
\bm{\Theta} = \mat{D}^{-1} (\mat{S}_{\bm{\Theta}} - \mat{C} \bm{\Phi}),
\label{eq:theta}
\end{equation}
and then form our discrete solution $\bm{\Psi} = \bm{\Phi} + \bm{\Theta}$. 

The $\mat{B} \mat{D}^{-1} \mat{C}$ term in \eref{eq:SGS} can be considered as a stabilisation term, where a number of approximations have been made to significantly decrease the cost of inverting $\mat{D}$ (see \cite{buchan_inner-element_2010, Dargaville_2014, Adigun2018}). Importantly, these approximations do not affect the conservation of our scheme, though we must ensure the construction/application of $\mat{D}^{-1}$ is $\mathcal{O}(n)$ in (possibly adapted) angle size (i.e., in $\eta_D^i$). 

To achieve this, we reduce $\mat{D}$ to being element local, by replacing the jump terms between elements in a DG formulation with a vacuum condition on all element boundaries (which we can apply scalably as detailed in \secref{sec:Wavelets}). Furthermore, we enforce a block-diagonal form, that blocks together each individual angular basis function present on an element, i.e., if we have a uniform angular discretisation, then the block for each angular basis function is of size $\eta_Q \times \eta_Q$. If however we have adapted in angle, the block size depends on the number of nodes in an element that each given angular basis function is present on. For example, if a given angular basis function (i.e., wavelet) is only present on 2 DG nodes in the element, the block is $2 \times 2$ for that basis function. With a uniform angular discretisation, this is very similar to the blocks constructed as part of a sweep algorithm on a DG mesh, though ours contain self-scatter and can be built with no dependency on a sweep ordering. The inversion of $\mat{D}$ is hence linear in both storage and work with the number of angles present. We can therefore construct this on the fly during our matrix-vector products, or as we do in this work for convenience, explicitly build this inverse and store it (with a storage cost of $\eta_Q$ copies of the fine angular flux, $\bm{\psi}$). 

We also scale our element blocks of $\mat{D}^{-1}$ by $\gamma$ ($0 < \gamma < 1$) defined in \cite{Ragusa2012}, to prevent locking in pure scatter regions. This has the effect of making our discretisation tend towards a standard CG discretisation, but only on elements with heavy scatter. As we do not have jump-terms in our discretisation, this scaling has no effect on the stencil of our discretisation, unlike in a standard DG formulation where $\gamma$ forces a transition to central differencing which decreases sweepability. 

If we are performing a matrix-free matrix-vector product in order to solve \eref{eq:SGS} we are relying on the scalable mapping from wavelet space to moment space in order to apply our scatter terms, by mapping our wavelet angular flux to moment space, applying our scatter coefficients and then mapping back. Furthermore, our wavelet mass matrix is diagonal, so we can easily compute our removal term directly in wavelet space. This is in contrast to explicitly forming the dense scatter/removal matrix, whose construction/application would scale quadratically with the number of angles present. Depending on the scatter order and number of angles present, the balance between explicitly forming our scatter/removal operator (at quadratic cost in angle size) and applying the scatter using a mapping (at quadratic cost in scatter order) may change. This is one disadvantage to solving \eref{eq:SGS} rather than \eref{eq:moment_bte} as the scatter order increases; we need to apply this mapping during every matrix-vector product, instead of twice for each outer iteration of a sweep/DSA algorithm. This extra cost may however be reduced by not having to solve a diffusion equation for each separate scatter moment. Furthermore, our simple hierarchical discretisation of the sphere features rings of elements with constant azimuthal angle; this is a key feature (shared by the astrophysical discretisation HEALPix mentioned above) as it reduces the cost of our mapping given the dependence on only the azimuthal angle in $z$ (see \cite{Mohlenkamp1999, Seljebotn2012}). This may make our algorithm competitive even with high order anisotropic scatter, though we leave the investigation of this for future work and only use isotropic scatter in this paper.

Given the remainder of \eref{eq:SGS} is constructed with standard finite-element components and the discussion in \secref{sec:Wavelets} regarding scalable wavelet components, it is easy to see how we can now perform a matrix-free matrix-vector product in wavelet space in linear time. Furthermore, we use linear basis functions in both the continuous and discontinuous spatial expansions given by \eref{eq:space}, hence we can often reuse temporary data during our matrix-free matrix-vector product, making our matvecs less expensive in practice. The benefit to solving \eref{eq:SGS} as opposed to standard DG formulation is that the static condensation (given the approximations applied to $\mat{D}$, our discretisation can be considered as formed from an approximate Schur-complement) allows us to solve for $\tilde{\bm{\Phi}}$ and then reconstruct $\tilde{\bm{\Psi}}$. Particularly in 3D, the size of $\tilde{\bm{\Phi}}$ on the CG mesh is much smaller than $\tilde{\bm{\Psi}}$ on the DG mesh, allowing us more flexibility in chosing iterative methods in these memory-constrained problems (see \secref{sec:Solver technology}). 
\section{Adaptivity algorithm}
\label{sec:Adaptivity algorithm}
We now formalise our iterative algorithm for the angular adaptivity. We begin the first adapt step by first solving the forward linear system with our coarse angular discretisation H$_1$ (possibly to a large tolerance), then solve the coarse adjoint linear system if goal-based adaptivity is used. We then compute the regular/goal-based error metric and perform refinement/coarsening. This is then followed by further adapt steps, up to some maximum refinement level.  Alternatively, if we are performing fixed angular refinement, we only need to refine our angular discretisation between given bounds and then perform a single forward linear solve. Coarsening cannot occur beyond a single scaling function per quadrant/octant (i.e., H$_1$). 

As mentioned, the direction space $\Omega$ in our adjoint problem is explicitly written as the negative of our forward angular domain (i.e., our adjoint angular domain is a reflection about the origin). Our error metric \eref{eq:gb_metric} ensures that refinement is triggered in areas important to both the forward and adjoint solutions. We do this as it simplifies our implementation, as we can then apply the ``same'' angular discretisation to our forward and adjoint problems. For simple forward/adjoint problems (e.g., a source/detector problem in a vacuum), this also means the ``same'' area of our angular domain is automatically refined for both problems, increasing locality of our adapted regions and data access during our adapts. A disadvantage of this approach is we may be applying more DOFs in angle than by performing adaptivity on the sphere separately for the forward and adjoint problems. In practice however, we find this is not significant.

Given our spatial discretisation described in \secref{sec:sub-grid}, the error metrics given in \secref{sec:Error metrics} are all computed using $\bm{\psi}$, which is formed from the sum of our coarse and fine solutions. This solution and hence our error metrics are computed on the fine mesh (i.e., the DG mesh), but we perform our angular adaptivity on the CG mesh. We therefore take the maximum error over the DG nodes that share their position with each CG node, to form an error metric, $\tilde{\mat{e}}$, on the CG mesh. For any wavelet on the CG spatial mesh with index $k$, we trigger refinement and add all its children to the discretisation if $\tilde{\mat{e}}_k \geq 1.0$. We trigger coarsening and remove wavelets on the CG spatial mesh if $\tilde{\mat{e}}_k < 0.01 $, subject to the constraint that we don't remove any wavelet that has children. The wavelets present on the DG nodes of the spatial mesh are then slaved to their CG counterparts and share the same angular discretisation.

The choice to adapt on the CG spatial mesh means that adjacent faces in our mesh share the same angular discretisation. For many neutronics/radiative transfer applications, ensuring matching adapted angular discretisations across faces is the sensible choice, given that the only source/interaction term (i.e., scatter) is restricted to the angular domain, meaning there is no need to interpolate to compute quantities within an element, regardless of the angular discretisation used. If a non-wavelet discretisation is used in angle, this means interpolation is not needed to compute DG jump terms. 

The choice is largely arbritrary in this work given that our wavelet discretisation means we would not need to perform interpolation even if we chose to adapt differently (e.g., element-wise, or allowing every DG node to adapt differently) or different source/interaction terms were present. One further benefit to our sub-grid discretisation is that if we used both a non-wavelet discretisation and allowed each DG node in our discretisation to adapt differently (and then slaved the CG nodes to a union of angles present in its surrounding DG nodes), we would still not be required to interpolate across faces, as our fine-scale does not require the computation of jump terms. This gives us considerable flexibility and also becomes significant in \secref{sec:Solver technology}. In common discretisations, care must be taken if interpolation is needed, to ensure conservation and stability (e.g., see \cite{Kophazi2015}). 

In intermediate adapt steps, to improve our runtimes we could easily reduce the tolerance of our linear solves, as only the final linear solve with the finest discretisation needs to be solved to a high tolerance. Often ``a few'' sweeps of an S$_n$/P$_0$ FEM discretisation are often used in the early adapt steps as a cheap surrogate. In our experience this is dangerous; we found our results varied drastically if we deliberately reduced our iterative tolerance too small in early adapt steps. In some cases, we required double the number of adapt steps to converge our discretisation, with great cost. As such, in this work we chose to only perform limited experimentation with reduced tolerance linear solves as part of the adaptivity algorithm, see \secref{sec:Results} for further details. 

One further benefit however, to using a wavelet discretisation is we could easily build robust tolerance critera for the linear solve at each adapt step. For example, with regular adaptivity, we could terminate our iterative method when we know that each wavelet coefficient $\mat{e}_k$ is correct to one decimal place, i.e., when we know whether each wavelet coefficient is smaller/bigger than 0.01 or 1.0. If we are using goal-based adaptivity and our error metric has a good effectivity index, then we could also use the error metric to define our linear solve tolerance. We leave investigation of this to further work.
\section{Linear solver}
\label{sec:Solver technology}
The iterative method we use is FGMRES \cite{saad_flexible_1993} preconditioned by a multigrid algorithm first presented by \cite{buchan_sub-grid_2012} to solve \eref{eq:SGS} matrix-free. Normally constructing Krylov vectors on a stable discretisation of the BTE is too memory intensive given the use of DG discretisations. We can use FGMRES with our discretisation given that we solve for the coarse scale of our discretisation, $\tilde{\bm{\Phi}}$, (i.e., the CG), reconstruct the fine-scale solution, $\tilde{\bm{\Theta}}$, then add these two solutions to obtain our solution. We therefore use more memory than moment mapping based methods, but it allows us to use the same iterative technology if we have source/interaction terms which cannot use the moment mapping, without resorting to building Krylov vectors on all the DG variables. By default we use a restart parameter of 30 in our FGMRES, though we can decrease this significantly with very little impact on our convergence. 

Our multigrid preconditioner is specifically designed to work with unstructured spatial grids and forms the lower multgrid levels by using element agglomeration algorithms (e.g., \cite{jones_amge_2001}). The key features of this multigrid include building both spatial tables and $\mat{D}^{-1}$ on the lower multigrid levels by using an agglomerate-local Galerkin projection on the top-grid element matrices. This is equivalent to performing a rediscretisation on the lower-grids. The lower-grid spatial nodes take the union all of the adapted angles present in the connectivity pattern of the prolongator for that node. Importantly, the fact that we do not use jump-terms in our discretisation means we do not need to worry about the non-straight element boundaries on the lower grids, formed from agglomerates of unstructured elements. This is similar to how boundary conditions are handled naturally by traditional multigrid algorithms which use Galerkin projection to form coarse-grid matrices. This allows us to perform matrix-free matrix-vector products on all multigrid levels, meaning we can use all of the scalable components mentioned above. 

Our multigrid projection/restriction operators are very simple; injection on coarse/fine spatial nodes and averaging (based on the spatial connectivity only) for the remaining spatial nodes (the restrictor is the transpose of the prolongator). These operators are the same for all angles. This allows us to easily apply them matrix-free and ensures that the memory used during our multigrid setup does not depend on the number of angles present, which is vital for scalability. The disadvantage to these simple operators is we do not expect them to be robust in the presence of strong advection, they should perform best in the heavy-scatter limit. We have investigated more robust operators, but designing them to ensure scalable work/memory consumption is difficult, we leave this to further work. Perhaps surprisingly, our multigrid still performs well in many problems with advection, see \secref{sec:Results} for more details. Our smoothers must also be entirely matrix-free and scalable; we use GMRES(3) preconditioned by an Jacobi method as our smoother on each multigrid level. We can therefore smooth matrix-free, and the diagonal on each level can be built scalably given the block-structure we use in $\mat{D}^{-1}$ (which allows us to easily form the diagonal contribution from $\mat{B} \mat{D}^{-1} \mat{C}$ in \eref{eq:SGS}) and the ability to compute the diagonal of our wavelet angular matrices scalably (as detailed in \secref{sec:Wavelets}). These very strong smoothers help compensate for our simple operators, though the use of GMRES as a smoother on the lower grids does impact our parallel performance. We do not examine this in this work, but note that we have had success in these problems building increasingly parallel smoothers (e.g., see \cite{baker_multigrid_2011}). 
\section{Results}
\label{sec:Results}
Outlined below are three examples we use to test our angular adaptivity algorithm. These three examples use fixed refinement in a given angular region, regular adaptivity, and goal-based adaptivity. For all problems, we start the adaptivity algorithm with a coarse uniform resolution of H$_1$ and we set a maximum level of refinement. We use a maximum of 15 levels of refinement in angle with two spatial dimensions, and 14 levels with three spatial dimensions (further refinement means the angle numbers discussed in \secref{sec:Data structures} cannot be represented as 32-bit integers). Each adapt step after H$_1$ increases the possible level of refinement used by the discretisation by one, up to the maximum level. Any subsequent steps stay at that maximum possible refinement level. Our adaptivity adds all children of a wavelet if refinement is triggered and only removes wavelets if coarsening is required and the wavelet does not have any children (i.e., a childless wavelet can only be removed in an adapt step after the one in which it was introduced). Memory use during the adapt process is profiled using massif (from valgrind) using default parameters, which measures peak heap usage. Very little memory in our simulations is not on the heap, so this provides an accurate measurement of our total peak memory use. 

It is worth noting that we are severely constrained by our ability to compute reference solutions of sufficient accuracy. In the problems shown, we could not compute solutions with uniform Haar wavelets with resolution of more than 6-7 levels of refinement in reasonable time. Indeed that is the point of this work, showing that angular adaptivity is practical in problems that require significant angular refinement. Although we have the capability to use different angular discretisations (e.g., P$_n$), the problems we are using in this work have strong discontinuities in angle, meaning high-order discretisations suffer from significant Gibbs oscillations or poor conditioning, limiting their use as reference solutions. Furthermore, each of our different angular discretisations is stabilised differently (the block-form of $\mat{D}$ described in \secref{sec:sub-grid} changes depending on our angular discretisation), and we are unable to feasibily converge our spatial discretisation in any of the problems tested. This also means we cannot use alternative low-order discretisations (e.g., P$_0$ FEM with a different discretisation of the sphere) as a reference, as they will converge to different solutions. 

The reason we cannot converge our spatial discretisation is related to the heavily anisotropic problems this work targets; ray effects produced by angular discretisations without rotational invariance (i.e., all non-P$_n$ discretisations) worsen with spatial refinement. This means in order to converge our phase-space discretisation, we would need to refine in space every time we refine in angle, until we reach the asymtotic regime in space/angle for each problem, which is not practical. Similarly, we cannot use Monte-Carlo references as the error from our under-resolved spatial discretisation rapidly dominates. 

As such, we are forced to use adapted results from our Haar wavelet discretisation as references, similar to \cite{Goffin2015, Goffin2015a}. We believe this is a ``best-effort'' approach, and we take significant care to ensure the adapted references we use are in the asymptotic regime for each problem. The standard and non-standard Haar decompositions are stabilised differently when adapted, so we use seperate reference solutions for each. We compare against the uniform Haar wavelet solutions where possible, performed convergence studies on the reference solutions with reduced thresholding tolerances, and ensure that the references are refined at least one order of refinement higher than any adapted solution we compare against. Even with our under-resolved spatial discretisation, some of these adapted references would require 1 \xten{13}--1\xten{14} DOFs to resolve uniformly, which helps show the challenge in computing true references in these problems. 

We present results in this Section from other angular discretisations for comparative purposes, namely uniform $P_n$, uniform LS P$_0$ FEM; and adapted linear octahedral wavelets \cite{Goffin2015, Goffin2015a}. This uniform LS P$_0$ FEM is different to the hierarchical P$_0$ FEM space our Haar wavelets are built upon, as they feature angular elements with the same azimuthal and polar length and equal area. The centre of these elements lays in the same position of the quadrature points in a level-symmetric S$_n$ method, in an attempt to compare against a low-order discretisation that doesn't suffer from the same clustering of elements around the poles as our uniform Haar discretisations. We denote this discretisation as the ``LS P$_0$ FEM''. To compute an angular mesh length for mesh convergence studies, we compute $2^{-l}\pi$ for the Haar discretisations, where $l$ is the maximum refinement level used throughout the domain (this gives the polar length of the angular elements across the equator of our angular discretisation). For the uniform LS P$_0$ FEM, we compute $(\pi/\textrm{no. angular DOFs})^{1/2}$. For the adapted linear octahedral wavelets, we use the same thresholding tolerance and compute error-metrics in the same fashion as our Haar wavelets. Again, given the under-resolved spatial discretisation, each of these different angular discretisation uses its own reference solution. Each of these methods uses the same multigrid-based iterative solver described in \secref{sec:Solver technology}. 

These other angular discretisations, along with the adapted Haar discretisation presented in this work have been implemented in FETCH2, the multi-physics, coupled Boltzmann transport code developed at the AMCG. There are some differences in the technology used by each of these discretisations that reflect the development process of this work. For example, we would not consider the linear octahedral wavelets \cite{Goffin2015, Goffin2015a} a scalable adapted angular discretisation, as fundamentally they are continuous within each octant. This makes it difficult to scalably compute our stabilisation as well as the eigendecomposition needed to apply boundary conditions. This restricts the linear octahedral wavelets to using a matrix-based iterative solver, which is also not scalable given the non-diagonal sparsity of the angular matrices. We find that in general, we cannot exceed more than 4-5 levels of refinement with reasonable runtimes/memory consumption, which is not sufficient for difficult problems. 

Similarly, the implementation of the standard Haar decomposition shown in this work is not completely scalable. Though it uses a matrix-free iterative method, it does assemble and use the adapted sparse angular matrices, instead of using the scalable FWT discussed in \secref{sec:Wavelets}. In an effort to reduce the memory consumption and make the standard decomposition more performanent, we construct these adapted sparse angular matrices only on a single reference octant that includes all active angles throughout the domain. We make use of the hierarchical structure of the angular matrices and only build new blocks in the matrices after each adapt step. With these modifications, we find the standard decomposition scales well up to 9-10 levels of refinement; after this, the increasing sparsity of the angular matrices severly affects the runtime/memory consumption.

Indeed it was this non-scalable implementation, the impact of the ``long, thin'' wavelets in 3D and our desire to simulate problems with highly anisotropic flux that lead to the development of the fully matrix-free, scalable FWT-based non-standard Haar decomposition. We decided to include these non-scalable implementations in the results as they show the impact of not considering every aspect of the discretisation/implementation/solver when building an angular adaptive algorithm.

We should note that for all the adapted simulations shown in this Section, the runtime shown includes all the adapt steps (i.e., all the linear solves performed, computation of error metrics, refinement/coarsening, etc) required to get to that order. For example, if we perform a regular adaptive simulation with 5 adapt steps, the runtime shown includes the time required to perform 5 linear solves. An equivalent goal-based simulation would include 9 linear solves; we do not solve the adjoint problem on the final adapt step. Unless otherwise noted, all linear solves were performed in serial to an absolute/relative tolerance of 1\xten{-10}.
\subsection{Brunner lattice problem}
\label{sec:Brunner lattice}
\begin{figure}[th]
\centering
\includegraphics[width=0.55\textwidth]{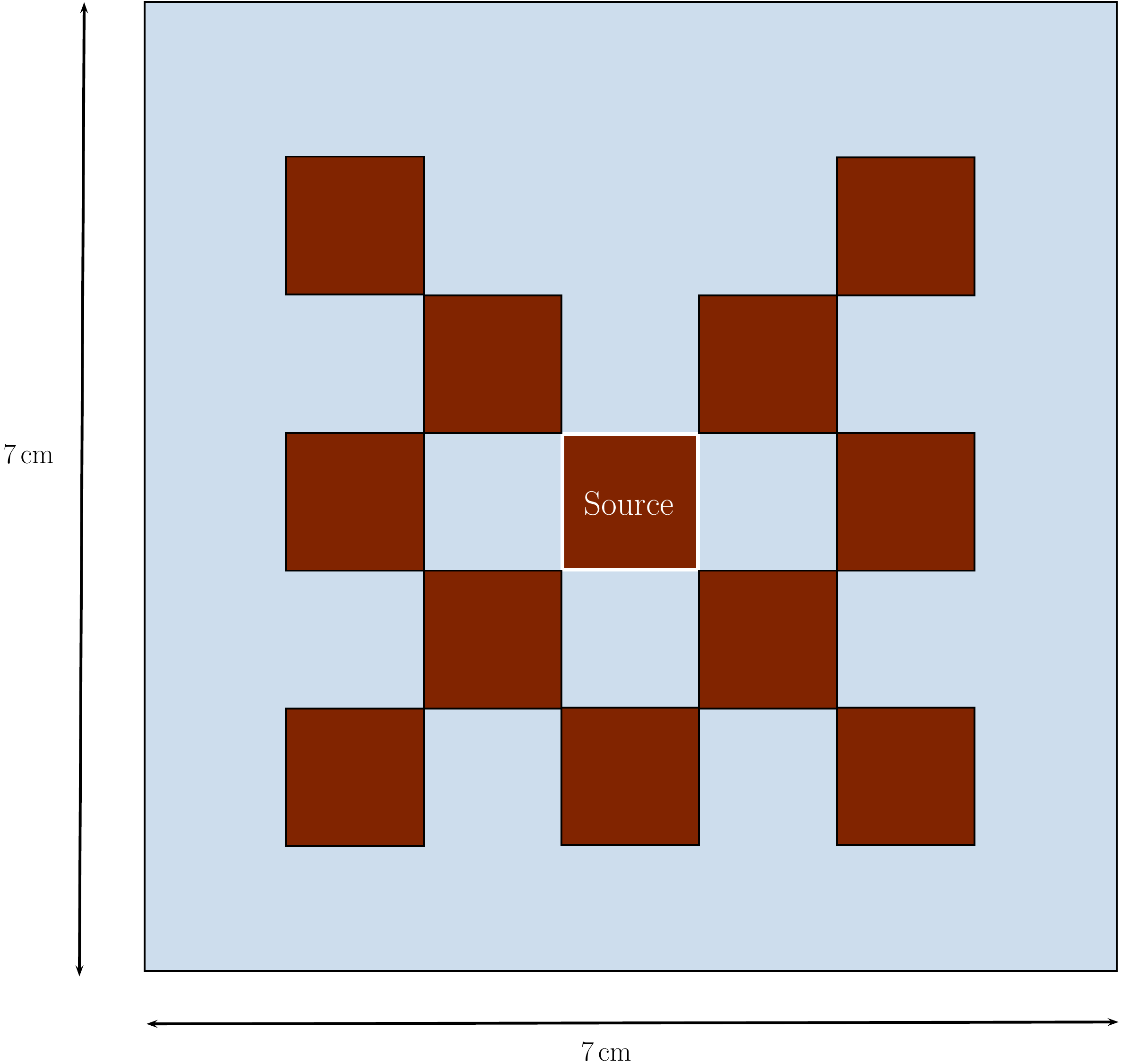}
\caption{Schematic of the Brunner lattice problem \cite{Brunner2002}. The red region is a
pure absorber (10 cm$^{-1}$), the blue region is pure scatter (1 cm$^{-1}$), with the
the white bordered region a source with strength 1. The origin is placed in the bottom left corner of the domain.}
\label{fig:brunner_schematic}
\end{figure}

The first example is the lattice problem from \cite{Brunner2002}, a schematic of which is shown in \fref{fig:brunner_schematic}. We discretise this problem in space with an unstructured triangular mesh with 3378 elements (1690 CG nodes and 10,134 DG nodes). This problem has regions of smoothness in angle, but still features discontinuities, particularly in the corners between the scattering and absorbing regions. We use regular adaptivity in this problem, as large regions of the phase-space are important to the final solution. The reference solution for the standard Haar decomposition used angular adaptivity with a maximum of 7 levels of refinement, 7 adapt steps and a thresholding tolerance of 1\xten{-7}, using 98M DOFs in the final adapt step (uniformly this would have required 190M DOFs). The non-standard Haar adapted reference used a maximum of 8 levels of refinement, 8 adapt steps, with a thresholding tolerance of 1\xten{-6}, using 263M DOFs in the final adapt step (uniformly this would have required 780M DOFs). The uniform LS P$_0$ FEM used a reference with 11,400 elements in angle (giving a discretisation similar to S$_{200}$), using 134M DOFs. The uniform P$_n$ used a P$_{101}$ reference, with 5253 DOFs in angle, using 60M DOFs. We are computing the relative error in the 2-norm of the scalar flux in this problem. This is a very forgiving and smoothly-varying metric in this problem, but it is a excellent test to ensure our wavelet thresholding scheme is behaving correctly (given the discussion in \secref{sec:Angular adaptivity}).

\begin{figure}[th]
\centering
\subfloat[][Standard Haar decomposition]{\label{fig:brunner_convg_standard}\includegraphics[width =0.47\textwidth]{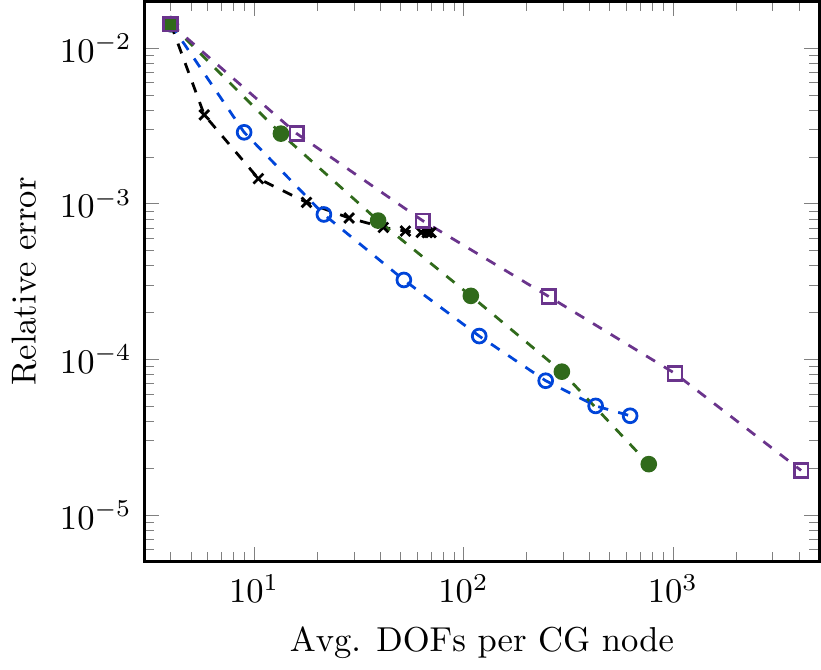}}
\subfloat[][Non-standard Haar decomposition]{\label{fig:brunner_convg_nonstand}\includegraphics[width =0.47\textwidth]{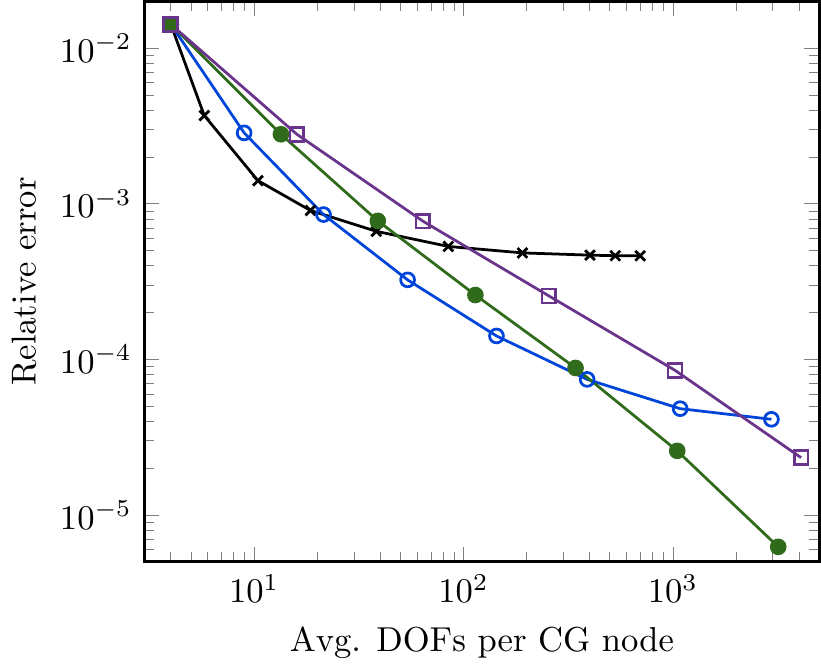}}\\
\caption{Convergence of the regular angular adaptivity, in the relative error of the 2-norm of the scalar flux across the domain, for the Brunner problem vs the average number of DOF per CG node of the spatial mesh. The x, \textcolor{matlabblue}{o} and \textcolor{foliagegreen}{\CIRCLE} markers use threshold coefficients 1\xten{-3}, 1\xten{-4} and 1\xten{-5}, respectively, with the \textcolor{gaylordpurple}{\Square} uniform (unadapted).}
\label{fig:brunner_convg_adapt}
\end{figure}

One of the key parameters in our regular adapt process is the thresholding tolerance we use; too small of a tolerance and the adapt process will add unecessary angles, too large and it will not add enough to reach a desired error. \fref{fig:brunner_convg_adapt} shows the error given three different thresholding tolerances, compared to uniform refinement with both the standard and non-standard Haar decompositions. We can see in both cases that too large a tolerance leads to the convergence stagnating. If the user desires a solution with large error however, it is more efficient to chose a  large tolerance, as it uses less angles in the region prior to stagnation, compared with the smaller tolerance. We can see that the smallest thresholding tolerance, 1\xten{-5} has resulted in relative errors at each step that is very close to that produced by uniform refinement, while using fewer DOFs in angle.  

\begin{figure}[th]
\centering
\subfloat[][Standard Haar decomposition]{\label{fig:brunner_time_standard}\includegraphics[width =0.47\textwidth]{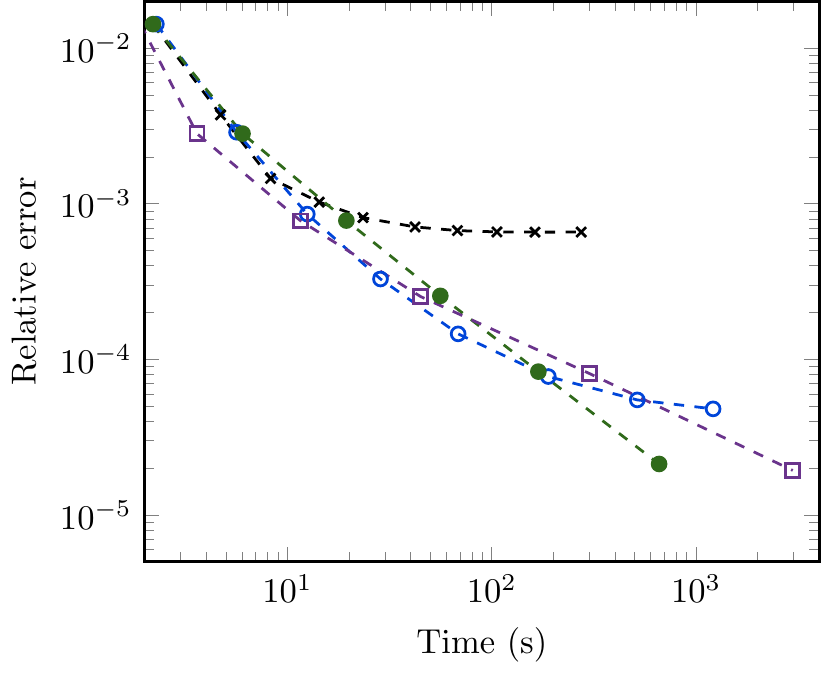}}
\subfloat[][Non-standard Haar decomposition]{\label{fig:brunner_time_nonstand}\includegraphics[width =0.47\textwidth]{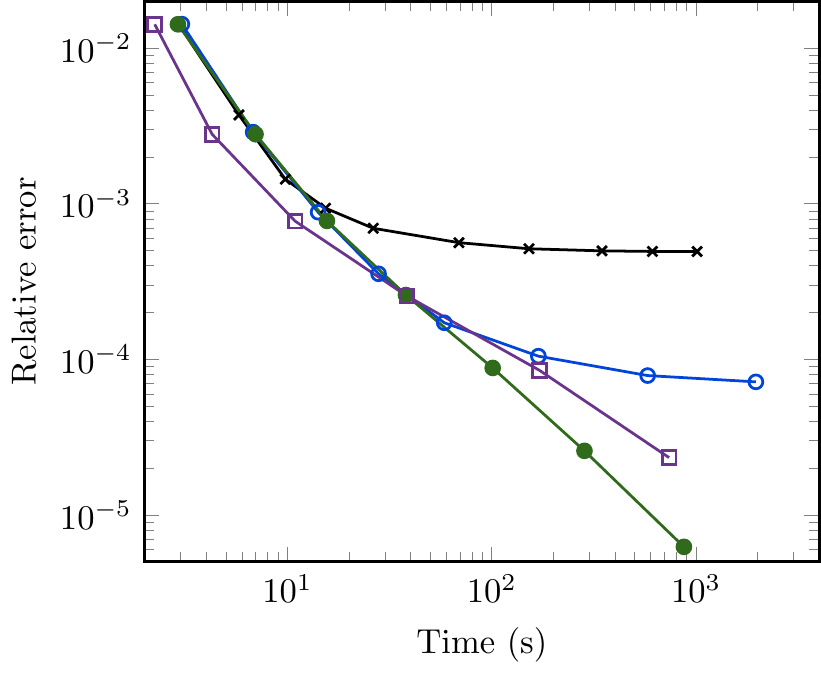}}\\
\caption{Convergence of the regular angular adaptivity, in the relative error of the 2-norm of the scalar flux across the domain, for the Brunner problem vs the total runtime. The x, \textcolor{matlabblue}{o} and \textcolor{foliagegreen}{\CIRCLE} markers use threshold coefficients 1\xten{-3}, 1\xten{-4} and 1\xten{-5}, respectively, with the \textcolor{gaylordpurple}{\Square} uniform (unadapted).}
\label{fig:brunner_time_adapt}
\end{figure}

\fref{fig:brunner_time_adapt} shows the total runtime for the standard and non-standard Haar decompositions shown in \fref{fig:brunner_convg_adapt}. We can see that these figures look very similar to those in \fref{fig:brunner_convg_adapt}, which shows strong evidence that our adaptive algorithm is scalable in this problem, with runtime related directly to the number of DOFs applied by our adaptive process. This encapsulates the scalability of all aspects of our algorithm, including our FWT-based matrix-free matrix-vector product, iterative solver, refinement process, calculation of error metric, etc. In particular, for both the standard and non-standard Haar decompositions, the adaptive simulation with thresholding tolerance 1\xten{-5} shows a fixed decrease in error with runtime. As mentioned previously, using large tolerances before stagnation could be benefitial in terms of DOFs applied, though both Figures \ref{fig:brunner_time_standard} \& \ref{fig:brunner_time_nonstand} show that the difference in runtime between the different tolerances for low levels of refinement are negligible. Using threshold tolerance of 1\xten{-3} and 1\xten{-4} in the second adapt step of \fref{fig:brunner_convg_nonstand} for example, results in an average of 5.78 and 8.96 DOF per CG node, respectively, though the runtimes are 5.79s and 6.79s. This is likely because of the element packing that occurs in our matrix-free matrix-vector product, combined with vectorisation, which for small angle sizes, results in similar runtimes. 
 
As might be expected, we see that the uniform simulations for 4 levels of refinement and less are quicker than any of the adaptive simulations, as each of the uniform simulations is only performing a single linear solve, compared with the adaptive simulations which are performing one per adapt step. Only once the difference in DOFs applied becomes significant for higher levels of refinement do we see the adapted simulations outperform the uniform in this problem. 

\begin{figure}[th]
\centering
\subfloat[][Standard Haar decomposition - 21M DOFs]{\label{fig:brunner_no_angles_1e-5_step_7_stand}\includegraphics[width =0.4\textwidth]{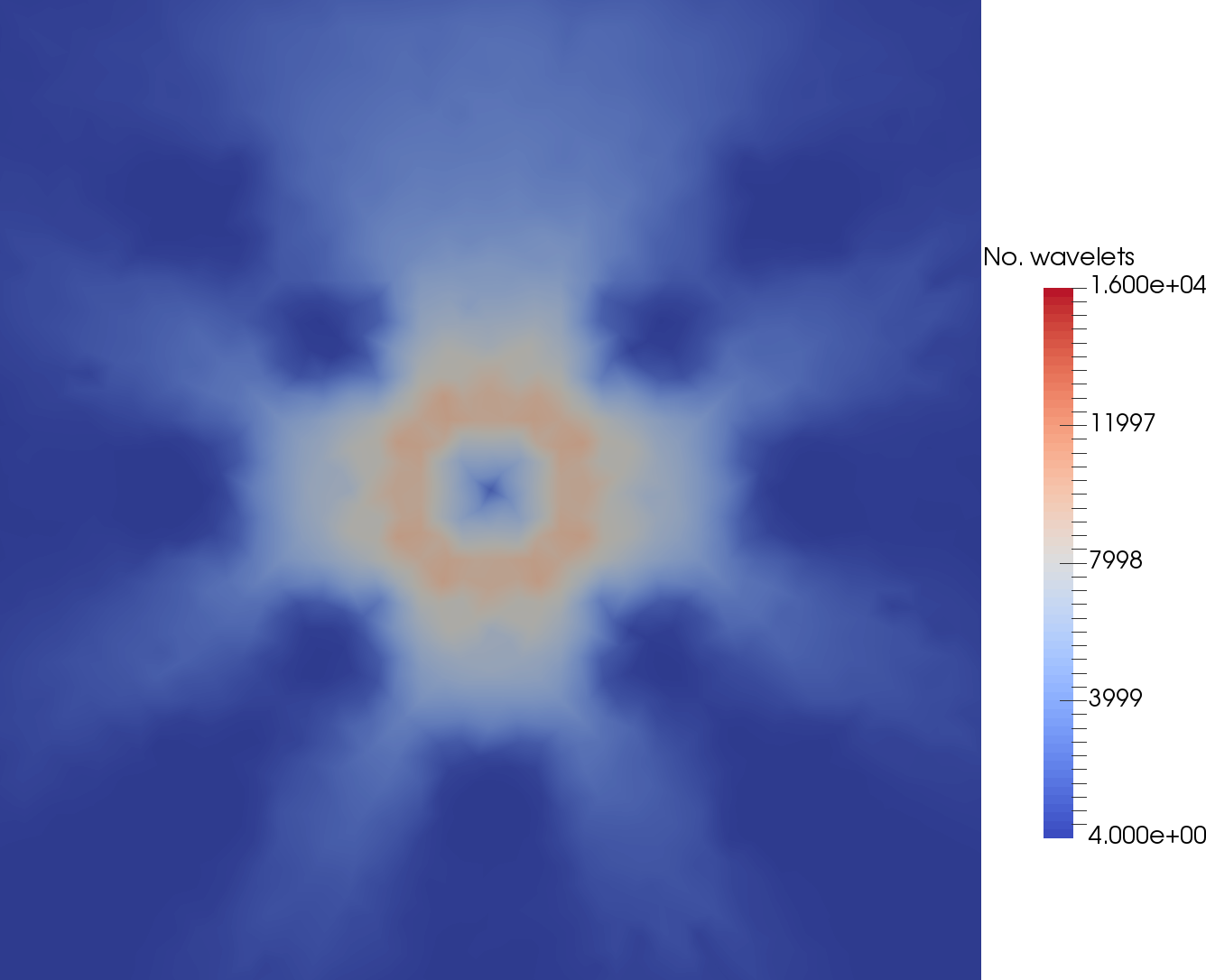}} \hspace{0.5cm}
\subfloat[][Non-standard Haar decomposition - 37M DOFs]{\label{fig:brunner_no_angles_1e-5_step_7_nonstand}\includegraphics[width =0.4\textwidth]{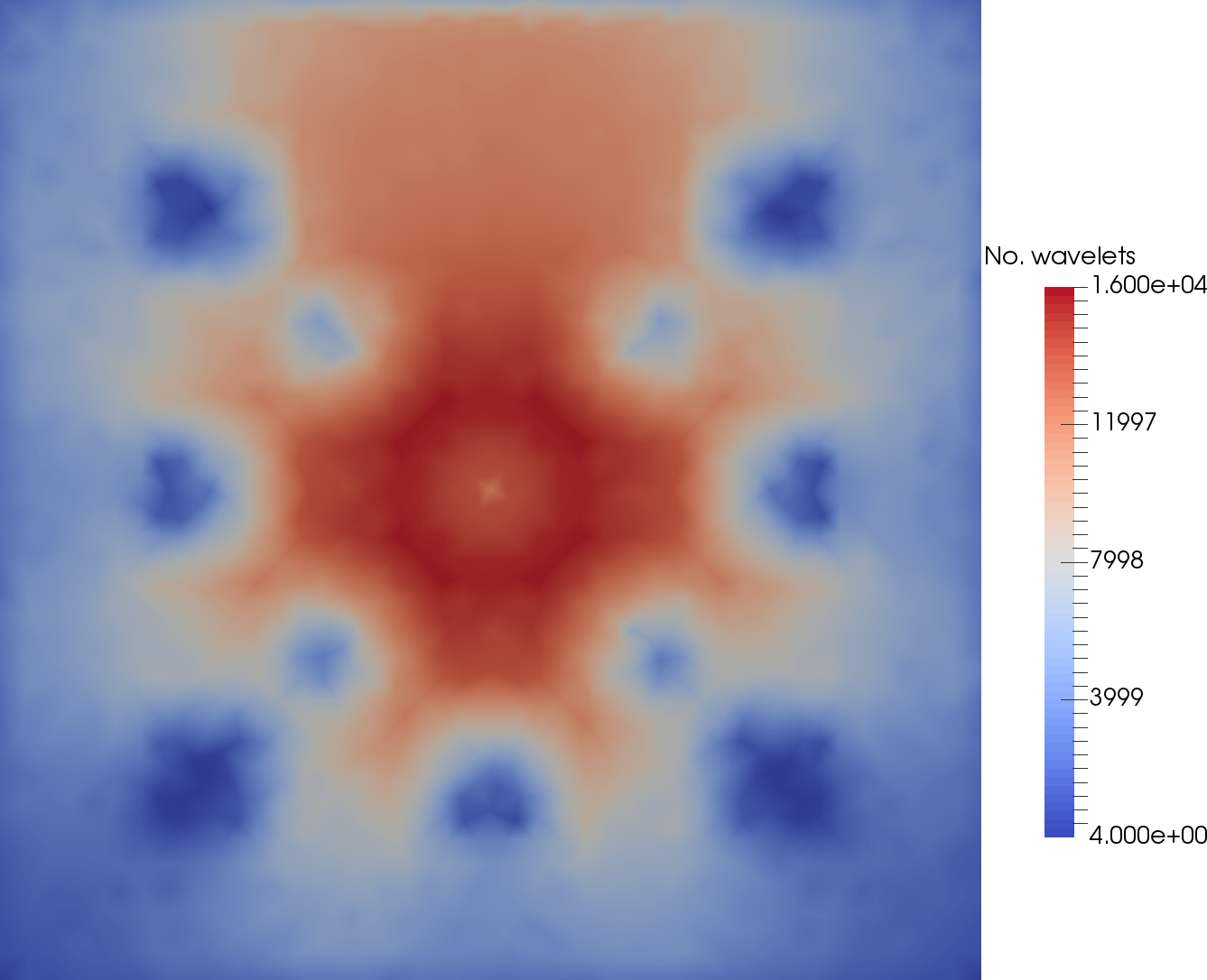}}\\
\caption{Number of wavelets across the spatial domain for the Brunner problem, plotted on the CG mesh, on the 7th step of regular angular adaptivity with threshold coefficient 1\xten{-5} (i.e., the last \textcolor{foliagegreen}{\CIRCLE} mark in Figures \ref{fig:brunner_convg} \& \ref{fig:brunner_time}).}
\label{fig:brunner_no_angles}
\end{figure}

\begin{figure}[th]
\centering
\subfloat[][$x=3, y=3.5$]{\label{fig:brunner_flux_leftcentre_5_5}\includegraphics[width =0.4\textwidth]{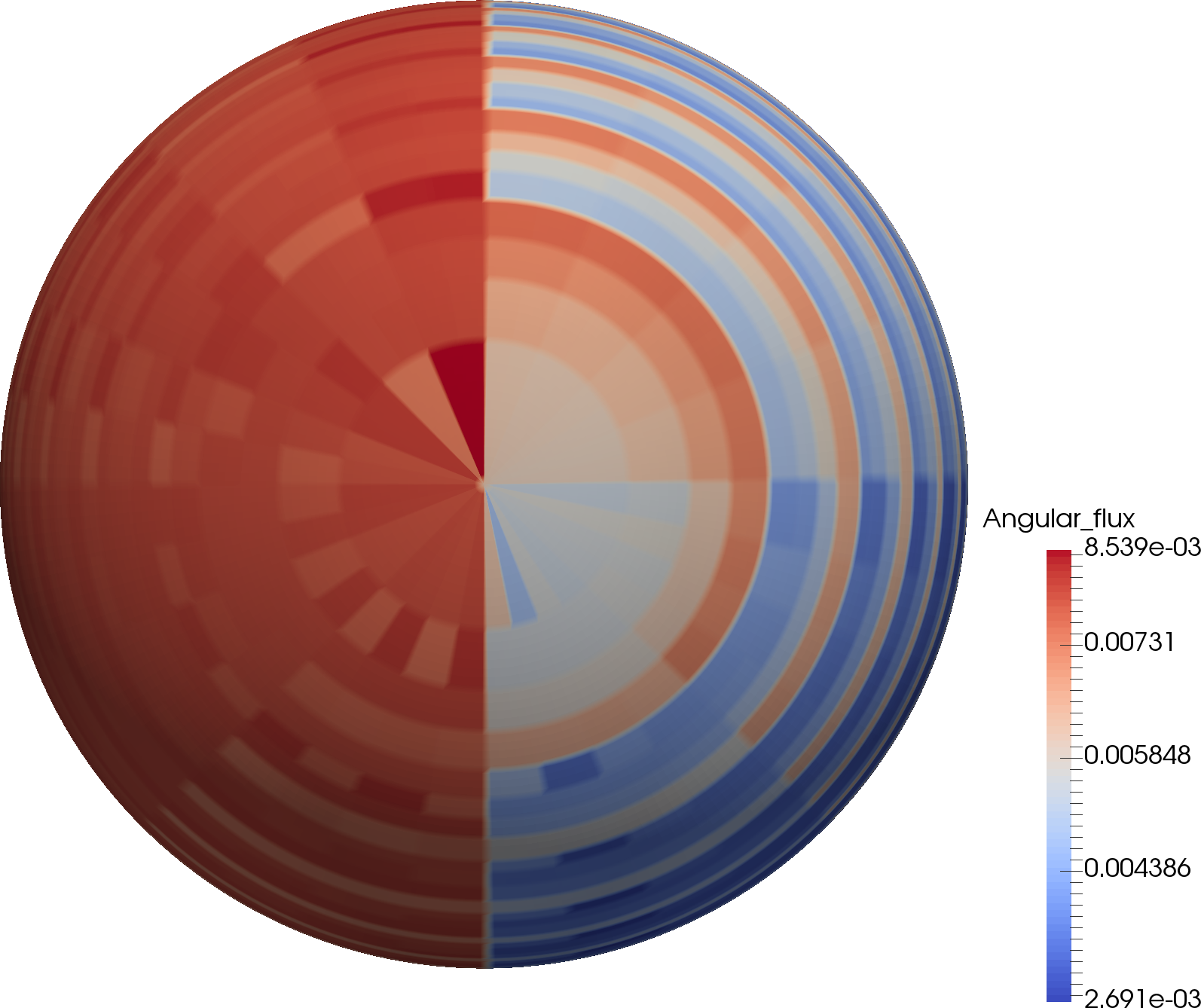}} \hspace{0.5cm}
\subfloat[][$x=2.5, y=2.5$]{\label{fig:brunner_flux_leftbottom_5_5}\includegraphics[width =0.4\textwidth]{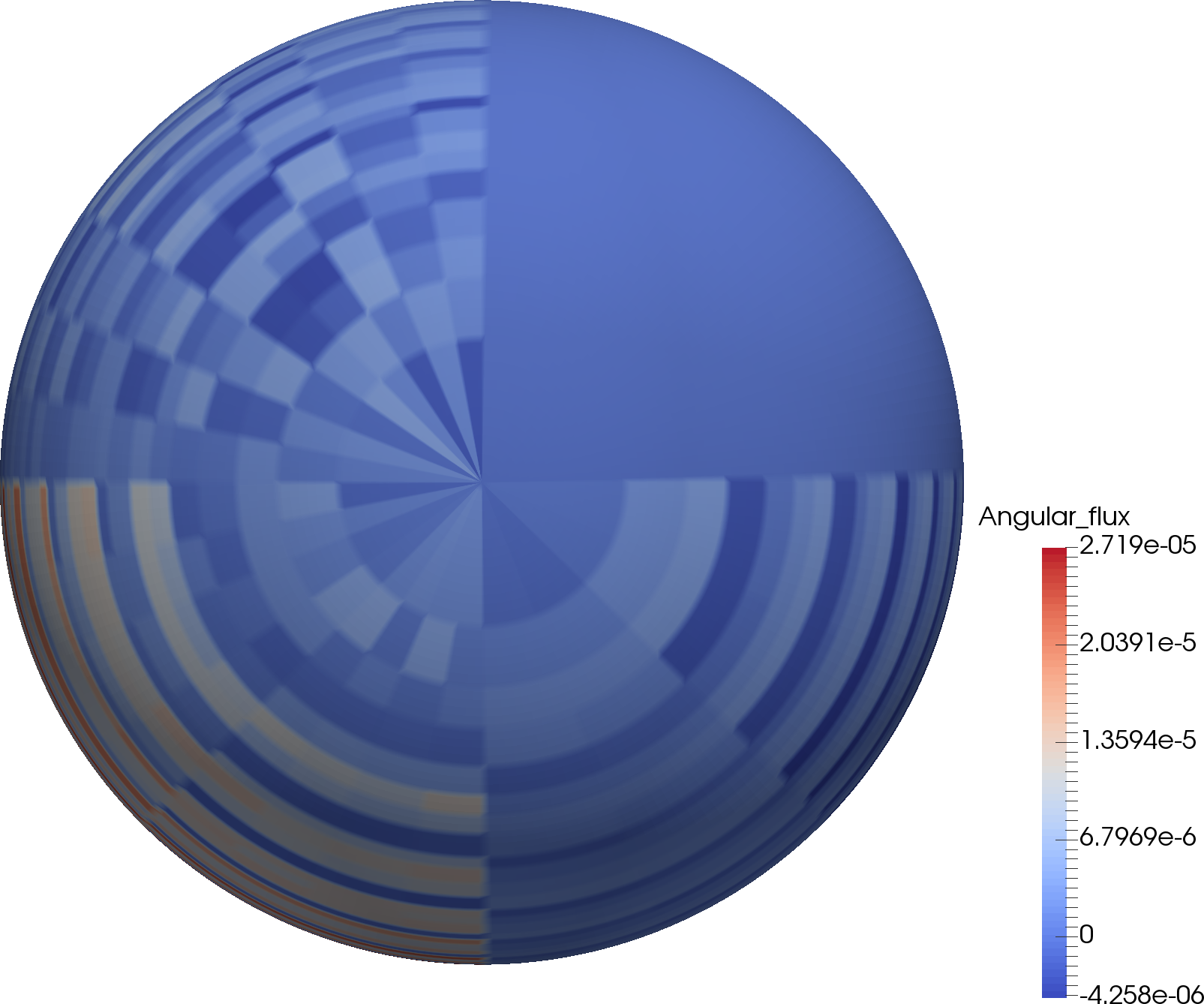}}\\
\caption{Angular flux in the Brunner problem at different spatial positions, with regular angular adaptivity given a standard Haar decomposition after 5 adapt steps, with threshold coefficient 1\xten{-5}. The camera is pointed in the $-z$ direction.}
\label{fig:brunner_flux}
\end{figure}

\fref{fig:brunner_no_angles} shows the spatial distribution of adapted angles applied by the standard and non-standard Haar decompositions, for the 7th adapt step with a thresholding tolerance of 1\xten{-5}. In \fref{fig:brunner_no_angles} we can see that both discretisations have applied the greatest number of angles in the areas where we could expect the angular flux to be large in this problem. We can also see that the standard decomposition, in general, has used less angles than the non-standard, given the same thresholding tolerance. This is to be expected in a problem with two spatial dimensions, as the ``long, thin'' wavelets can capture the symmetry in $z$ more effectively than the non-standard wavelets with fixed suport. This is shown in \fref{fig:brunner_flux}, where the angular flux after 5 adapt steps with tolerance 1\xten{-5} is plotted at two different points in space, for the standard Haar decomposition. Firstly, we can see that both points in space have adapted in angle differently, anistropically across the sphere. This is particularly evident in \fref{fig:brunner_flux_leftbottom_5_5}, which is the angular distribution in one of the absorber regions to the bottom-left of the source region. We can see that the octant pointing back towards the source region has not adapted at all, as very little flux is directed back towards the source, with the highest flux, and hence most adapted region, in the octant pointing away from the source. Furthermore, the symmetry in $z$ is noticeable, particularly in \fref{fig:brunner_flux_leftcentre_5_5}, where long, thin bands of high flux are visible, which can be captured by the standard decomposition with fewer DOF than the non-standard.

\begin{figure}[th]
\centering
\subfloat[][Error vs CDOFs]{\label{fig:brunner_convg}\includegraphics[width =0.47\textwidth]{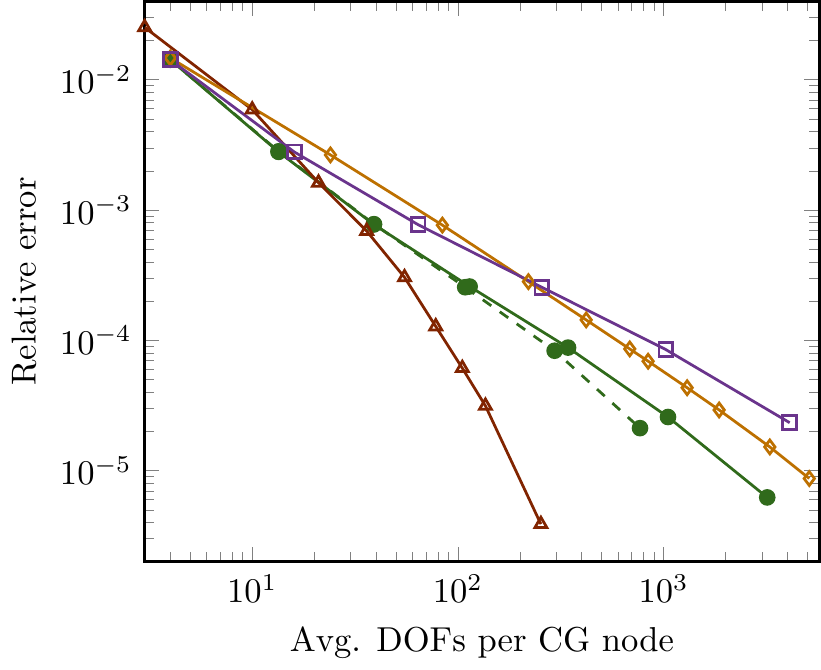}}
\subfloat[][Error vs mesh length]{\label{fig:brunner_convg_mesh_length}\includegraphics[width =0.47\textwidth]{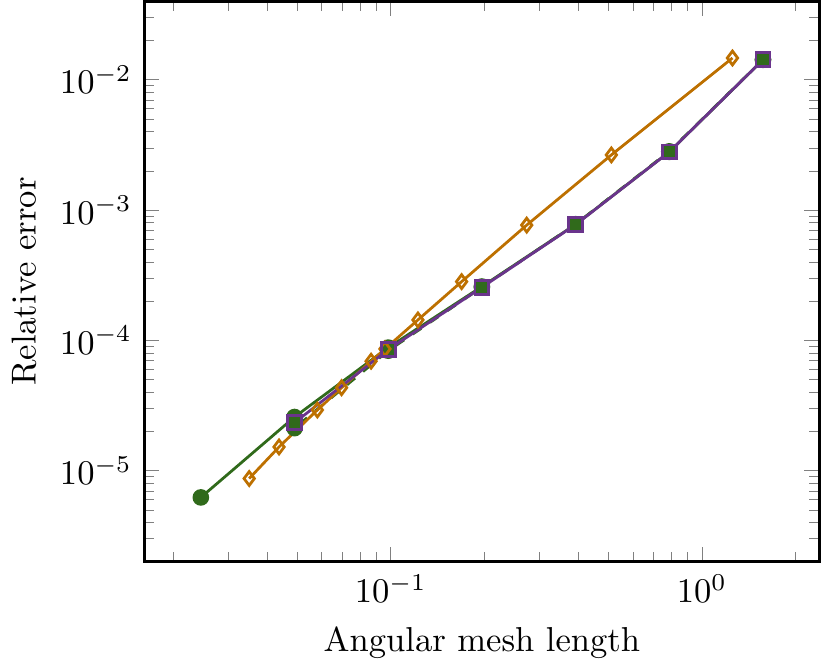}}\\
\subfloat[][Error vs total runtime]{\label{fig:brunner_time}\includegraphics[width =0.47\textwidth]{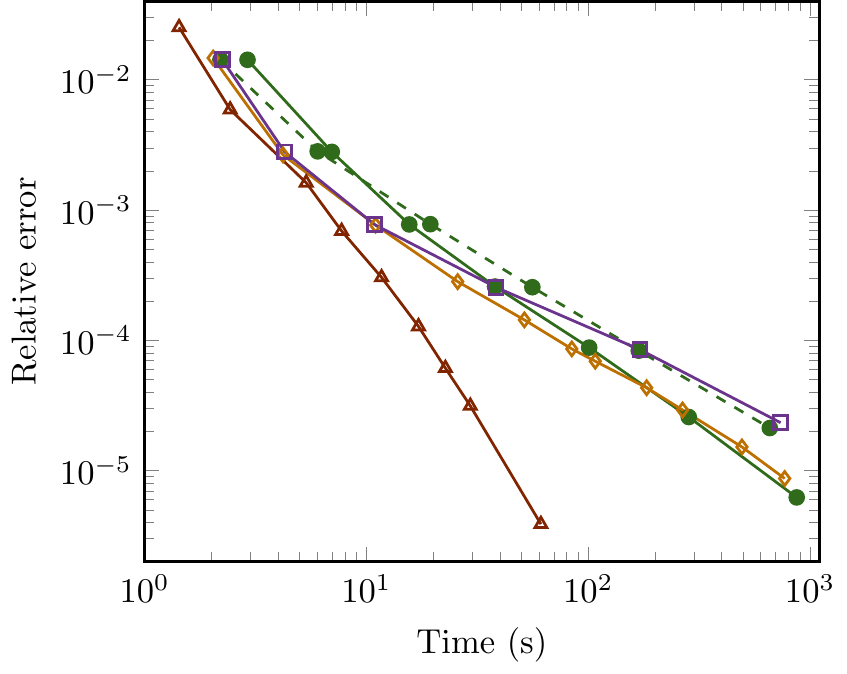}}
\caption{Comparison of the relative error of the 2-norm of the scalar flux across the domain, for different angular discretisations, for the Brunner problem. The \textcolor{foliagegreen}{\CIRCLE} are regular Haar adapts with threshold coefficient 1\xten{-5}. The dashed line is the standard Haar decomposition and the solid the non-standard (from Figures \ref{fig:brunner_convg_adapt} \& \ref{fig:brunner_time_adapt}). The \textcolor{fireenginered}{$\bigtriangleup$} is uniform P$_n$ and \textcolor{deludedorange}{$\diamond$} is uniform LS P$_0$ FEM.}
\label{fig:brunner_result}
\end{figure}

Finally, \fref{fig:brunner_result} shows a comparison between the standard and non-standard Haar adapts, with thresholding tolerance 1\xten{-5} against other angular discretisations. We can see in \fref{fig:brunner_convg} that P$_n$ is well suited to this problem, converging per DOF much faster than the other discretisations, indicating this problem posseses sufficient regularity that the high-order spectral nature of P$_n$ is beneficial. P$_n$ discretisations are spectral, so we would not expect the runtime to scale linearly with the number of DOF, but we can see in \fref{fig:brunner_time}, that due to the high-order convergence, the P$_{27}$ discretisation still outperforms the low-order methods in runtime. As expected, we see that the low-order methods all converge at much the same rate per DOF. Similarly, in \fref{fig:brunner_convg_mesh_length} we can see that the convergence vs angular mesh length for the low-order methods is also similar. Importantly, the uniform Haar, and both the adapted standard and non-standard Haar decompositions all converge at the same rate. Least-squares fiting a linear curve to the results in \fref{fig:brunner_convg_mesh_length} gives the uniform S$_n$ and the non-standard adapted Haars converging at orders 2.07 and 1.79, respectively.

Interestingly, we see in \fref{fig:brunner_convg} that although the standard Haar decomposition outperforms the non-standard in DOFs used, \fref{fig:brunner_time} shows that the non-standard has a smaller runtime with larger number of adapt steps. This is the impact of constructing/using the sparse angular matrices for the standard decomposition, as opposed to using the scalable FWT as the non-standard, as discussed above (and this effect only becomes more pronounced as we increase the maximum refinement level). \fref{fig:brunner_time} also shows that for a large number of adapt steps, the adapted non-standard Haars is faster than the uniform LS P$_0$ FEM, even though it is performing a larger number of linear solves and the uniform Haar angular mesh clusters elements around the pole. This is one of the advantages of adapting on such a simple angular subdivision scheme; even if the uniform equivalent does not provide a ``good'' mesh on the sphere, as is evidenced in \fref{fig:brunner_convg}, where the uniform LS P$_0$ FEM has lower error per DOF than the uniform Haar wavelet discretisation, an adapted mesh can. Clearly however, this problem is smooth enough that using high-order methods like P$_n$ is ideal. 
\begin{table}
\centering
\begin{tabular}{ l c c c c c c c}
\toprule
Adapt step: & 1 & 2 & 3 & 4 & 5 & 6 & 7\\
\midrule  
Cum. runtime ($\mu$s) per final DOFs: & 63.7 & 42.3 & 31.2 & 25 & 21.6 & 19.6 & 19.5 \\
Peak memory use: & 150.1 & 56.9 & 33.3 & 22.7 & 20.4 & 19.2 & 18.2\\
\bottomrule  
\end{tabular}
\caption{Runtime and peak memory used for the Brunner problem, for the non-standard Haar decomposition with threshold coefficient 1\xten{-5}. Peak memory use is on the heap (measured by massif) scaled to the size of the angular flux. The runtime is the cumulative runtime of all adapt steps up to that level, scaled by the NDOFs in the final adapt step.}
\label{tab:brunner_memory}
\end{table}

Finally, \tref{tab:brunner_memory} shows the runtime and memory usage for the non-standard Haar decomposition shown in \fref{fig:brunner_result} during the adapt process. We can see that the cumulative runtime per final DOF stabilises at around 19 $\mu$s with higher number of adapt steps, decreasing from a high of 63 $\mu$s with the coarsest angular discretisation. This is likely because of the increasing effect of vectorisation as our angle size increases. Importantly, this runtime is the cumulative runtime over all the adapt steps up to that level, scaled by the number of DOFs present in the final adapt step, not the cumulative number of DOFs, nor simply the time taken by the final adapt step. This means that even as we increase the number of adapt steps, the need to perform all the adapt steps prior to the maximum desired level does not impact our scalability; \tref{tab:brunner_memory} shows that the cumulative runtime per DOF as we increase the refinement level is fixed. 

Similarly, \tref{tab:brunner_memory} shows that the memory usage in early adapt steps is high, around 150 copies of the angular flux. This is because the size of the angular flux is small and proportionally, several setup stages (e.g., the setup of our matrix-free multigrid) uses considerable memory. However, these setup stages do not change as adaptivity is performed, and as the adapt process adds DOFs, the setup uses an increasingly small fraction of the peak memory. We can see that after 7 adapt steps, the peak memory use has settled to around 18 copies of the angular flux. Importantly, this is the \textit{adapted} angular flux, which shows that our memory use is not growing with adapt step and hence is scalable. Furthermore, this memory usage can be easily reduced. As a simple example, changing the restart parameter on our top-level FGMRES (see \secref{sec:Solver technology}) from 30 to 5 and removing the explicit storage of $\mat{D}^{-1}$ on the top level reduces this memory use to 12.2 copies of the angular flux, with minor differences in the converge of the iterative method (no difference in the iteration count) and a moderate increase in the runtime. 
\subsection{2D dogleg problem}
\label{sec:2D Dogleg}
\begin{figure}[th]
\centering
\includegraphics[width=0.45\textwidth]{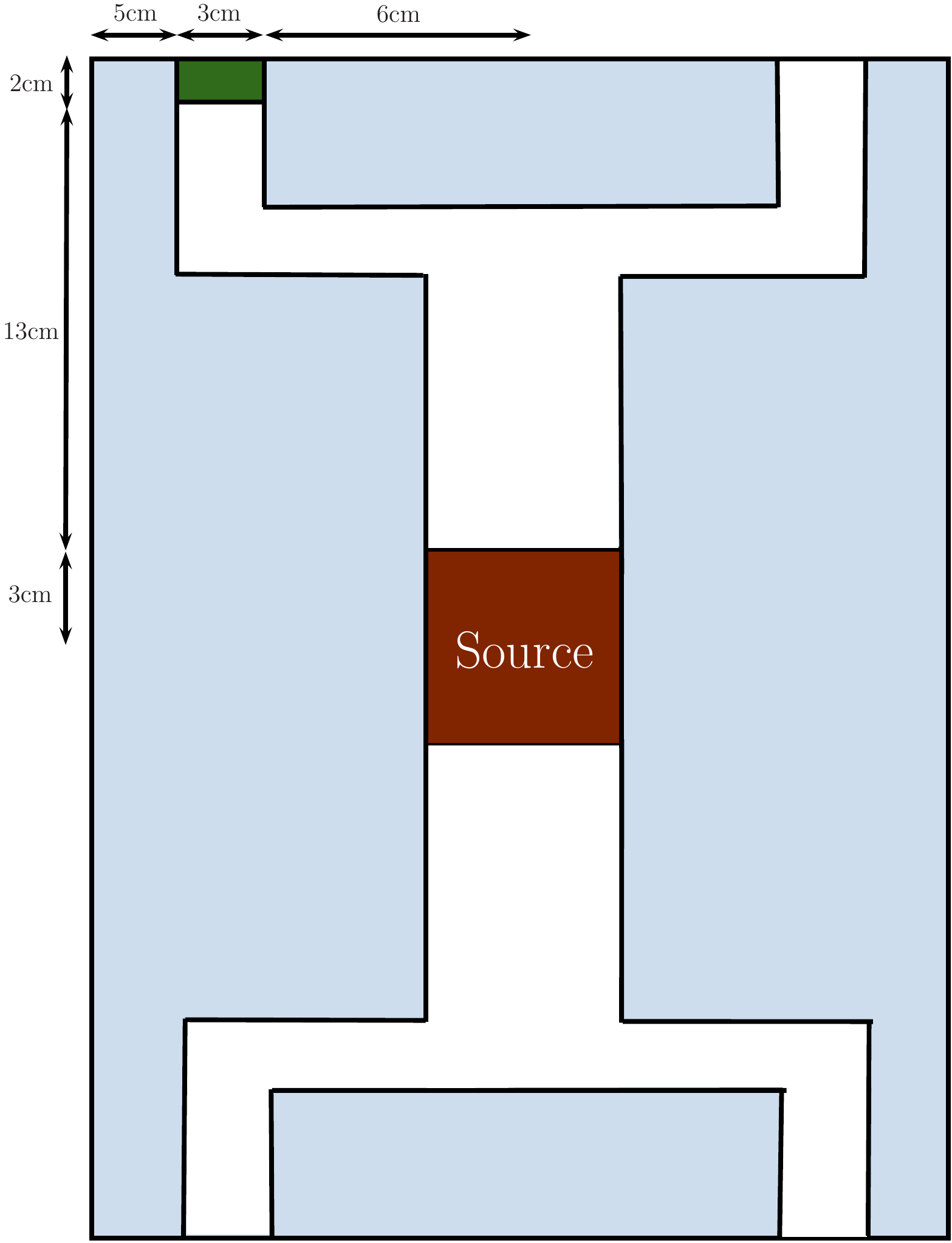}
\caption{Schematic of the 2D dogleg problem \cite{Goffin2015a}. The red region is source with strength 1 and a pure absorber (0.5 cm$^{-1}$), the blue region is a pure absorber (0.5 cm$^{-1}$), with the the white and green regions pure absorbers (0.005 cm$^{-1}$). The origin is placed in the middle of the source region.}
\label{fig:dogleg_schematic}
\end{figure}
The second example we use is a 2D duct problem \cite{Goffin2015a}, a schematic of which is shown in \fref{fig:dogleg_schematic}. We discretise this problem in space with an unstructured triangular mesh with 2824 elements (1477 CG nodes and 8472 DG nodes). This problem features highly anisotropic flux throughout the duct regions. We use goal-based adaptivity in this problem, with the goal being the average flux throughout the green region in \fref{fig:dogleg_schematic}. This metric is much more difficult to converge than that used in \secref{sec:Brunner lattice}, as it is sensitive to ray-effects, and hence is more representative of the types of problems our wavelet scheme would be suited to. The reference solution for the standard Haar decomposition used angular adaptivity with a maximum of 12 levels of refinement, 14 adapt steps and a target error of 1\xten{-5}. The non-standard Haar adapted reference (which was also used as the reference for the adapted linear octahedral wavelets) used angular adaptivity with a maximum of 15 levels of refinement, 17 adapt steps and a target error of 1 \xten{-5}, using 153M DOFs in the final adapt step (uniformly this would have required 1 \xten{13} DOFs). The uniform LS P$_0$ FEM used a reference with 61,600 elements in angle (giving a discretisation similar to S$_{350}$), using 612M DOFs. The uniform P$_n$ used a P$_{91}$ reference, with 4278 DOFs in angle, using 42M DOFs.
\begin{figure}[th]
\centering
\subfloat[][Standard Haar decomposition]{\label{fig:dogleg_convg_standard}\includegraphics[width =0.47\textwidth]{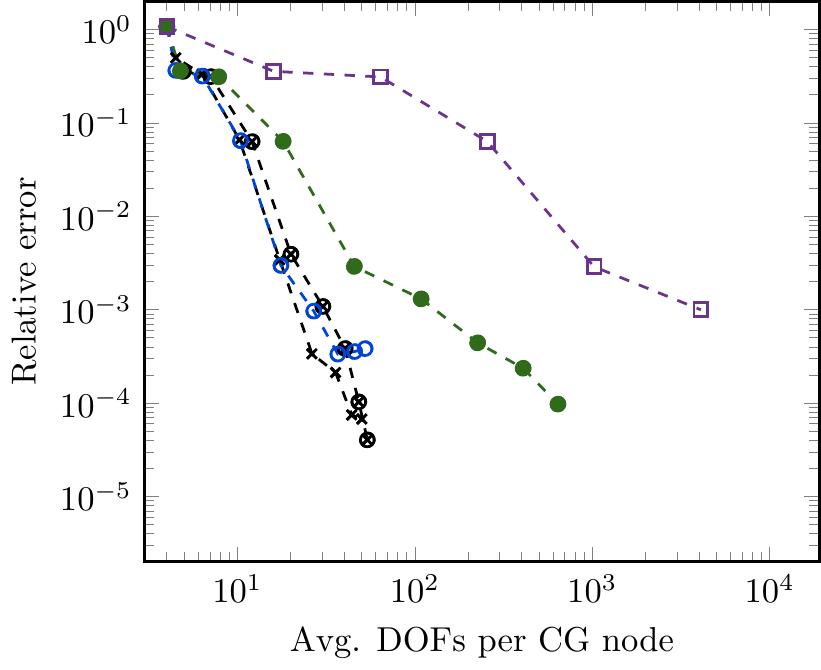}}
\subfloat[][Non-standard Haar decomposition]{\label{fig:dogleg_convg_nonstand}\includegraphics[width =0.47\textwidth]{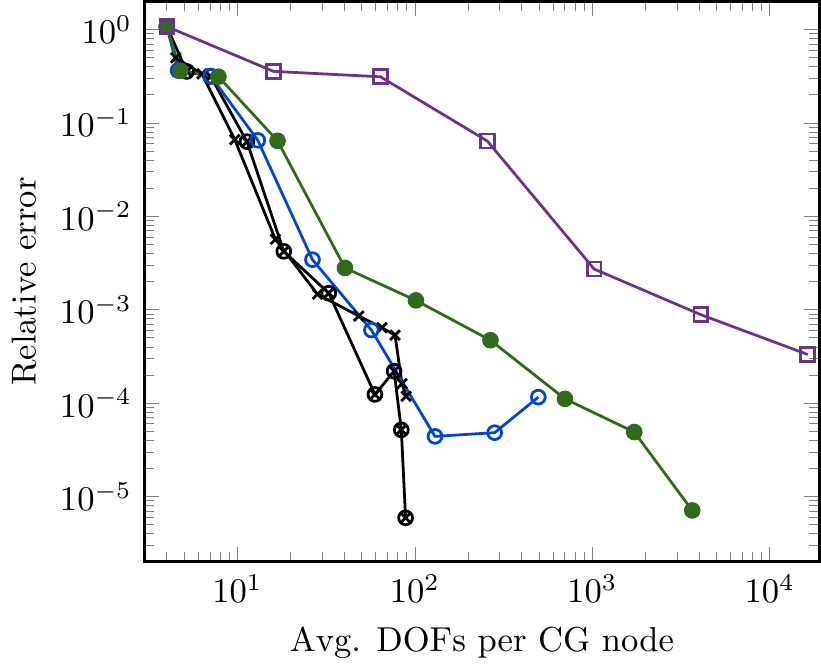}}\\
\caption{Convergence of the goal-based angular adaptivity, in the relative error of the detector response, for the 2D dogleg problem vs the average number of DOF per CG node of the spatial mesh. The x, \textcolor{matlabblue}{o} and \textcolor{foliagegreen}{\CIRCLE} markers use target errors 1\xten{-3}, 1\xten{-4} and 1\xten{-5}, respectively, with the \textcolor{gaylordpurple}{\Square} uniform (unadapted). The $\otimes$ marker uses target error 1\xten{-3} but performs reduced tolerance linear solves then takes an extra adapt step after reaching the maximum order with tight linear solve tolerance.}
\label{fig:dogleg_convg_adapt}
\end{figure}
\begin{figure}[th]
\centering
\subfloat[][Standard Haar decomposition]{\label{fig:dogleg_time_standard}\includegraphics[width =0.47\textwidth]{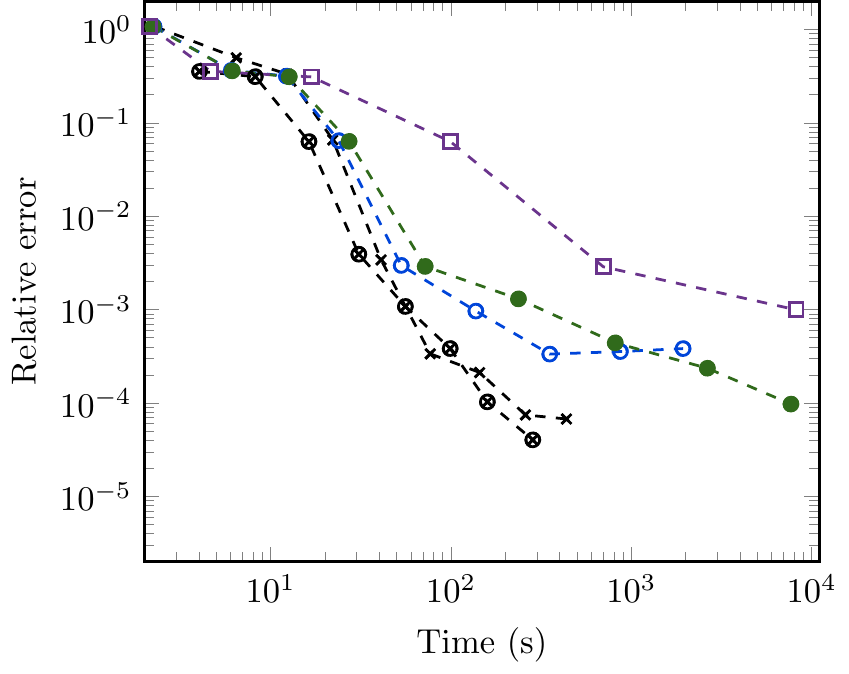}}
\subfloat[][Non-standard Haar decomposition]{\label{fig:dogleg_time_nonstand}\includegraphics[width =0.47\textwidth]{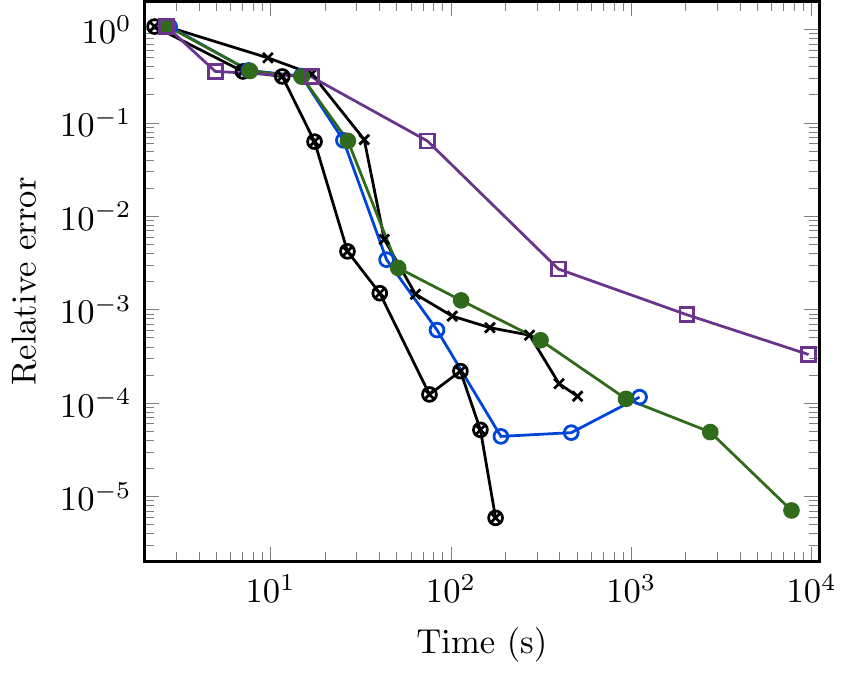}}\\
\caption{Convergence of the goal-based angular adaptivity, in the relative error of the detector response, for the 2D dogleg problem vs the total runtime. The x, \textcolor{matlabblue}{o} and \textcolor{foliagegreen}{\CIRCLE} markers use target errors 1\xten{-3}, 1\xten{-4} and 1\xten{-5}, respectively, with the \textcolor{gaylordpurple}{\Square} uniform (unadapted). The $\otimes$ marker uses target error 1\xten{-3} but performs reduced tolerance linear solves then takes an extra adapt step after reaching the maximum order with tight linear solve tolerance.}
\label{fig:dogleg_time_adapt}
\end{figure}

\fref{fig:dogleg_convg_adapt} shows the impact of modifying the target error of our goal-based scheme (which affects the thresholding). We can see that for both the standard and non-standard Haar decompositions, we again see the stagnation of our adapt process when the target error is too large. Our adapted discretisations with a target error of 1\xten{-5}, uses at least 2 orders of magnitude fewer DOF in this problem when compared to the uniform to achieve the same error, though it appears the adapted methods are starting to converge more rapidly as we increase the levels of refinement. Using a target error of 1\xten{-5} in both discretisations produces an error roughly equivalent to the uniform discretisation after each step of refinement, and seems to provide reliable convergence. We can see however for large target errors, that our adapted methods no longer converges smoothly, due to the presence of significant ray-effects. We would consider these simulations to hit the asymptotic limit after 5 levels of refinement, as we can see the convergence rate for the uniform discretisations stabilises. 

\begin{figure}[th]
\centering
\subfloat[][Standard Haar decomposition - 500K DOFs]{\label{fig:dogleg_no_angles_1e-3_step_9_stand}\includegraphics[width =0.4\textwidth]{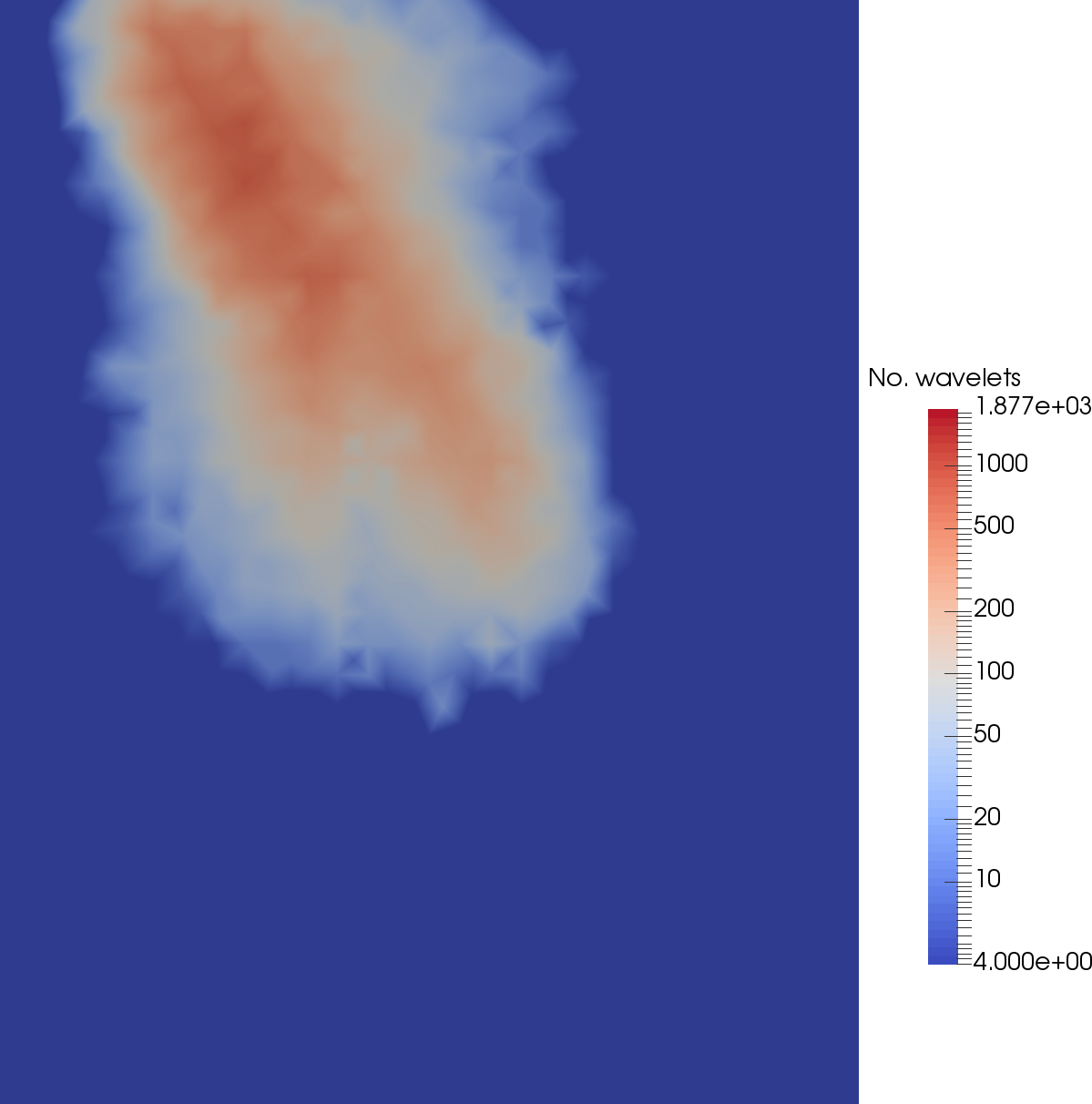}} \hspace{0.5cm}
\subfloat[][Non-standard Haar decomposition - 770K total CDOFs]{\label{fig:dogleg_no_angles_1e-3_step_9_nonstand}\includegraphics[width =0.4\textwidth]{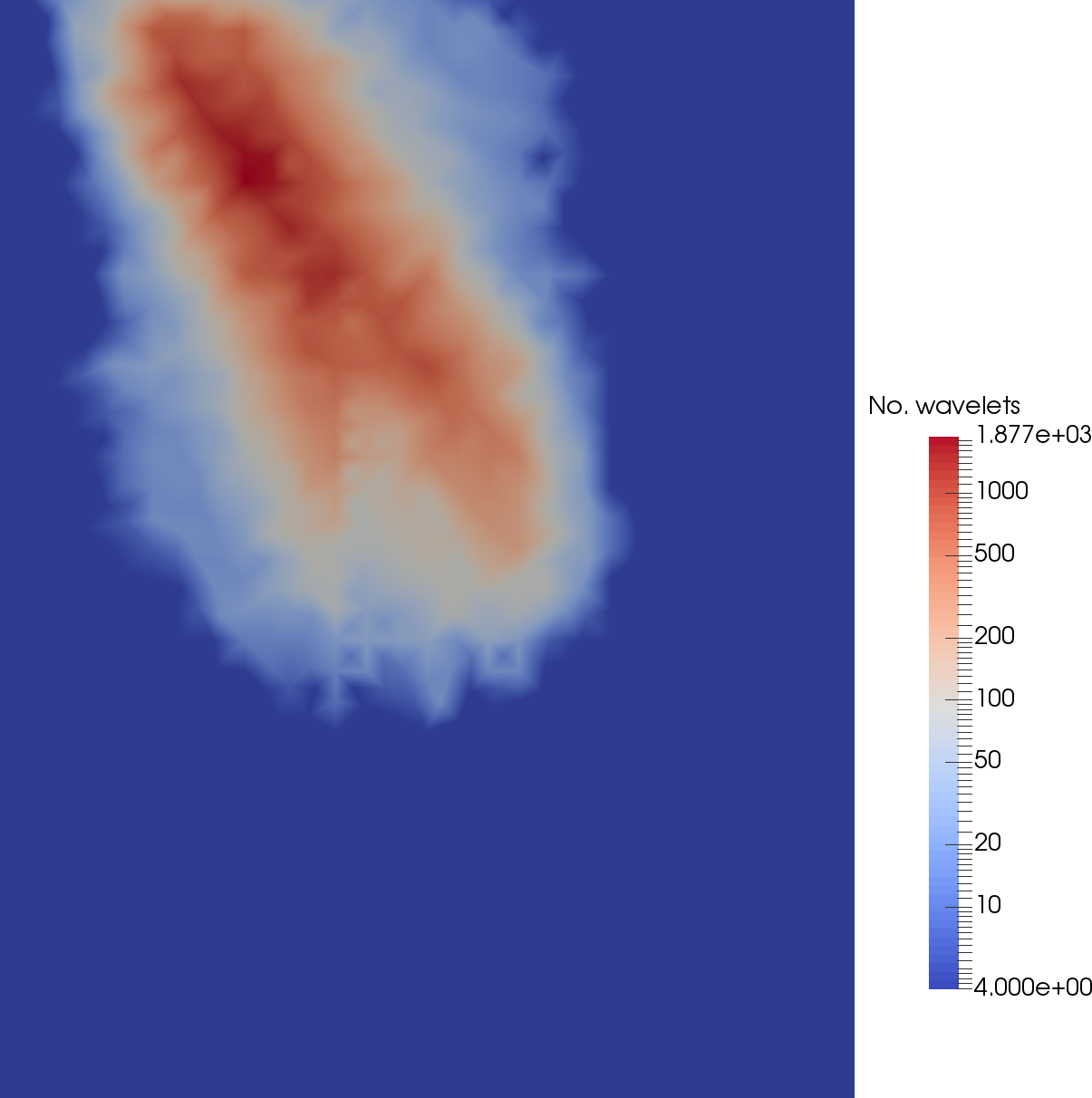}}\\
\caption{Number of wavelets across the spatial domain for the 2D dogleg problem, plotted on the CG mesh, on the 9th step of goal-based angular adaptivity with error target 1\xten{-3}.}
\label{fig:dogleg_no_angles}
\end{figure}

We found that for this problem, it can be benefitial to allow the adapt process to ``settle-down'' once it has hit the maximum level of refinement, by allowing an extra adapt step at that maximum order. \fref{fig:dogleg_convg_adapt} also shows the results from using a large error target of 1\xten{-3}, but allowing this extra step, combined with reducing the tolerance on our iterative method in all steps except this final settling step (in an attempt to decrease the runtime, this is discussed further below). For the non-standard case in \fref{fig:dogleg_no_angles_1e-3_step_9_nonstand}, the results from this show excellent convergence, producing a relative error of $\sim5.9$\xten{-6} with many orders of magnitude fewer DOF than a uniform simulation would require, and even using roughly two orders of magnitude less DOF than the adapted case with target error 1\xten{-5}. The convergence however, is not as reliable as when using the target error of 1\xten{-5} and the same number of adapt steps/max refinement level. There are two plausible reasons we are seeing benefit in allowing this settling process with a larger target error. The first is that streaming paths at a set level of refinement trigger refinement in other spatial regions, that can only be detected after that level of refinement has been reached. The second thought is that our error metric with small target errors (e.g, 1\xten{-5}) is too conservative and is applying angular DOFs that are unnecessary. We will discuss this further below. 

\fref{fig:dogleg_time_adapt} shows the runtimes from the adapted simulations in \fref{fig:dogleg_convg_adapt}. Again we see that the adapted simulations outperform the uniform simulations considerably, and that the convergence vs runtime looks very similar to that against the DOF. In particular, the non-standard Haar decomposition shown in \fref{fig:dogleg_time_nonstand} with the extra settling step takes only 176s to perform 11 adapt steps, with a maximum refinement of 10 levels (this gives a minimum solid angle of 5.9\xten{-6} sr). In both the standard and non-standard cases, we can see the reduction in runtime for early adapt steps resulting from the reduced tolerance linear solve as part of the settling case. We chose not to experiment any further with different permutations on our adapt, e.g., using more settlings steps, reducing the iterative method tolerance further (or tying it directly to our goal-based error metric), or even increasing the maximum refinement level by more than one after each adapt step. These modifications could likely allow even quicker adapted simulations, but we leave their investigation to future work; our key contribution here is showing scalable adapts, as even using a (possibly overly) small target error still outperforms uniform simulations considerably.
\begin{figure}[th]
\centering
\includegraphics[width=0.45\textwidth]{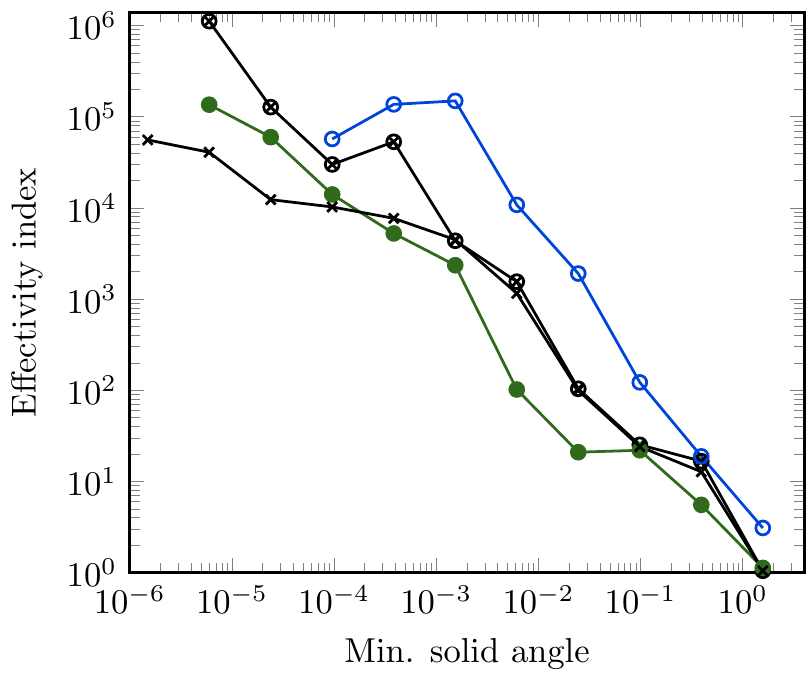}
\caption{Absolute value of the effectivity index of the forward, goal-based error metric vs solid angle of the minimum angular element present after each adapt step (the adapt process moves from right to left on the $x$ axis). The x, $\otimes$, \textcolor{matlabblue}{o} and \textcolor{foliagegreen}{$\CIRCLE$} markers are the goal-based non-standard Haar adapts from Figures \ref{fig:dogleg_convg_nonstand} \& \ref{fig:dogleg_time_nonstand}}
\label{fig:dogleg_effec}
\end{figure}

As mentioned above and in \secref{sec:Goal-based adaptivity}, although our goal-based error metric is triggering refinement correctly in regions that contribute to the error in our goal (and hence the error in our adaptive simulations is decreasing with low enough target error), it is likely that the effectivity index of our metric is poor/conservative. \fref{fig:dogleg_effec} shows the absolute value of our forward effectivity index plotted for the non-standard Haar decompositions from \fref{fig:dogleg_convg_nonstand}, given the minimum solid angle present in the adapted discretisation. We can see that our computed effectivity index is pathologically conservative (the adjoint effectivity index matches the forward almost exactly, as it should). Our effectivity index is always greater than one, indicating it is conservative, but ideally it should stay very close to one. 

This pathological behaviour is because the true error reached by our adapted simulations is decreasing (again see \fref{fig:dogleg_convg_nonstand}), while the approximate error we compute is not. Our approximate error converges to a fixed value given \eref{eq:disc_resid}, which simply results in a scaled sum of the wavelet coefficients of our adapted solution $\bm{\psi}$, and we know that in the limit of infinite refinement, the coarse wavelets dominate this sum (given the coefficient of highly refined wavelets will be small). Of course, this behaviour is expected given the poor reduced accuracy residual we use in Section \ref{sec:Error metrics}. Perhaps the surprising fact is that our error metric still works so well in identifying regions that require refinement/coarsening. We leave the improvement of this error metric to future work, but emphasise that scalability must come before effectivity; a perfectly effective error metric would be useless in difficult problems as we require an arbritrary number of adapt steps and hence must calculate the metric many times. 

\begin{figure}[th]
\centering
\subfloat[][$x=0, y=18$]{\label{fig:dogleg_flux_centre_source_6}\includegraphics[width =0.4\textwidth]{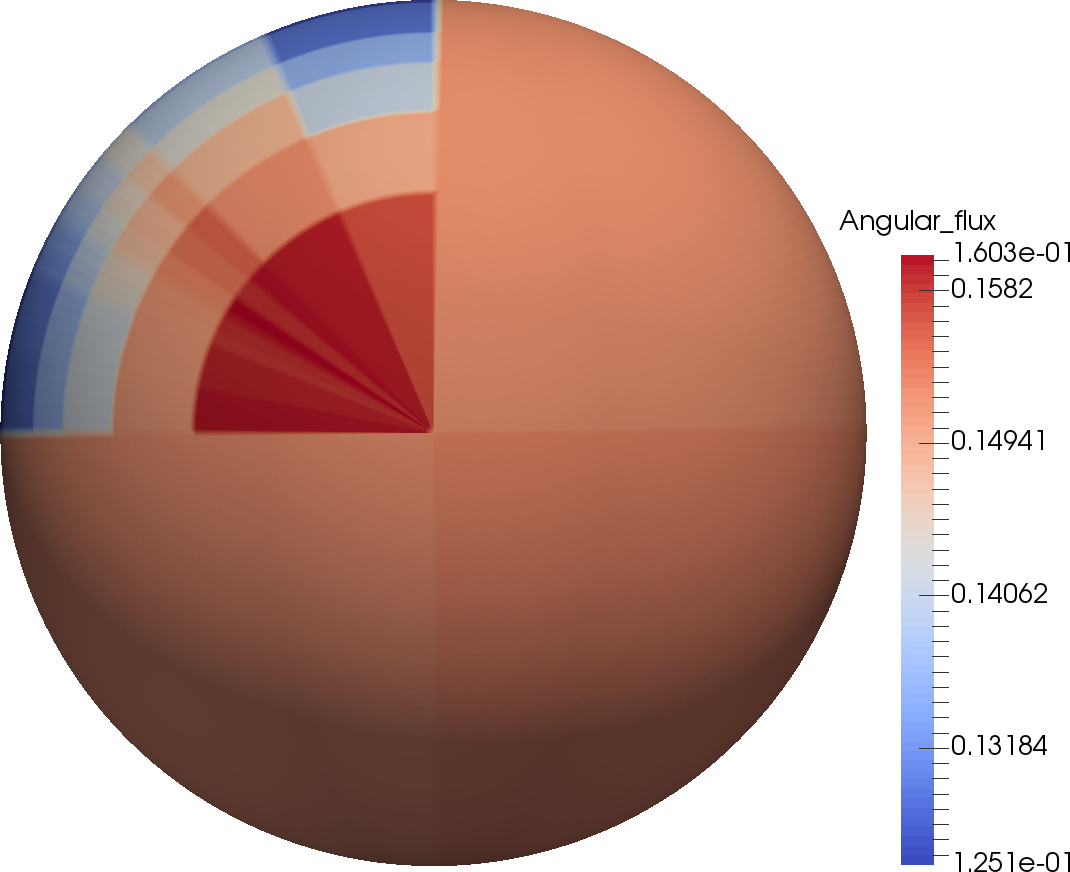}} \hspace{0.5cm}
\subfloat[][$x=-6, y=30$]{\label{fig:dogleg_flux_near_end_6}\includegraphics[width =0.4\textwidth]{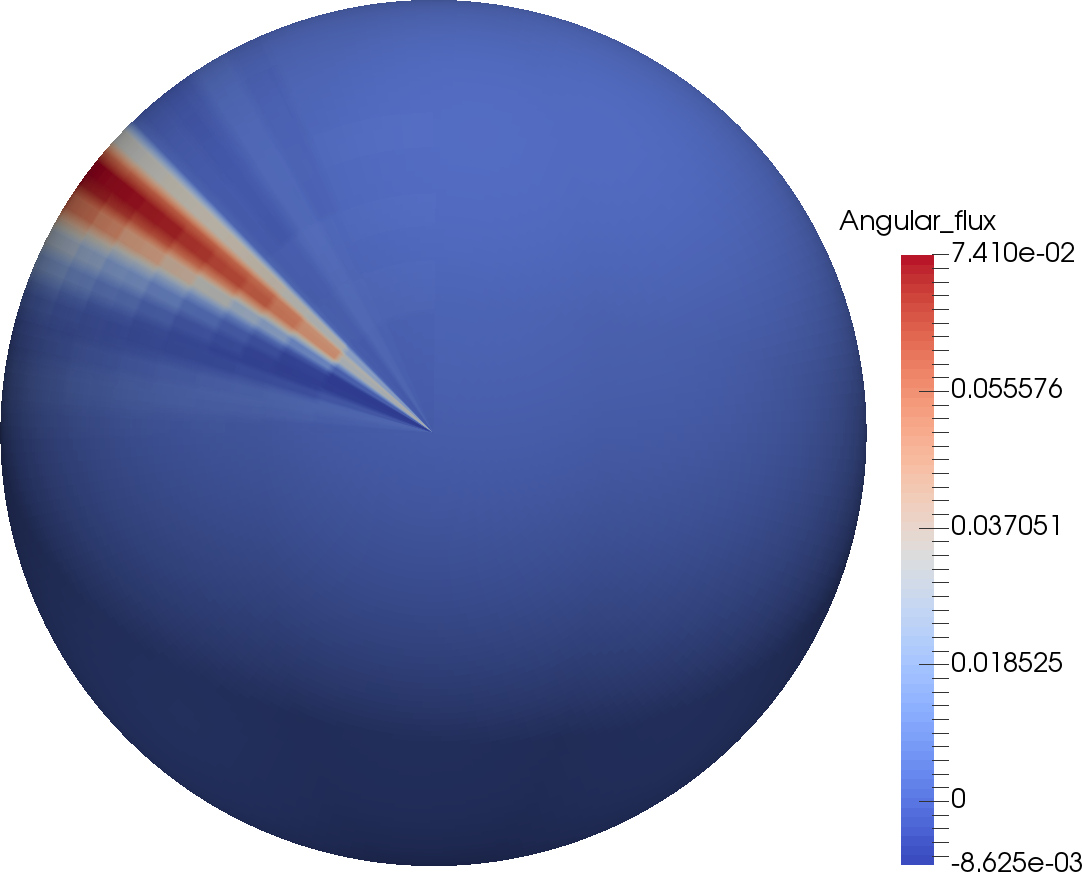}}\\
\caption{Forward angular flux in the 2D dogleg problem at different spatial positions, with goal-based angular adaptivity given a standard Haar decomposition after 6 adapt steps, with error target 1\xten{-3}. The camera is pointed in the $-z$ direction.}
\label{fig:dogleg_flux}
\end{figure}

\fref{fig:dogleg_no_angles} shows the spatial distribution of adapted angles, for both the standard and non-standard Haar decompositions after 9 adapt steps with an error target of 1\xten{-3}. We can see that both methods have placed angles in the streaming path between the source and the detector, as expected. Similar to \secref{sec:Brunner lattice}, the standard decomposition has used less DOF with the same target error, again this is due to the suitability of the ``long, thin'' wavelets in representing the angular symmetry that occurs two spatial dimensions. We have also plotted the angular flux after 6 adapt steps, in two different spatial positions for the standard decomposition in \fref{fig:dogleg_flux}.  The first position shown in \fref{fig:dogleg_flux_centre_source_6} is within the source and we can see that very little refinement has been triggered, as the flux is fairly uniform (any asymmetry in the flux comes from our coarse, unstructured spatial mesh). In contrast, \fref{fig:dogleg_flux_near_end_6} shows the flux within the duct region, and we can see heavily anisotropic distribution, and hence heavy angular refinement to capture this. Three of the visible quadrants of the sphere have not refined at all. 
\begin{figure}[th]
\centering
\subfloat[][Error vs CDOFs]{\label{fig:dogleg_convg}\includegraphics[width =0.47\textwidth]{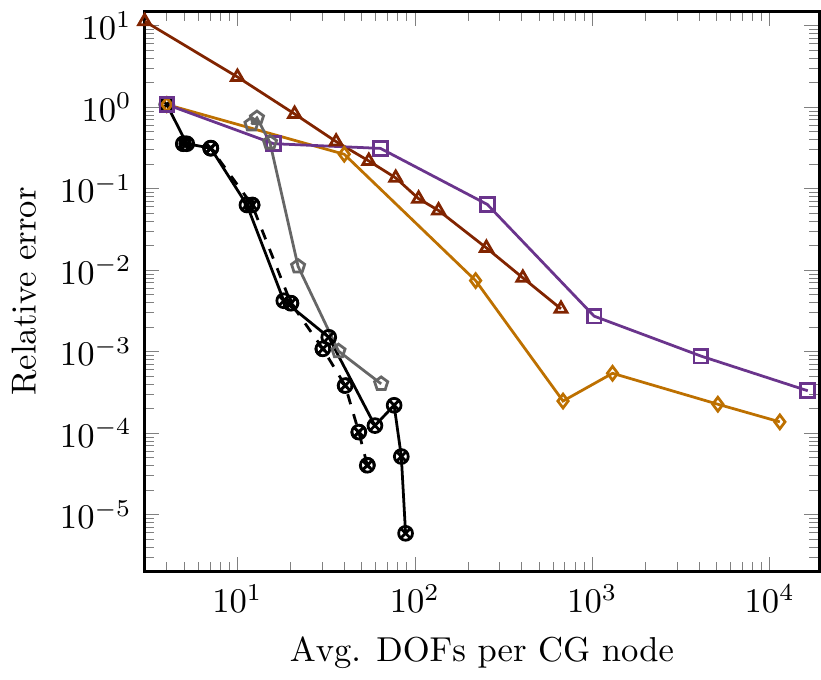}}
\subfloat[][Error vs mesh length]{\label{fig:dogleg_convg_mesh_length}\includegraphics[width =0.47\textwidth]{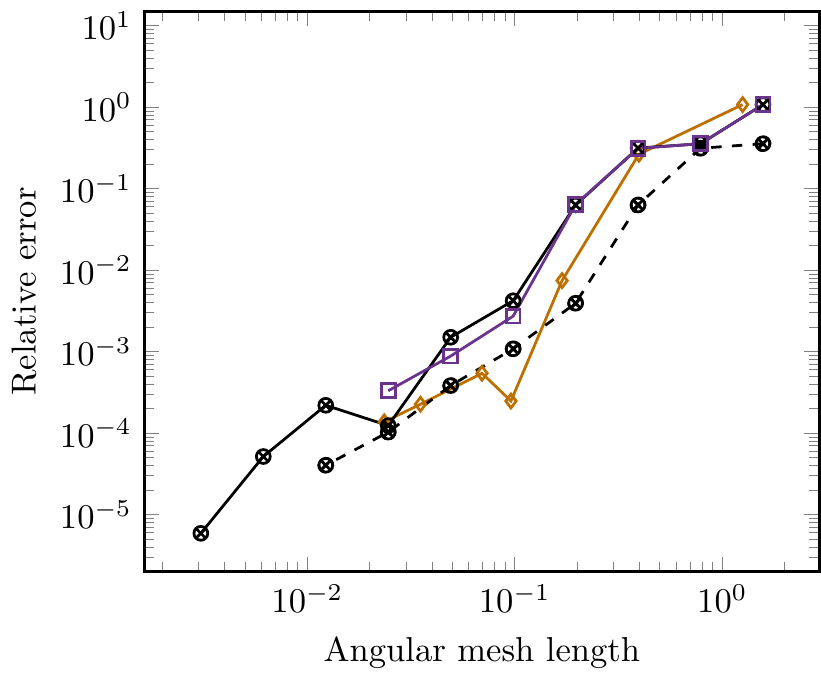}}\\
\subfloat[][Error vs total runtime]{\label{fig:dogleg_time}\includegraphics[width =0.47\textwidth]{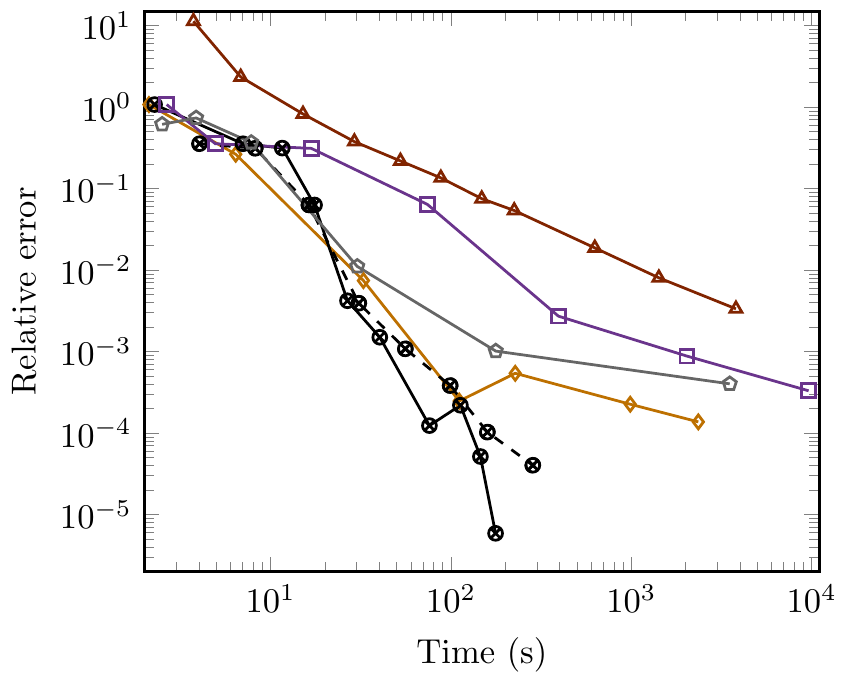}}
\caption{Comparison of the relative error of the detector response, for different angular discretisations, for the 2D dogleg problem. The $\otimes$ markers are goal-based Haar adapts with error target 1\xten{-3} and one extra adapt step. The dashed line is the standard Haar decomposition and the solid the non-standard (from Figures \ref{fig:dogleg_convg_adapt} \& \ref{fig:dogleg_time_adapt}). The \textcolor{fireenginered}{$\bigtriangleup$} is uniform P$_n$, \textcolor{deludedorange}{$\diamond$} is uniform LS P$_0$ FEM and \textcolor{dark}{$\pentagon$} is goal-based adapts with linear octahedral wavelets and error target 1\xten{-3} \cite{Goffin2015, Goffin2015a}.}
\label{fig:dogleg_result}
\end{figure}

Finally, \fref{fig:dogleg_result} shows a comparison with other angular discretisation methods for this problem. We have chosen to plot only the most effective standard and non-standard Haar simulations, which in both cases was the extra settling step/reduced iterative tolerance with target error of 1\xten{-3}. In both DOF and runtime, our adaptive Haar method outperforms all other angular discretisations. Again we see that a uniform LS P$_0$ FEM outperforms our uniform Haar discretisation, but our adapted Haar results converge with many orders of magnitude fewer DOF. As expected, \fref{fig:dogleg_convg_mesh_length} shows that all the low-order methods converge similarly with the angular mesh length. In contrast to \fref{fig:brunner_convg_mesh_length} in the Brunner problem, we can see that the different Haar simulations converge slightly differently given the same angular mesh size. This may be due to ray-effects in the solution, as the presence/absence of some angular mesh elements can lead to differences in error due to the oscillatory solution surrounding each ``ray'' (and subsequent cancellation in integral quantities like our detector response). Least-squares fits to the results in \fref{fig:dogleg_convg_mesh_length} however, do show similar order of convergence, with uniform S$_n$ at 2.51, and non-standard Haars at 1.98. The contrast between \fref{fig:dogleg_convg} and \fref{fig:dogleg_convg_mesh_length} emphasises that our adaptive method is merely placing DOF optimally in our problem, it is not increasing the order of convergence when compared to other low-order methods. 
\begin{table}
\centering
\begin{tabular}{ l c c c c c c c c c c}
\toprule
Adapt step: & 1 & 2 & 3 & 4 & 5 & 6 & 7 & 8 & 9 & 10\\
\midrule  
Cum. runtime ($\mu$s) per final DOF: & 57 & 130 & 144 & 134 & 121 & 97.2 & 94.6 & 102 & 119 & 136 \\
Peak memory use: & 254.6 & 199.2 & 155.9 & 95.3 & 66.4 & 48.8 & 35.2 & 31.3 & 30.72 & 30.67\\
\bottomrule  
\end{tabular}
\caption{Runtime and peak memory used for the 2D dogleg problem, for the non-standard Haar decomposition with extra adapt step and error target 1\xten{-3}. Peak memory use is on the heap (measured by massif) scaled to the size of the angular flux. The runtime is the cumulative runtime of all adapt steps up to that level, scaled by the NDOFs in the final adapt step.}
\label{tab:dogleg_memory}
\end{table}

This highly anisotropic problem shows the need for low-order adaptive methods, as we can see in \fref{fig:dogleg_convg} that the convergence of a spectral P$_n$ method has degraded to roughly that of the uniform P$_0$ methods per DOF. We should also note that as expected, the conditioning of the linear systems produced when using P$_n$ in this problem is worsening significantly with refinement, requiring many more iterations to solve, whereas the low-order methods remain fairly consistant. \fref{fig:dogleg_time} shows this, as even coarse P$_n$ simulations takes longer to compute than the low-order methods. 

Once again, we can see in \fref{fig:dogleg_convg} that the standard Haars use less DOF than the non-standard, but have larger runtimes with high levels of refinement. In particular, the runtime of the last adapt step of the standard Haars shown in \fref{fig:dogleg_time} is showing a different gradient to the gradient seen with DOF in \fref{fig:dogleg_convg}. This is the effect of the non-scalable sparse angular matrices at level 9 refinement and above. Furthermore, we also tested goal-based, adaptive linear octahedral wavelets in this problem \cite{Goffin2015, Goffin2015a}. \fref{fig:dogleg_convg} shows that we do not see any benefit from the linear wavelets per DOF, and importantly, we see the effect of the non-scalable components used as part of their construction in the runtime. \fref{fig:dogleg_time} shows that after only 6 levels of refinement, the runtime starts to increase significantly, making them impractical for problems with heavily anisotropic features. 

\tref{tab:dogleg_memory} shows the runtime and peak memory use during our adapt process, and again we can see that our memory use starts proportionally high, but settles down to around 30 copies of the adapted angular flux after 10 adapt steps. This is slightly less than double the memory use of the Brunner example (see \tref{tab:brunner_memory}), as we have to solve both the forward and the adjoint problem,  and we can reuse some of our data for both. Again, the memory use shown in \tref{tab:dogleg_memory} has not been heavily optimised (the same modification to the restart parameter and $\mat{D}^{-1}$ mentioned in the Brunner problem gives memory use at 19.6 copies), though \tref{tab:dogleg_memory} still shows that our memory use does not grow with adapt steps, giving scalable memory use with our goal-based adaptive algorithm. 

The cumulative runtime per final DOF starts low for the coarse angular discretisation, but \tref{tab:dogleg_memory} shows it stabilises at around 100--130$\mu$s. This is higher than that shown in \tref{tab:brunner_memory}, but the runtime in this problem includes the time spent solving adjoint problems. Also our iterative method takes more iterations to solve in this problem with angular refinement, in part because of the presence of heavy advection, but also because prior to reaching the asymptotic region, the initial condition provided to an adapt step from the previous adapt step may not be ``good'' given the presence of significant ray-effects, meaning the iterative method does not benefit from previous adapt steps. Even so, we can see that the cumulative runtime in \tref{tab:dogleg_memory} is fairly constant even with increasing number of adapt steps and hence our angular adaptivity in this problem is scalable.
\subsection{3D void problem}
\label{sec:3D duct}
\begin{figure}[th]
\centering
\includegraphics[width=0.5\textwidth]{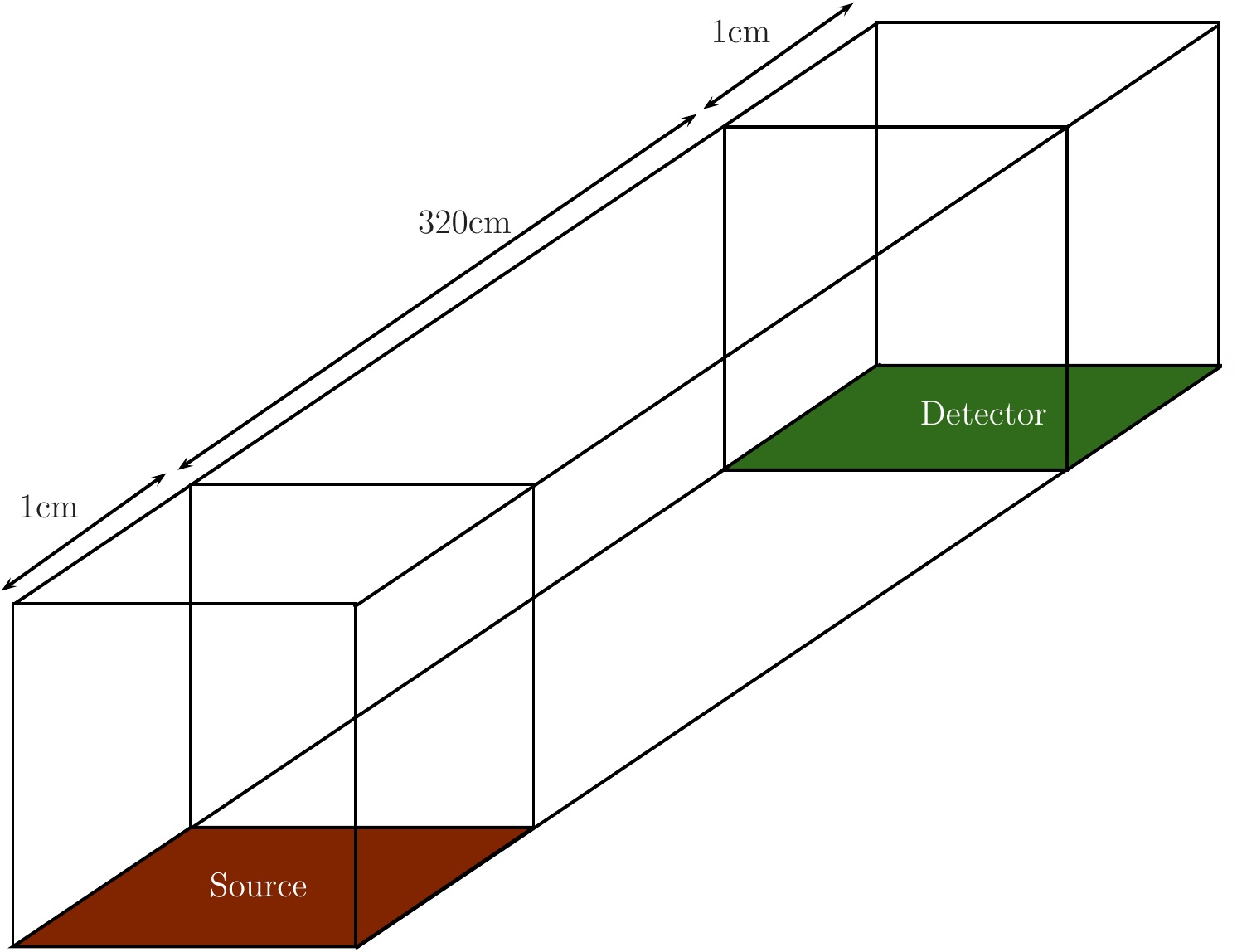}
\caption{Schematic of a source/detctor problem in a vacuum in 3D (the length of the duct is in the $y$ direction). All regions are pure vacuum (0 cm$^{-1}$), with the entire red cube region also a volumetric source of strength 1.}
\label{fig:duct_3D}
\end{figure}

The final example is both a challenging problem for anisotropic angular adaptivity and somewhat trivial, it features pure streaming down a long duct, a schematic of which is shown in \fref{fig:duct_3D}. We discretise this problem in space with a structured hexahedral mesh with 25,712 elements (40,200 CG nodes and 205,696 DG nodes). We used a structured mesh in this problem only to ensure the alignment of spatial nodes down the centre of the duct. This problem is challenging in that increasing the length of the duct, we require increasing levels of refinement to obtain a non-zero response in the detector. At a width:length ratio of 1:320, geometrically we require a minimum of 10 levels of refinement. We used fixed angular refinement in this problem, as goal-based adaptivity would fail, given the discussion in \secref{sec:Goal-based adaptivity}. Angular elements were refined identically across all points in space, to ensure the source could ``see'' the detector, giving refinement between $\mu \in [-0.005, 0.005]$ and $\omega \in [1.5707, 1.5709]$. In this sense, the problem is trivial, in that we know \textit{a priori} where angular refinement must occur, and given the lack of scatter, this problem could be solved with a single sweep. As such, we consider this problem an excellent test of the scalability of our adapt process, rather than the scalability of our iterative method, or the suitability of goal-based metrics for problems featuring pure streaming.

Given the highly anistropic flux, we did not run uniform angular discretisations in this problem, nor did we run the standard Haar decomposition, given that both the ``long, thin'' wavelets in 3D and the use of sparse angular matrices make this method non-scalable in this problem. The reference solution for the non-standard Haar decomposition used fixed angular refinement with a maximum of 14 levels of refinement, using 639M DOFs (uniformly this would have required 1 \xten{14} DOFs). Given we are using fixed angular refinement in this problem, we do not need to perform intermediate adapt steps to ``build-up'' the refined angular mesh; the results presented at each level of refinement require only a single linear solve at that level. Also, we solve this problem in parallel with 20 cores; we do not need to perform load balancing given we have the same number of angles at every spatial node. 

\begin{figure}[th]
\centering
\subfloat[][Absolute value of the scalar flux in the 3D duct along the length of the duct, from (0.5, 0.5, 0.5) to (0.5, 321.5, 0.5). The solid lines are shaded blue when they are entirely positive down the length of the tube. The pink line corresponds to the solution with 14 levels of refinement.]{\label{fig:tube_3D_flux}\includegraphics[width =0.47\textwidth]{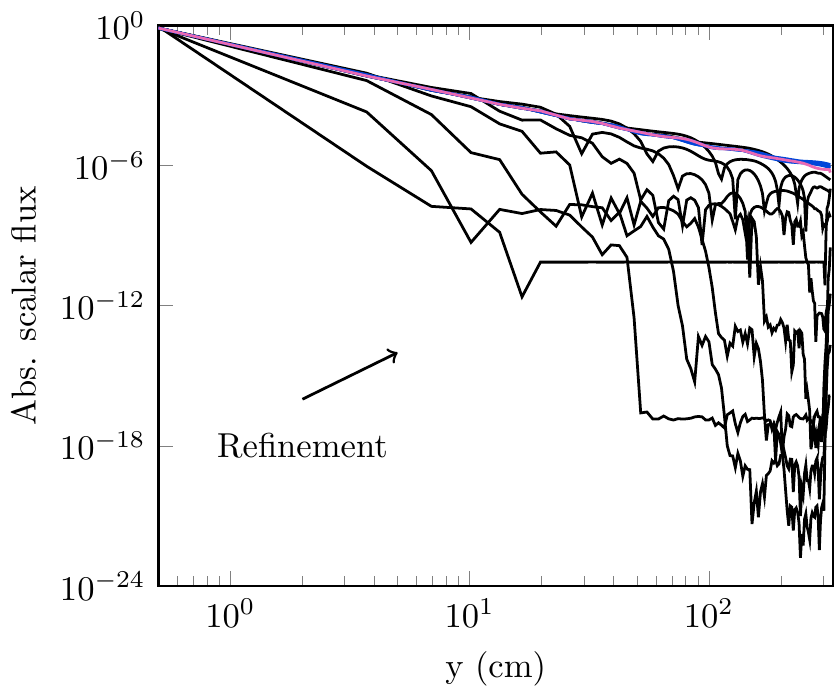}} \hspace{0.25cm}
\subfloat[][Absolute value of the average flux in the detector, wih the dotted line the analytic solution from a point source approximation.]{\label{fig:tube_3D_response}\includegraphics[width =0.47\textwidth]{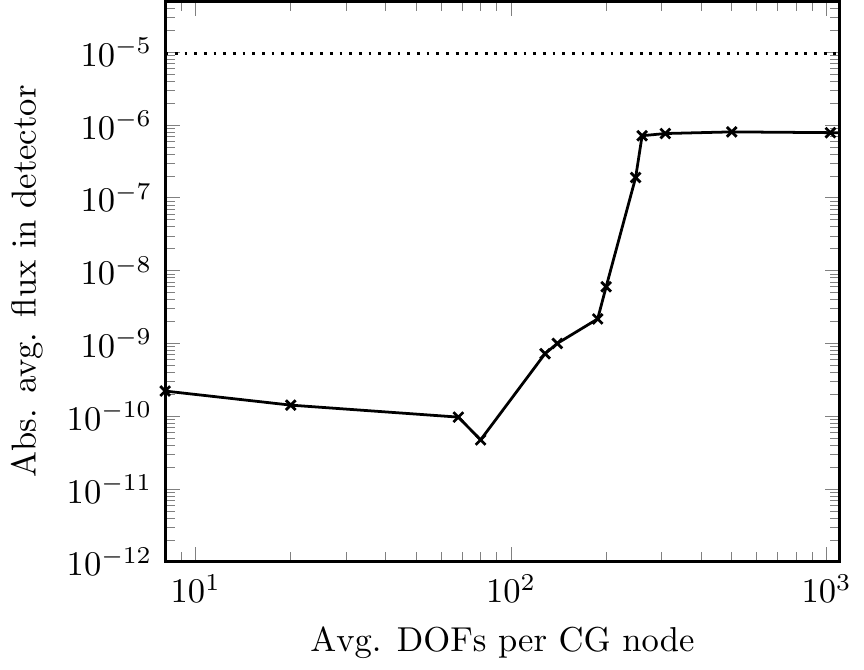}}\\
\caption{Solution of the 3D duct problem, given non-standard Haar decomposition, with fixed angular refinement in a given angular region performed identically across space, from level 1 to 14.}
\label{fig:tube_3D_result}
\end{figure}

\fref{fig:tube_3D_result} shows the results from the fixed angular refinement, with \fref{fig:tube_3D_flux} showing the absolute value of the scalar flux down the centre of the duct. We can see heavy ray effects, which require many levels of angular refinement to resolve. The first level of refinement in which the solution is positive everywhere is the 10th level, which matches well with the geometric argument of when our angular discretisation can first ``see'' the detector. \fref{fig:tube_3D_flux} also highlights the flux with 14 levels of refinement, where the scalar flux is still slightly oscillatory at the end of the duct; least squares fitting a straight line to this solution gives a gradiant of -2.085, which matches the expected analytic radiative ``drop-off'' of $1/r^2$ well. \fref{fig:tube_3D_result} also plots the absolute value of the average flux in the detector; again this shows that we only reach the asymptotic regime after 10 levels of refinement. We have also plotted an ``analytic'' solution for the average flux on \fref{fig:tube_3D_result}, derived from integrating $1/r^2$ over the detector volume, assuming a point source positioned at $(0.5, 0.5, 0.5)$. Our numerical detector response is $\sim$12.5 smaller than the approximate analytic solution, and as \fref{fig:tube_3D_result} shows, does not continue converging towards this solution as our angular mesh is refined. Given our scalar flux down the duct behaves qualitatively correct with angular refinement, and as in previous examples, we have an under-resolved spatial discretisation combined with the error in our point source approximation, we believe our adapted scheme is behaving as expected.
\begin{figure}[th]
\centering
\subfloat[][Error vs CDOFs]{\label{fig:tube_3D_convg}\includegraphics[width =0.47\textwidth]{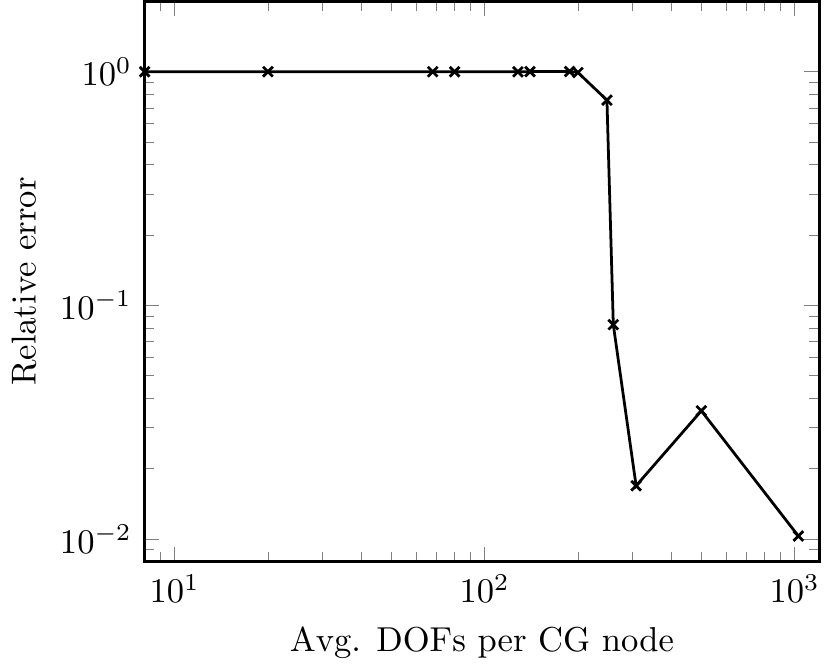}} \hspace{0.25cm}
\subfloat[][Error vs total runtime]{\label{fig:tube_3D_time}\includegraphics[width =0.47\textwidth]{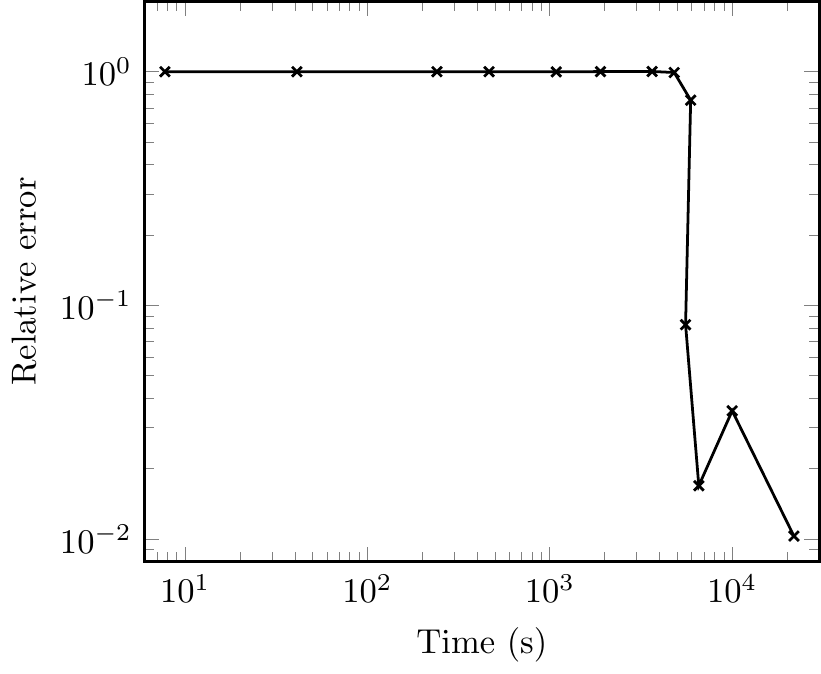}}\\
\caption{Convergence of fixed angular refinement in the relative error in our detector response, for the 3D duct problem, given non-standard Haar decomposition. Angular refinement is performed uniformly across space, from level 1 to 13.}
\label{fig:tube_3D}
\end{figure}

\fref{fig:tube_3D} shows the convergence of our fixed angular refinement towards our reference solution, with \fref{fig:tube_3D_convg} showing the convergence vs the number of angular DOFs per CG node, with \fref{fig:tube_3D_time} showing the converge vs the total runtime. Again we can see the necessity of angular refinement in this problem, as 13 levels of refinement only produces solutions with a relative error of $\sim$1 \xten{-2} in the detector response. We can see in \fref{fig:tube_3D_time} that although our runtime plot again looks similar to \fref{fig:tube_3D_convg}, in this case that shows that our runtime is increasing significantly with angular refinement in this problem, as we are only performing a single linear solve given the fixed angular refinement.
\begin{table}
\centering
\begin{tabular}{ l c c c c c c c c c c c c c c}
\toprule
Refinement level: & 1 & 2 & 3 & 4 & 5 & 6 & 7 & 8 & 9 & 10 & 11 & 12 & 13 & 14\\
\midrule  
DOFs per node & 8 & 20 & 68 & 80 & 128 & 140 & 188 & 200 & 248 & 260 & 308 & 500 & 1028 & 2600 \\
Time (s) & 7.8 & 41.0 & 240.5 & 464.1 & 1084 & 1897 & 3638 & 4810 & 5918 & 5552 & 6557 & 10034 & 21833 & 62094 \\
Runtime ($\mu$s) per DOF & 3.9 & 8.3 & 14.4 & 23.6 & 34.4 & 55.1 & 78.7 & 97.8 & 97.0 & 86.8 & 86.6 & 81.6 & 86.4 & 97.1 \\
Iterations & 25 & 47 & 91 & 141 & 211 & 331 & 460 & 553 & 555 & 486 & 490 & 471 & 485 & 501 \\
Rel. runtime per iter. & 1 & 1.13 & 1.002 & 1.06 & 1.04 & 1.06 & 1.09 & 1.12 & 1.11 & 1.13 & 1.12 & 1.1 & 1.13 & 1.23 \\
\bottomrule  
\end{tabular}
\caption{Performance of our iterative method for each adapt step shown in \fref{fig:duct_3D}. As we used fixed refinement in this problem, the runtimes shown are for a single linear solve. Number of iterations is shown to solve to absolute tolerance of 1\xten{-8}. The relative runtime per iteration computes runtime/(DOFs $\times$ Iterations) for each individual linear solve, relative to the linear solve at refinement level 1.}
\label{tab:duct_3D}
\end{table}

\tref{tab:duct_3D} shows the performance of our iterative method in this problem, when applying the fixed angular adaptivity. We can see that although our runtime per DOF settles down to around 90 $\mu$s with higher levels of refinement, this is around 30 times larger than the runtime per DOF for the coarsest angular discretisation. We can see that it is the increase in the number of iterations our iterative method requires that causes the increased runtime. Indeed our iterative method goes from 25 iterations to $\sim$500 with high levels of refinement; this is hardly surprising given our simple multigrid algorithm, whose performance would be expected to degrade in the present of pure advection. The relative runtime per iteration however, shows that our angular adaptivity scales very well with both number of DOFs and levels of refinement; each of the iterations of our iterative solver, at level 14 refinement, with 2600 angular elements per spatial node, only requires 1.23 times longer per DOF than the uniform, level 1 discretisation. This is a considerable achievement, given the uniform, level 1 discretisation allows for uniform memory access. This shows the importance of considering factors like those discussed in \secref{sec:Data structures} in order to build an efficient algorithm in practice. 
\section{Conclusions}
This paper has presented an angular adaptivity algorithm, based on a hierarchical P$_0$ FEM discretisation of the 2D angular domain, built with two different Haar wavelet decompositions. This algorithm was tested on several problems with two and three spatial dimensions, with a range of regularity. The angular discretisation was allowed to vary across all points in space independently, and refinement was performed not only in a fixed angular region, but also by using regular and goal-based error metrics. Both the regular and goal-based adaptive algorithms showed scalable runtimes and memory use, proportional to the number of DOFs applied by the adaptivity, when a non-standard 2D Haar wavelet discretisation was used. We believe this is a signficiant advance, as there is no existing work that shows evidence of scalable angular adaptivity in such a general setting. With that said, there is significant work remaining until we have a truly general and robust angular adapativity algorithm. The last example in this work shows one of the disadvantages of foregoing a sweep based algorithm, that is an substantial increase in iteration count in pure-advection problems, which is to be expected given our iterative method (a multigrid method with simple operators). The proportional cost of performing our adaptivity and a single iteration of our solver remains scalable, but overall our adaptive algorithm in this case could not be considered scalable. We have also not investigated the parallel performance of our iterative method, or the load-balancing that would be required to efficiently solve these adapted discretisations in parallel. Future work will investigate these issues, and also show the impact of using non-symmetric operators as part of our multigrid algorithm, which are specifically designed to tackle problems with strong advection. 

Our scalable goal-based error metric performs refinement/coarsening in correct regions of the angular domain, however the use of a simple diagonal matrix-vector product when computing a ``reduced-accuracy'' residual means that the effectivity index of our error metric is pathological; it will converge to a fixed value even as the true error in the discretisation decreases. Another disadvantage of our goal-based metric is that if a signal does not reach the detector/functional at coarse resolutions (e.g., due to ray effects), our metric will not trigger refinement. This is a problem faced by all goal-based metrics, and is often tackled in Boltzmann-tranport problems by using a surrogate solution method to ``bootstrap'' the adaptivity algorithm. Using a diffusion approximation (which does not suffer from ray effects) to produce a surrogate solution is common when performing spatial adaptivity, but this is not suitable for use with angular adaptivity. A diffusion approximation would produce an isotropic flux in angle and even a P$_0$ angular discretisation could reproduce this exactly with no angular refinement. 

Some of the methods mentioned in \secref{sec:Introduction} are sometimes used (e.g., first-collision source approximations) as a surrogate, though they all suffer from a lack of generality. Without a goal-based metric that is robust in the presence of ray-effects, regardless of the parameter regime, geomtry/alignment, etc, there is no guarantee that all signficiant streaming paths in a given problem will be resolved, effectively defeating the point of using goal-based angular adaptivity. Future work will show how careful choice of surrogate solutions to bootstrap our wavelet-based adapt can produce robust goal-based metrics.

Even with these shortcomings, we have shown evidence of scalable angular adaptivity, in both runtime and memory consumption for the first time in challenging problems. We adapt up to 15 levels of refinement, producing angular elements with a solid angle as small as 5 \xten{-9} sr. Performing these simulations with uniform angular resolution would require $\sim$1 billion angular elements on every spatial node in the problem, resulting in a linear system with $10^{13}$ DOFs in total. Perhaps the key message of this manuscript is that scalable angular adaptive algorithms for Boltzmann transport problems must be built on hierarchical, discontinuous FEM discretisations in angle with a compatible wavelet discretisations for refinement criteria that include both importance and smoothness. Using low-order basis functions lets us use quadratures which integrate general interaction/source terms exactly on each angular element. We can then refine elements to capture discontinuities, without introducing Gibbs oscillations from high-order interpolation. Errors in our discretisation of interaction/source terms and streaming operators can both be captured simultaneously and drive our adapt. The use of FWTs means that solving in wavelet space does not change the scalability of the underlying algorithm. If desired, only the error can be calculated in wavelet space, with the FWT allowing the system to be solved in the original DG FEM space and hence traditional sweeps/moment mappings algorithms can be used. Indeed, this work provides an easy migration path for performing angular adaptivity in existing nuclear codes that use sweeps and low-memory algorithms, while also providing a general purpose ``low-memory'' adaptive algorithm for use in other fields that use different source/interaction terms.

In a general sense, scalable angular adaptivity algorithms only form part of the picture in adequately resolving difficult Boltzmann transport problems. Combined space/angle/energy adaptivity is required in many problems, particularly given that discretisation errors affect all the spaces simultaneously (e.g., ray effects worsen with spatial refinement). Performing adaptivity in the entire phase-space is an ambitious goal, but would allow us to simulate problems that are simply not feasible using current deterministic technology. 
\section*{Acknowledgements}
The authors would like to acknowledge the support of the EPSRC through the funding of the EPSRC grant EP/P013198/1.





\section*{References}
\bibliographystyle{model1-num-names}
\bibliography{bib_library}







\end{document}